\documentclass[twocolumn]{aastex62}

\usepackage{amsmath}
\usepackage{lineno}

\shorttitle{Wave-Mean Flow Interactions on Tidally Locked Planets}
\shortauthors{Hammond \& Pierrehumbert}

\begin{document}

\title{Wave-Mean Flow Interactions in the Atmospheric Circulation of Tidally Locked Planets}

\correspondingauthor{Mark Hammond}
\email{mark.hammond@physics.ox.ac.uk}

\author{Mark Hammond}
\affiliation{University of Oxford}
\author{Raymond T. Pierrehumbert}
\affiliation{University of Oxford}


\begin{abstract}

We use a linear shallow-water model to investigate the global circulation of the atmospheres of tidally locked planets. Simulations, observations, and simple models show that if these planets are sufficiently rapidly rotating, their atmospheres have an eastward equatorial jet and a hot-spot east of the substellar point. We linearize the shallow-water model about this eastward flow and its associated height perturbation. The forced solutions of this system show that the shear flow explains the form of the global circulation, particularly the hot-spot shift and the positions of the cold standing waves on the night-side.

 We suggest that the eastward hot-spot shift seen in observations and 3D simulations of these atmospheres is caused by the zonal flow Doppler-shifting the stationary wave response eastwards, summed with the height perturbation from the flow itself. This differs from other studies which explained the hot-spot shift as pure advection of heat from air flowing eastward from the substellar point, or as equatorial waves travelling eastwards. We compare our solutions to simulations in our climate model Exo-FMS, and show that the height fields and wind patterns match.

We discuss how planetary properties affect the global circulation, and how they change observables such as the hot-spot shift or day-night contrast. We conclude that the wave-mean flow interaction between the stationary planetary waves and the equatorial jet is a vital part of the equilibrium circulation on tidally locked planets.

\end{abstract}



\section{Introduction}\label{sec:intro}

Tidally locked planets always present the same face to the star they orbit, so only receive starlight on one side. If such a planet has an atmosphere, its circulation transports heat from the permanent day-side to the permanent night-side. Simulations of such atmospheres show they have an eastward jet on the equator and a hot-spot to the east of their substellar point, rather than at the substellar point as might be expected. They also show a pair of cold low pressure lobes on the night-side, peaked in the mid-latitudes and symmetrically disposed about the equator. In this paper, we use a shallow-water model linearized about the equatorial jet to investigate this circulation, which has previously been interpreted to be composed of stationary planetary-scale waves \citep{showman2011superrotation} \citep{tsai2014three}. We show that the interaction between this jet and the stationary waves is crucial to understanding the global circulation.

 Previous studies suggested that the eastward hot-spot shift is caused by propagating Kelvin waves \citep{showman2011superrotation}, advection by the equatorial jet \citep{zhang2017dynamics}, or a Doppler-shift of the stationary Rossby waves \citep{tsai2014three}. \citet{showman2011superrotation} and \citet{heng2014analytical} investigated the global circulation using a linear shallow-water model with zero background flow, focusing on the eastward jet and standing waves. \citet{showman2011superrotation} compared their linear shallow-water model to their GCM simulations, but found the shallow-water model only matched the GCM runs a few days after they were spun up from rest, and not when they reached equilibrium. We suggest this was because the model in \citet{showman2011superrotation} does not include the effect of the zonal flow in equilibrium.  \citet{tsai2014three} included a uniform zonal flow $\bar{U}(y)=\bar{U}_{0}$, and showed how this Doppler-shifted the forced response, explaining the position of the maximum height -- the hot-spot.

 We built on these previous studies with a shallow-water model linearized about both a non-uniform equatorial jet $\bar{U}(y)$ and its associated height perturbation $\bar{H}(y)$. We used a pseudo-spectral method to solve these equations on an equatorial beta-plane and on a sphere, to find the stationary wave response to a day-night forcing. Our solutions agreed with the interpretation of \citet{tsai2014three} that the hot-spot shift is caused by the eastward jet Doppler-shifting the stationary waves eastwards.

We introduce the linearized shallow-water equations in Section \ref{sec:sw-equations}. In Section \ref{sec:forced-system} we use a pseudo-spectral method to solve the shallow-water equations linearized about $\bar{U}(y)$ and $\bar{H}(y)$ on an equatorial beta-plane \citep{boyd2000spectral} \citep{boyd2017equatorial}. We find the free modes of the system and its response to stationary forcing. This lets the response pattern be understood in terms of the spectrum of waves excited by the forcing. We explain how the jet shifts the wave response eastwards and changes the pattern of planetary standing waves. Our results match our simulations much better than previous linearized shallow-water calculations in zero or uniform background flow.

In Section \ref{sec:sphere-solutions} we solve the shallow-water equations linearized about $\bar{U}(\phi)$ and $\bar{H}(\phi)$ in a spherical coordinate system, which exactly represents the latitudinal direction, unlike the beta-plane. We compare this to our solutions on the beta-plane, and show that the solutions are similar despite the approximations of the beta-plane

Section \ref{sec:scaling-relations} shows how observables such as the hot-spot shift and day-night contrast depend on the properties of the planet and atmosphere. We derive simple one-dimensional scaling relations on the equator, and compare them to the effects of changing the parameters of the full two-dimensional shallow-water model.

Section \ref{sec:gcm-results} compares our linear shallow-water solutions to simulations in our General Circulation Model (GCM) Exo-FMS. We decompose its output to isolate the Rossby and Kelvin components for comparison with the linear shallow-water model. Our results show how the standing waves are Doppler-shifted as the zonal jet forms during the spin-up from rest, and match the linear shallow-water model when the GCM reaches equilibrium.

We conclude that linearizing about the eastward jet is vital to the shallow-water model of a tidally locked atmosphere. Our model gives a new interpretation of the main processes controlling the circulation on such planets.

\newpage
\section{The Atmospheric Circulation of Tidally Locked Planets}\label{sec:tl-atmospheres}

Planets such as gas giants in short-period orbits (``Hot Jupiters''), close-in terrestrial planets around G-stars (``lava planets''), and a range of terrestrial planets around M-dwarfs are all expected to be tidally locked. The atmospheres of these planets are very different to those in the Solar System. Their permanent day-side receives much more radiation than the permanent night-side, which drives a strong global circulation. Articles such as \citet{showman2012review}, \citet{heng2015review}, and \citet{pierrehumbert2018review} review the atmospheric circulation of tidally locked planets.

Shallow-water models \citep{showman2011superrotation} \citet{tsai2014three}, GCM simulations \citep{showman2012review}, and observations \citep{louden2015winds} \citep{crossfield2015observations} \citep{parmentier2017handbook} suggest that tidally locked atmospheres have a superrotating eastward equatorial jet and an eastward hot-spot shift.  \citet{showman2011superrotation} set out a theory of the global circulation of atmospheres of tidally locked planets. They focused on the initial forcing and planetary waves which formed an equatorial jet, and did not consider the effect of the jet on the waves after it had formed.

\citet{tsai2014three} discussed the effect of a uniform mean flow $\bar{U}_{0}$ on the equatorial waves, and identified how the jet Doppler-shifts the waves eastwards \citep{arnold2012superrotation}. They suggested that the hot-spot shift is caused by this Doppler-shift, rather than free wave propagation or advection of air from the substellar point. We build on their work by linearizing the shallow-water equations about a shear flow $\bar{U}(y)$ and a height perturbation $\bar{H}(y)$, which are in geostrophic balance or gradient wind balance. \citet{boyd1978shearI} and \citet{wang1996shear} solved the linear shallow-water equations for planetary waves in shear flows, and showed how the shear affects the latitudinal extent of the waves. The shear flow in our linear solutions affects both the latitudinal and longitudinal structure of the forced response. We reach the same conclusions as \citet{tsai2014three} about the mechanism of the hot-spot shift, but our linear model matches our GCM results more closely, as the sheared flow and the jet height perturbation have a large effect on the response to stationary forcing.

\begin{figure*}
 \gridline{\fig{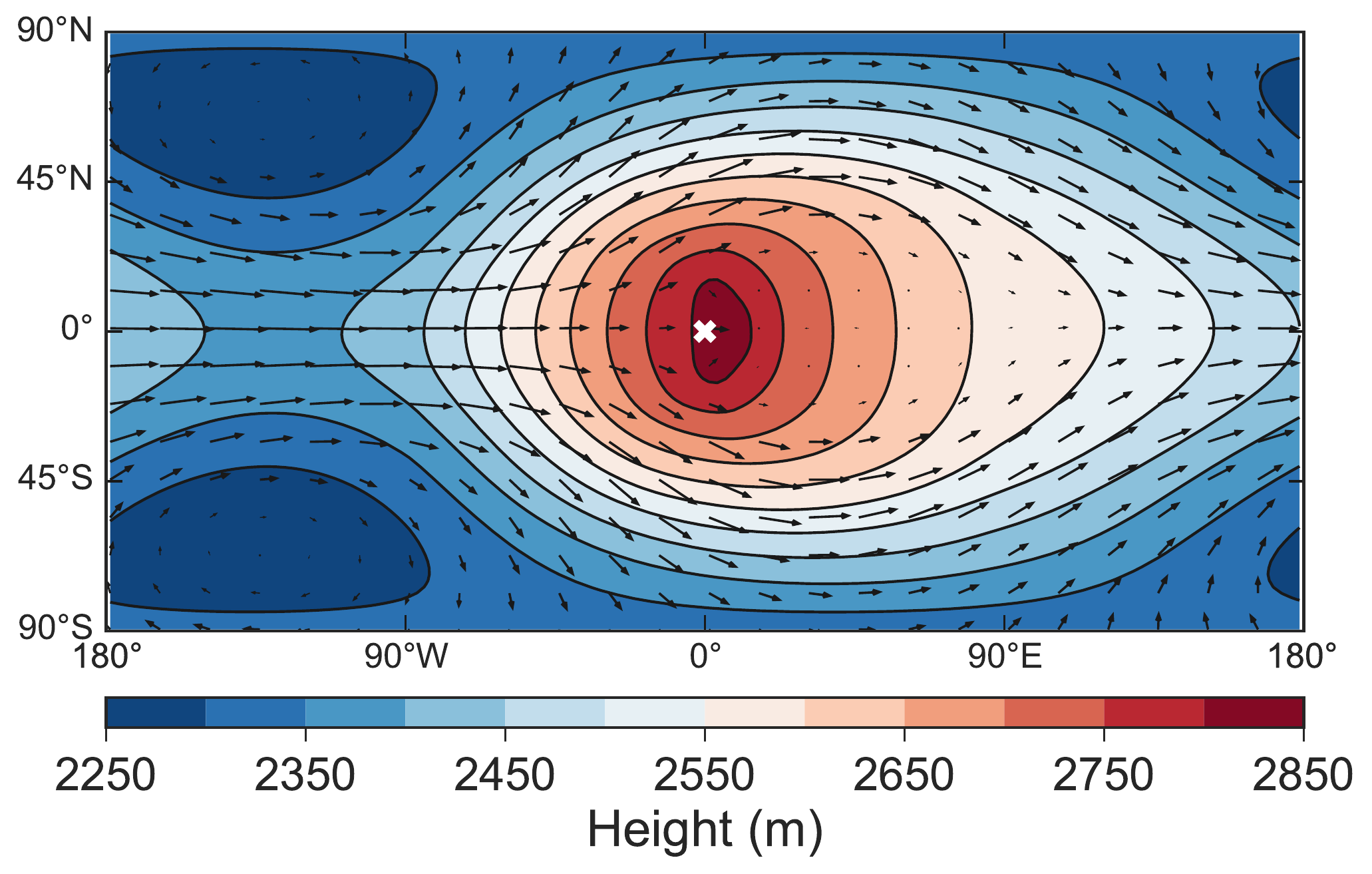}{0.33\textwidth}{GCM simulation of the atmosphere of a tidally locked planet, showing the hot-spot shift in the mid-atmosphere height field.}
 \fig{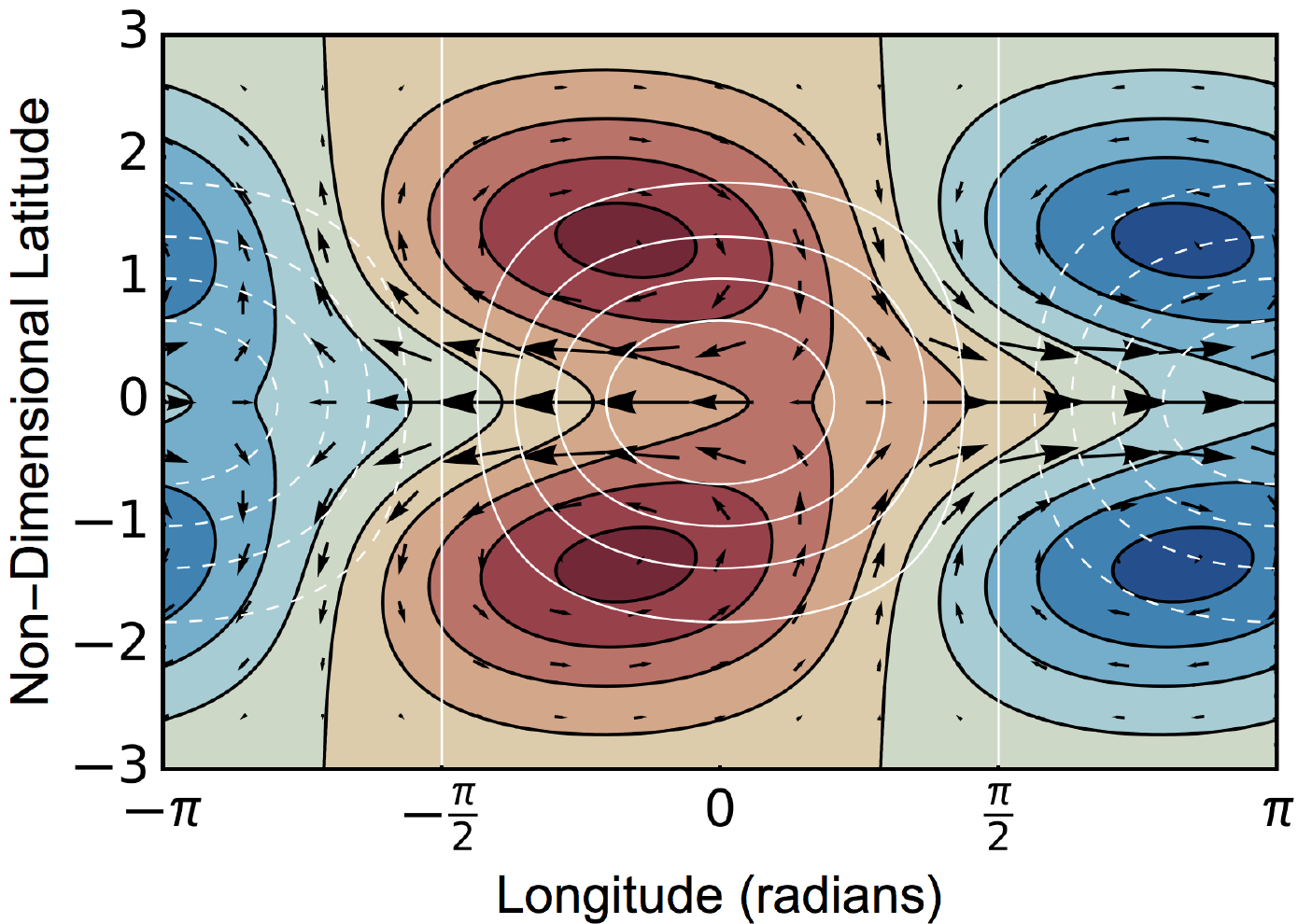}{0.33\textwidth}{The height field in the analytic linear model in \citet{showman2011superrotation}, which explained the eastward equatorial acceleration.}
 \fig{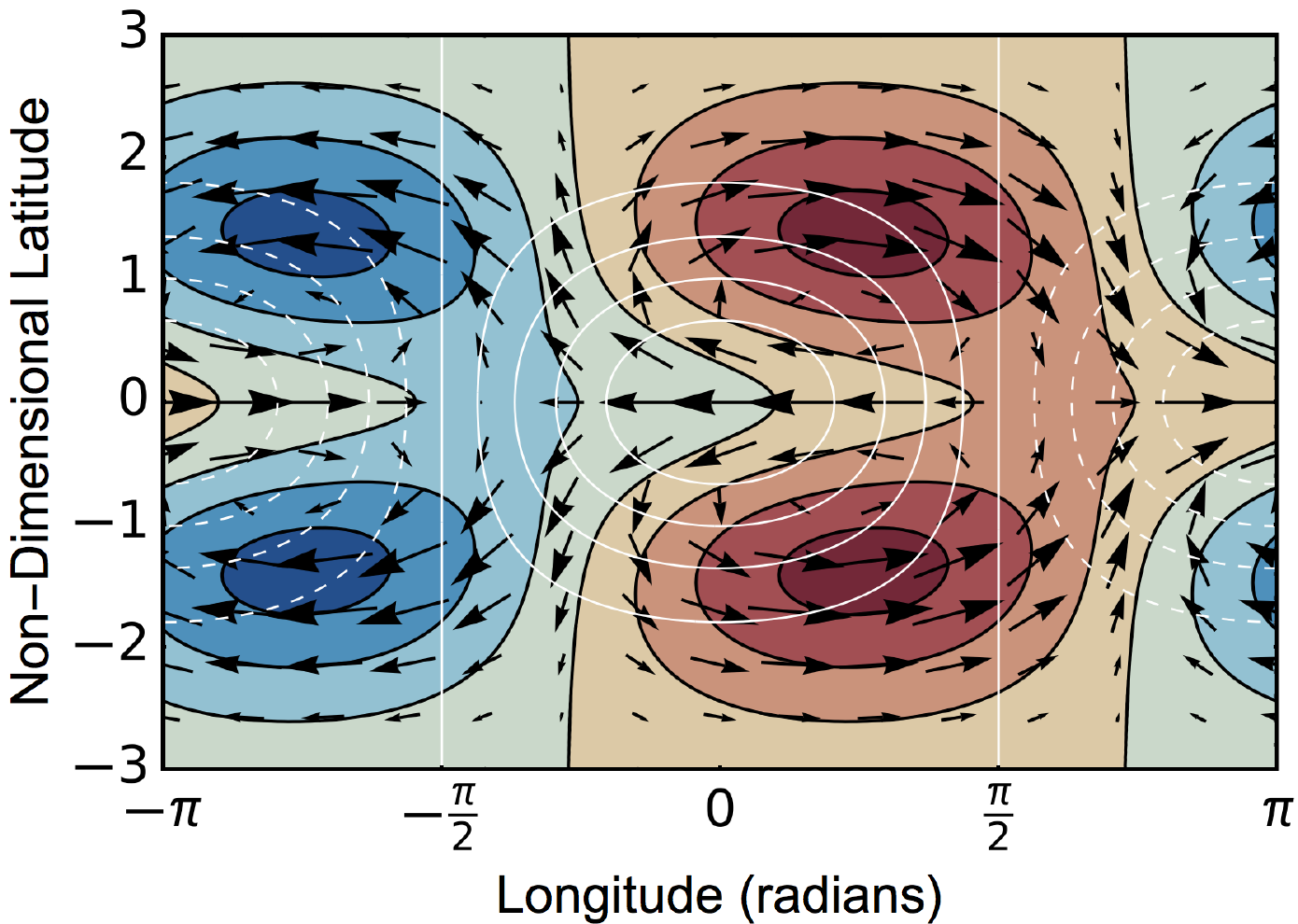}{0.33\textwidth}{The height field in the linear model in \citet{tsai2014three} which showed how a uniform flow $\bar{U}_{0}=1.0$ shifts the forced response.}
}
\caption{Summary of previous work on this topic, showing how previous linear shallow-water models (the second and third plots) did not fully explain the equilibrium circulation in simulations of such planets. Areas of high height are red, and areas of low height are blue. In the first plot, the substellar point is the white cross at $0 \degr$ latitude and $0 \degr$ longitude. In the second and third plots, the forcing $Q$ is overlaid in white, with solid contours for positive forcing and dashed contours for negative forcing.}\label{fig:motivating-plot}
\end{figure*}

Figure \ref{fig:motivating-plot} summarises the problem that we aim to solve in this study. The first plot shows typical GCM results of the mean height (analogous to temperature here) for a tidally locked Earth-sized planet. The global circulation is similar to that seen in Hot Jupiter simulations, e.g. \citet{showman2015circulation}. The second plot reproduces the linear shallow-water model of \citet{showman2011superrotation}. This linear model did not match equilibrated GCM results similar to those in Figure \ref{fig:motivating-plot}, but did match their GCM a short time after it was started from rest. Specifically, the model did not match the positions and wind directions of the hot (high height) and cold (low height) parts of the GCM results. The third plot reproduces the linear shallow-water model in \citet{tsai2014three} which showed how a uniform background flow $U_{0}$ could shift the maxima and minima of the forced response. This Doppler-shifted response matches the position of the height maximum (i.e. the hot-spot in the GCM). Our shallow-water model linearized about a shear flow $\bar{U}(y)$ and associated height perturbation $\bar{H}(y)$ matches  the overall form of our equilibrated GCM results. We use the new model to predict how the global circulation depends on the planetary parameters like rotation rate or damping rate.

Understanding the mechanism behind the circulation on these planets is crucial to interpreting observations of them. \citet{komacek2016daynightI}, \citet{komacek2017daynightII}, and \citet{zhang2017dynamics} developed a simplified model of the circulation on a tidally locked planet using a one-dimensional balance of advection and damping, to predict how the day-night temperature difference and the hot-spot shift scale with planetary parameters. They tested this against GCM results and observations, and found that their scaling worked well. The advective model of these studies has the potential to explain the equatorial hot-spot shift, but does not contain the wave dynamics required for the off-equator response. This off-equator response can significantly affect the phase curve -- for example, the coldest parts of the circulation at the level of the jet are often the night-side off-equator Rossby lobes. In this paper, we suggest that the wave dynamics are important, as in our linear model, the hot-spot shift is caused by a Doppler-shift of the the wave response (see Section \ref{sec:forced-system}). In Section \ref{sec:scaling-relations} we discuss how our wave-based approach leads to some of the same scaling relations identified in \citet{zhang2017dynamics}, but with a different physical basis.

 The global circulation on tidally locked planets has been shown to affect habitability and climate stability \citep{kite2011instability} \citep{yang2013clouds} \citep{kopparapu2017moist}. We hope to show that simple atmospheric models are key to understanding the processes driving the global circulation on these planets, and to interpreting observations of them.

\section{Linearized Shallow-Water Equations in Zero Flow on the Beta-Plane}\label{sec:sw-equations}

We used the linear shallow-water equations on a one-layer equatorial beta-plane to model the atmosphere of a tidally locked planet. These equations describe the motion of a single layer of fluid of constant density where the horizontal scale of its flow is much greater than the depth of the fluid. The linear form of these equations describe small perturbations to this layer \citep{vallis2006book}. We model the atmosphere of a tidally locked planet with a similar shallow-water model to \citet{showman2011superrotation}. The model corresponds to an active upper layer following the single-layer shallow water equations, above a quiescent layer which can transport mass and momentum to and from the upper layer. The forcing due to stellar heating is represented by $Q$, a relaxation to the radiative equilibrium height field.

In this section, we derive the wave response to stationary forcing on the beta-plane \citep{matsuno1966quasi}. The beta-plane system approximates the Coriolis parameter as linear, which is only accurate at low latitudes but leads to more intuitive and useful solutions than the full spherical geometry. We solve the equations in a spherical geometry in Section \ref{sec:sphere-solutions}, and show that the beta-plane approximation leads to very similar solutions, as in other studies of the atmospheres of tidally locked planets \citep{showman2011superrotation} \citep{heng2014analytical}.

All the quantities are linearized as the sum of a zonally mean background value $F(y)$  and a perturbation with the form  $f(y) e^{i( k_{x} x-\omega t)}$ (unlike \citet{matsuno1966quasi}, who uses the less conventional form  $f(y) e^{i( k_{x} x+\omega t)}$). Throughout this paper, we will use capital letters for mean zonal quantities such as $\bar{U}$ and $\bar{H}$, and lower-case letters for perturbations to this background, such as $u$ and $h$ (unless otherwise specified, such as the forcing $Q$). The shallow-water equations for these perturbations with zero background flow are:

\begin{equation}\label{eqn:sw-eqns-1}
  \begin{gathered}
     \frac{\partial u}{\partial t} - \beta y v +\frac{\partial h}{\partial x} = 0 \\
      \frac{\partial v}{\partial t} + \beta y u + \frac{\partial h}{\partial y} = 0 \\
    \frac{\partial h}{\partial t} +c^{2}(\frac{\partial u}{\partial x} + \frac{\partial v}{\partial y}) = 0 \\
  \end{gathered}
\end{equation}

where $h$ is the height, $c = \sqrt{gH}$ is the gravity wave speed \citep{matsuno1966quasi}, and there is no friction or damping. Non-dimensionalizing with time scale $\sqrt{1/c \beta}$ and length scale $\sqrt{c/\beta}$ (the equatorial Rossby radius of deformation $L_{R}$), and assuming all quantities have the form $f(y) e^{i( k x-\omega t)}$, the free equations are:

\begin{equation}\label{eqn:sw-eqns-2}
  \begin{gathered}
      - i \omega u - y v + i k_{x} h = 0 \\
      - i \omega v + y u + \frac{\partial h}{\partial y} = 0 \\
      - i \omega h + i k u + \frac{\partial v}{\partial y} = 0 \\
  \end{gathered}
\end{equation}

\subsection{Free Solutions in Zero Flow}

The dispersion relation for the free modes is:

\begin{equation}\label{eqn:matsuno-dispersion}
\omega ^ { 2} - k_{x} ^ { 2} - \frac { k_{x} } { \omega } = 2n + 1
\end{equation}

and the free modes $\xi$ are:

\begin{equation}\label{eqn:matsuno-solutions}
  \begin{pmatrix}
    v \\
    u \\
    h
\end{pmatrix}  = \begin{pmatrix}
  i(-\omega_{nl}^{2}-k_{x}^{2})\psi_{n} \\
  \frac{1}{2}(-\omega_{nl}-k_{x})\psi_{n+1}+n(-\omega_{nl}+k_{x})\psi_{n-1}\\
  \frac{1}{2}(-\omega_{nl}-k_{x})\psi_{n+1}-n(-\omega_{nl}+k_{x})\psi_{n-1}
  \end{pmatrix}
\end{equation}

where $\psi_{n} = e^{-\frac{1}{2}y^{2}}H_{n}(y)$ (the parabolic cylinder functions, see Appendix \ref{sec:app-beta}). In these solutions, $l$ marks the three roots of each mode denoted by $n$ \citep{matsuno1966quasi}.

\subsection{Forced Solutions in Zero Flow}

\citet{matsuno1966quasi} derives the stationary response to forcing $Q$ with a damping rate $\alpha$. We introduce forcing and damping terms and then impose a sinusoidal dependence on $x$ as before:

\begin{equation}\label{eqn:sw-eqns}
  \begin{gathered}
  \alpha_{dyn} u - yv + ik_{x} h = F_{x} \\
  \alpha_{dyn} v + yu + \frac{\partial h}{\partial {y}} =  F_{y} \\
  \alpha_{rad} h + ik_{x}u +  \frac{\partial v}{\partial y} = Q \\
  \end{gathered}
\end{equation}

for constant non-dimensional forcing $\sigma = (F_{x},F_{y},Q)$, dynamical damping $\alpha_{dyn}$, and radiative damping $\alpha_{rad}$. \citet{matsuno1966quasi} sets $\alpha_{dyn}=\alpha_{rad}=\alpha$ and imposes a Gaussian forcing on the height $h$ to give an exact solution:

\begin{equation}
 \sigma = \begin{pmatrix}
 0  \\
 0 \\
Q_{0} e^{-\frac{1}{2} y^{2}}
 \end{pmatrix}
\end{equation}

The boundary conditions are:

\begin{equation}
  u, v, h \rightarrow 0 \quad \textrm{for} \quad y \rightarrow \pm \infty
\end{equation}

\citet{matsuno1966quasi} shows that a forced solution $\chi$ can be expressed as a sum of the free solutions $\xi$:

\begin{equation}\label{eqn:forced-sw-solutions}
  \chi = \Sigma a_{m} \xi_{m}
\end{equation}

where

\begin{equation}\label{eqn:sw-eqns-projection}
  \begin{gathered}
    a_{m} = \frac{1}{\alpha - i  \omega_{m}} b_{m} \\
    b_{m} = \frac{\int \xi_{m}(y) \sigma (y) dy}{\int \lvert \xi_{m} (y) \rvert ^{2} dy}
  \end{gathered}
\end{equation}

The second plot in Figure \ref{fig:motivating-plot} shows this forced response, given the sum of the free modes weighted by their response coefficients $a_{m}$. We used an equal radiative and dynamical damping rate $\alpha = 0.2$ (this equality is relaxed later on), and forcing magnitude $Q_{0}=1$  to match \citet{matsuno1966quasi}. This is also consistent with the range of values used in \citet{showman2011superrotation} to represent a Hot Jupiter.

The even parity forcing $Q$ only excites even modes in $u$ and $h$, and odd modes in $v$, which is $\pi$ out of phase with $u$ and $h $ in the shallow-water equations. The Rossby and Kelvin modes dominate the forced response as they have the lowest frequencies, so are quasi-resonant with the stationary forcing with zero frequency.

Equation \ref{eqn:project-coeff-flow} shows that a uniform background flow can Doppler-shift the frequency $\omega_{m}$, changing the imaginary part of the wave response which shifts its longitude. The third plot in Figure \ref{fig:motivating-plot} shows how the forced response in the second plot changes when Doppler-shifted by an eastward zonal flow.

\subsection{Formation and Effect of Zonal Flow}\label{sec:form-effect-zonal-flow}

These standing waves can transport momentum through the atmosphere. The zonal momentum equation predicts that the zonal acceleration depends on the gradient of the stationary wave response \citep{andrews1976acceleration}. \citet{showman2011superrotation} show that this acceleration is:

\begin{equation}\label{eqn:sw-jet-accn}
\begin{gathered}
    \frac{\partial \bar{U}(y)}{\partial t} = \bar{v}^{*}(f-\frac{\partial \bar{u}}{\partial y})-\frac{1}{\bar{h}} \frac{\partial ( (\overline{hv)^{\prime}u^{\prime}})}{\partial y}\\ + (\frac{1}{\bar{h}} \overline{u^{\prime}Q^{\prime}} + \overline{R}_{u}^{*}) - \frac{\bar{u}^{*}}{\tau_{\mathrm{drag}}} - \frac{1}{\bar{h}}\frac{\partial (\overline{h^{\prime}u^{\prime}})}{\partial t}
\end{gathered}
\end{equation}

where overbars are zonal means, primes are deviations from zonal means, and the ``thickness-weighted zonal average of any quantity A'' is $\overline { A } ^ { * } \equiv \overline { h A } / \overline { h }$ \citep{showman2011superrotation}. $R_{u}$ is the zonal component of the vertical momentum transport $\mathbf{R}$:

\begin{equation}
\mathbf { R } = \left\{ \begin{array} { l l } { - \frac { Q \mathbf { v } } { h } , } & { Q > 0 } \\ { 0 , } & { Q < 0 } \end{array} \right.
\end{equation}

\begin{figure}
 \plotone{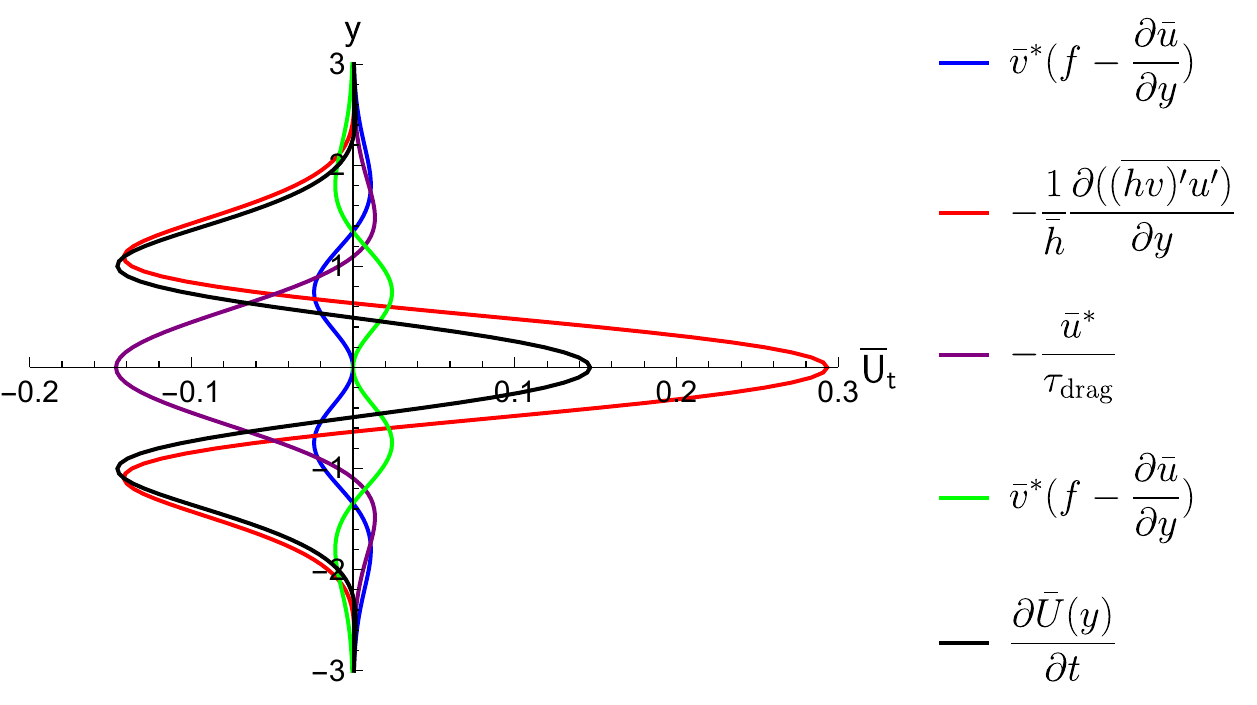}
\caption{Zonal acceleration profile in zero background flow, calculated using equation \ref{eqn:sw-jet-accn}, showing the non-dimensional eastward acceleration at the equator. The black line shows the sum of the four coloured lines.}\label{fig:forced-response-accn}
\end{figure}

\citet{showman2011superrotation} show why the tilted wave pattern (such as the first plot in our Figure \ref{fig:shear-2D}) transports eastward zonal momentum towards the equator. The tilts result in velocities ``such that $u'v' < 0$ in the northern hemisphere and $u'v' > 0$ in the southern hemisphere'', so the term in $u'v'$ in Equation \ref{eqn:sw-jet-accn} produces an eastward acceleration.

Figure \ref{fig:forced-response-accn} shows the latitudinal profile of this jet acceleration, calculated from the forced response in the second plot of Figure \ref{fig:motivating-plot}. The black line is $\partial \bar{U}(y) / \partial t$, which is the sum of the first term (blue), second term (red), third term (purple), and fourth term (green) in Equation \ref{eqn:sw-jet-accn}. Given our length and time scales \citep{matsuno1966quasi}, the equatorial acceleration of 0.1 implies a dimensional acceleration of $0.01\ \textrm{ms}^{-2}$ on an Earth-like planet or $0.1\ \textrm{ms}^{-2}$ on a Hot Jupiter. GCM simulations show that the jets of order $100\ \textrm{ms}^{-1}$ on Earth-like planets (see Section \ref{sec:gcm-results} and jets of order $1000\ \textrm{ms}^{-1}$ on Hot Jupiters form from rest in about 20 days. The linear shallow-water model on the beta-plane therefore predicts an acceleration about 20 times too large. We will show later that this is because the beta-plane model requires an unrealistically large forcing in order to balance the height perturbation due to the jet (as the jet height perturbation scales linearly with the jet speed on the beta-plane). The model in spherical geometry gives a more realistic equatorial acceleration, as the jet height perturbation scales quadratically (correctly) with the jet speed, so is balanced by a smaller forcing $Q$.

The net equatorial acceleration in Figure \ref{fig:forced-response-accn} is eastward, but there is a westward acceleration further from the equator \citep{showman2011superrotation}. Our GCM simulations do not show significant zonal-mean westward flows at the pressure level of the standing Rossby waves, although some simulations in different studies with different planetary parameters show strong westward jets \citep{showman2015circulation}. The westward acceleration predicted by the linear model may be opposed by an eastward acceleration from a Hadley circulation in the GCM, which is not represented in the linear model. Eastward flow may also take place in the lower or upper atmosphere of a real planet or a GCM simulation, which is not represented by our single-layer model.

The eastward flow from this acceleration affects the forced wave solutions. A zonally uniform jet modifies the projection coefficients in Equation \ref{eqn:project-coeff-flow} \citep{tsai2014three}, Doppler-shifting the waves eastward to a maximum of $\pi/2$ ahead of the forcing. The flow shifts the frequency $\omega$ of the free modes by $k_{x} \bar{U}$, which then affects the imaginary part of the coefficients projecting the free modes onto the forced solution:

\begin{equation}\label{eqn:project-coeff-flow}
  a_{m} = \frac{1}{\alpha - i  (\omega_{m} - k_{x} \bar{U})}
\end{equation}

A positive eastward flow $\bar{U}$ shifts the wave response eastwards. The third plot in Figure \ref{fig:motivating-plot} shows how a uniform flow shifts the maximum of the forced response towards $+90\degr$ longitude. For a significant shift, the frequency shift $k_{x}\bar{U}$ must be on the order of $\omega_{m}$, i.e. the jet speed $\bar{U}$ must be on the order of $\omega_{m} / k_{x}$. The frequencies are the solutions of $\omega^{2}-k_{x}^{2}-k_{x}/\omega=2n+1$ \citep{matsuno1966quasi} for zonal wavenumber $k_{x}$ ($k_{x}=1$ in our periodic system) and mode $n$. So, the Doppler shift of the $n=1$ Rossby wave is significant when $k_{x}\bar{U} \approx 0.25$. The scale of the zonal wavenumber of these planetary waves is set by the planet's radius: $k \sim 1 / a \sim 1$. The dimensional velocity is $[U]  = c = (gH)^{1/2}$, so the critical jet speed at which $\omega_{m} \sim k_{x} \bar{U}$ and the Rossby mode is significantly shifted is $\bar{U}_{crit} = (gH)^{1/2} / 4$.

For an Earth-like planet, $g \sim 10\ \textrm{ms}^{-2}$ and $h_{e} \sim 5\ \textrm{km}$, so $\bar{U}_{crit} = 50\ \textrm{ms}^{-1}$. GCM simulations show equilibrium jet speeds of order $100\ \textrm{ms}^{-1}$, so the Doppler shift should be significant. A Hot Jupiter has $g \sim 20\ \textrm{ms}^{-2}$ and $h_{e} \sim 200\ \textrm{km}$ \citep{showman2011superrotation}, so $\bar{U}_{crit} = 500\ \textrm{ms}^{-1}$. GCMs and observations show equilibrium jet speeds of order $1000\ \textrm{ms}^{-1}$, so the Doppler shift should again be significant. In Section \ref{sec:forced-system}, we will linearize the shallow-water equations about a non-uniform jet.


\section{Linearized Shallow-Water Equations in Shear Flow on the Beta-Plane}\label{sec:forced-system}

In this section, we discuss the main results of this paper -- our solutions to the shallow-water equations linearized around a zonally uniform shear flow $\bar{U}(y)$ and height $\bar{H}(y)$  \citep{boyd2017equatorial}. $\bar{H}(y)$ and $\bar{U}(y)$ satisfy the second line of Equation \ref{eqn:sw-eqns-1}, so are geostrophically balanced ($\bar{H}_{y}(y)=-y\bar{U}(y)$). For our Gaussian jet $\bar{U}(y)=U_{0}e^{-y^{2}/2}$, the height perturbation is $\bar{H}(y)=U_{0}e^{-y^{2}/2}$.

As before, we use a forcing magnitude $Q_{0}=1$ and equal radiative and dynamical damping rates $\alpha_{rad}=\alpha_{dyn} = 0.2$ \citep{matsuno1966quasi}. We will discuss the effect of unequal radiative and dynamical damping rates in Section \ref{sec:effect-of-damping}. As explained in Section \ref{sec:form-effect-zonal-flow}, the forced response is significantly shifted eastwards by zonal flows of order unity, so we vary our non-dimensional velocity $U_{0}$ from 0 to 1.

The forcing magnitude $Q_{0}$ must be comparable to the velocity and height produced by the jet, which are both of order $U_{0}=1$, so we set $Q_{0}=1$. This large forcing results in a zonal jet acceleration of order $ 0.1$ in Figure \ref{fig:forced-response-accn}, which implies that the jet will accelerate about ten times faster than in our GCM. This is due to the large forcing required on the beta-plane discussed previously. In Section \ref{sec:sphere-solutions} we will show that a more realistic forcing magnitude in a spherical geometry produces a more physically reasonable acceleration.

The new background of $\bar{U}(y)$ and $\bar{H}(y)$  changes the shallow-water equations in Section \ref{sec:sw-equations} to:

\begin{widetext}
  \begin{equation}\label{eqn:shear-sw-equations}
      \begin{gathered}
        \frac{\partial u}{\partial t} +  \alpha_{dyn} u + \frac{\partial \bar{U}(y)u}{\partial x} +(\frac{\partial \bar{U}(y)}{\partial y} - y)v + \frac{\partial h}{\partial x} = 0 \\
        \frac{\partial v}{\partial t} +  \alpha_{dyn} v + \frac{\partial \bar{U}(y)v}{\partial x} + y u + \frac{\partial h}{\partial y} = 0 \\
        \frac{\partial \bar{H}' u}{\partial x} + \bar{H}'\frac{\partial v}{\partial y} - y\bar{U}(y) v +\frac{\partial h}{\partial t} +  \alpha_{rad} h + \frac{\partial \bar{U}(y) h}{\partial x} = Q(y)\\
          \bar{H}' = 1+\bar{H}(y)
      \end{gathered}
  \end{equation}
\end{widetext}

The solutions have the form $A(y) e^{i(k_{x}x - \omega t)}$. To consider the free (unforced) modes of the system, we set $Q(y)=0$ and $\partial /\partial t = -i \omega$, giving the free linear system:

\begin{widetext}
  \begin{equation}\label{eqn:free-sw-shear}
    \begin{gathered}
      \begin{pmatrix}
      \alpha_{dyn} + i k_{x}\bar{U}(y) & \frac{\partial\bar{U}(y)}{\partial y}-y & i k_{x} \\
      y & \alpha_{dyn} + i k_{x}\bar{U}(y) & \frac{\partial}{\partial y} \\
      i k_{x} \bar{H}' & -y \bar{U}(y) + \bar{H}' \frac{\partial}{\partial y} & \alpha_{rad} + k_{x}\bar{U}(y)
      \end{pmatrix}
      \begin{pmatrix}
      u \\
      v \\
      h
      \end{pmatrix}
      =
      i \omega
      \begin{pmatrix}
      u \\
      v \\
      h
      \end{pmatrix} \\
        \bar{H}' = 1+\bar{H}(y)
    \end{gathered}
  \end{equation}
\end{widetext}

To find the stationary response to steady forcing (in this case, a day-night heating difference) we set $Q(y)=Q_{0}e^{-y^{2}/2}$ \citep{matsuno1966quasi} and $\partial / \partial t = 0 $, giving the forced linear system:

\begin{widetext}
\begin{equation}\label{eqn:forced-sw}
  \begin{gathered}
    \begin{pmatrix}
    \alpha_{dyn} + i k_{x}\bar{U}(y) & \frac{\partial\bar{U}(y)}{\partial y}-y & i k_{x} \\
    y & \alpha_{dyn} + i k_{x}\bar{U}(y) & \frac{\partial}{\partial y} \\
    i k_{x}\bar{H}' & -y \bar{U}(y) + \bar{H}' \frac{\partial}{\partial y} & \alpha_{rad} + k_{x}\bar{U}(y)
    \end{pmatrix}
    \begin{pmatrix}
    u \\
    v \\
    h
    \end{pmatrix}
    =
    \begin{pmatrix}
    0 \\
    0 \\
    Q(y)
    \end{pmatrix} \\
      \bar{H}' = 1+\bar{H}(y)
  \end{gathered}
\end{equation}
\end{widetext}

Appendix \ref{sec:app-ps-method} explains how we solved both of these sets of equations using a pseudo-spectral method, by expanding the solutions in terms of parabolic cylinder functions. Appendix \ref{sec:app-beta} shows the accuracy of this method, demonstrating that the pseudo-spectral method identifies the exact solution in the case with no background jet, and that the solution with a background flow changes by less than 1 part in 10,000 for any modes past $n=30$. The pseudo-spectral method finds $N_{m}$ solutions (the number of modes used in the calculation) for the eigenvalue equation governing the free modes. Many of these are spurious, but we can distinguish them from the physical modes by inspecting their eigenvalues. The pseudo-spectral method only produces a single solution for the forced linear system, which is simpler to interpret.

The mean shallow-water height $H_{0}$ (non-dimensionalized to unity above) is determined by the vertical modes excited by the vertical heating profile of the atmosphere.  \citet{tsai2014three} showed that in tidally locked atmospheres, the vertical heating profile excites a continuum of vertical modes, leading to a continuum of horizontal modes. Most of the energy is contained in one of these vertical modes, so it is sufficient to only consider this vertical mode and associated horizontal mode with height $H_{0}$.

\subsection{Free Solutions in Shear Flow}\label{sec:free-solutions-subsection}

\begin{figure}
 \plotone{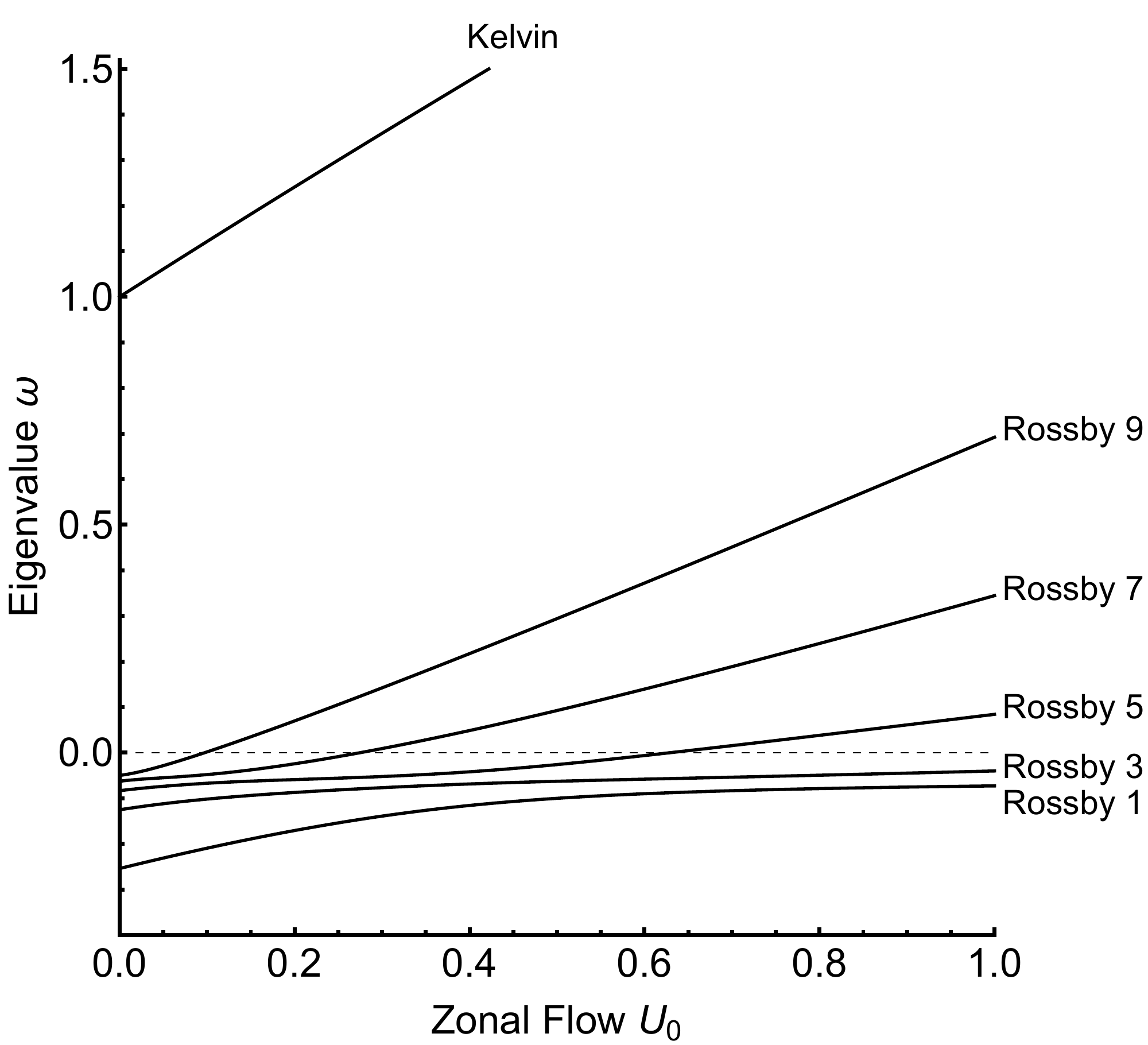}
\caption{The real parts of the eigenvalues of the lowest-order symmetric free modes of Equation \ref{eqn:free-sw-shear}, for different values of $U_{0}$ in a shear background flow $\bar{U}(y)=U_{0}e^{-y^{2}/2}$. The position of the maximum of each mode in the forced response depends on the real part of its eigenvalue and the magnitude of the damping in the system.}\label{fig:shear-flow-eval-shift}
\end{figure}

\begin{figure*}
 \gridline{\fig{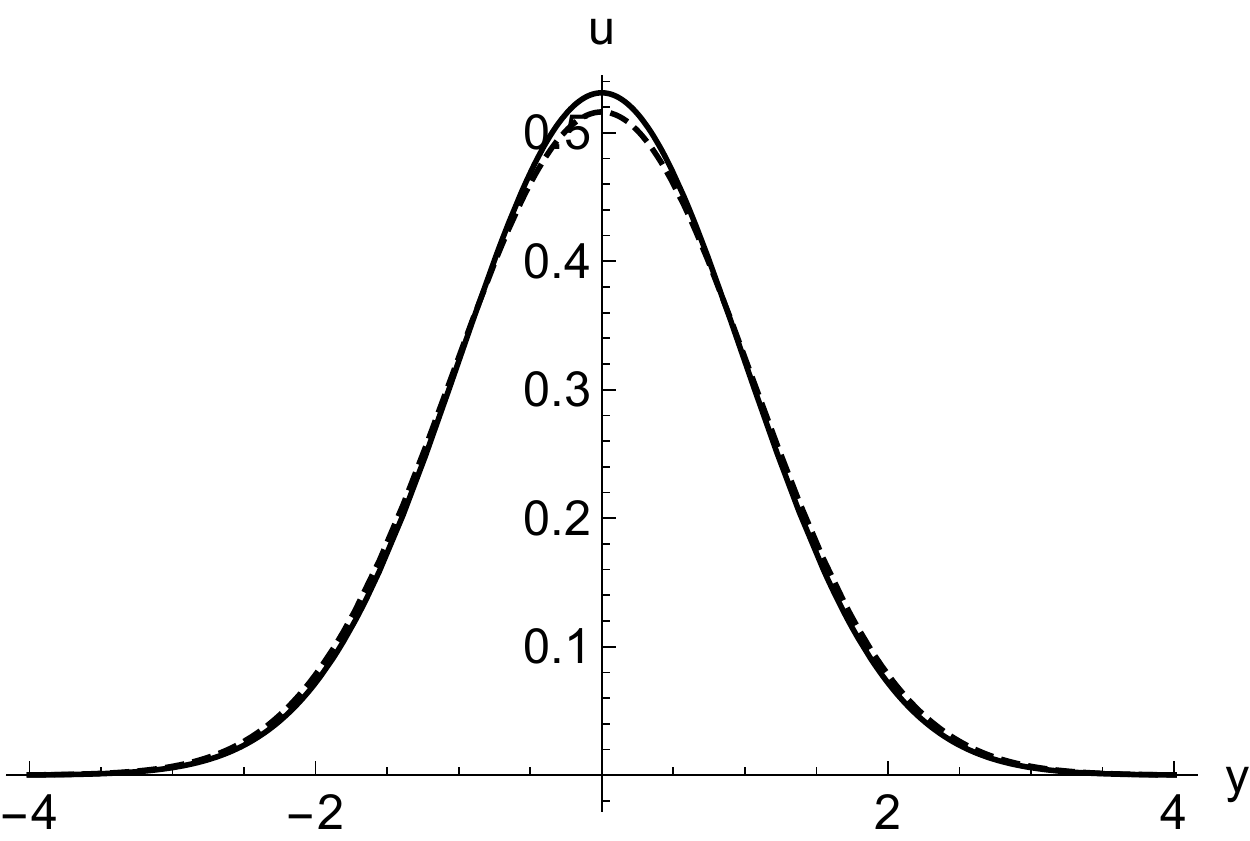}{0.32\textwidth}{Zonal velocity $u$.}
 \fig{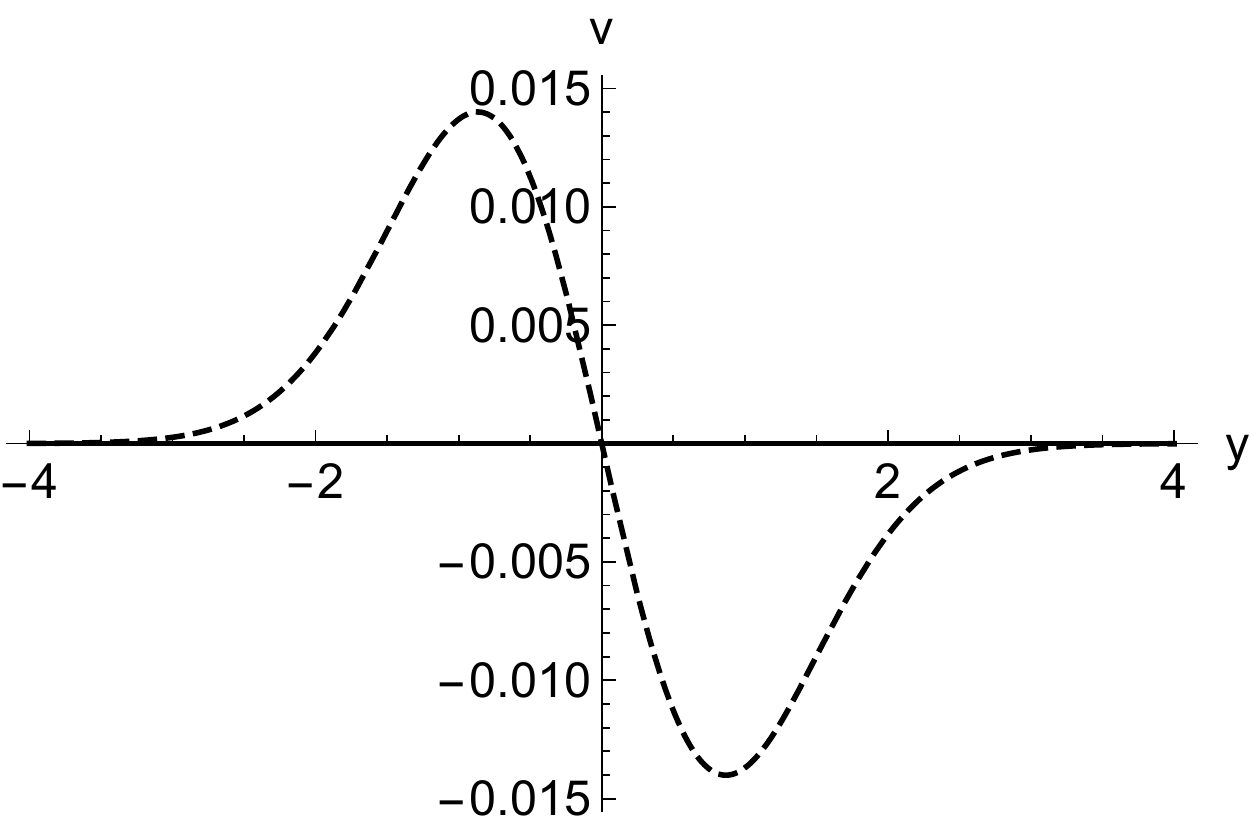}{0.32\textwidth}{Meridional velocity $v$}
 \fig{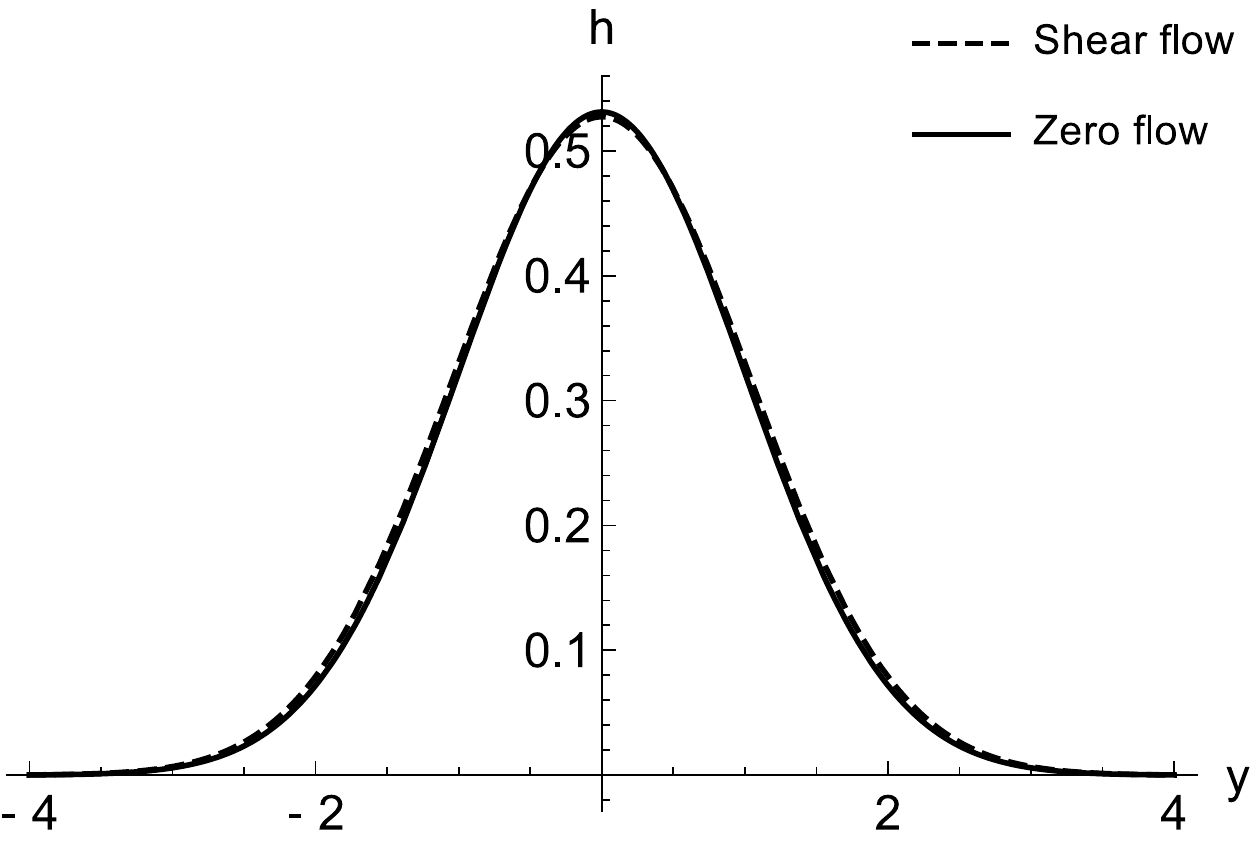}{0.32\textwidth}{Height $h$}
           }
\caption{The meridional structure of the free Kelvin mode of Equation \ref{eqn:free-sw-shear}, with and without a background shear flow $\bar{U}(y)=0.1 e^{-y^{2}/2}$.}\label{fig:free-shear-meridional-kelvin}
\end{figure*}

\begin{figure*}
 \gridline{\fig{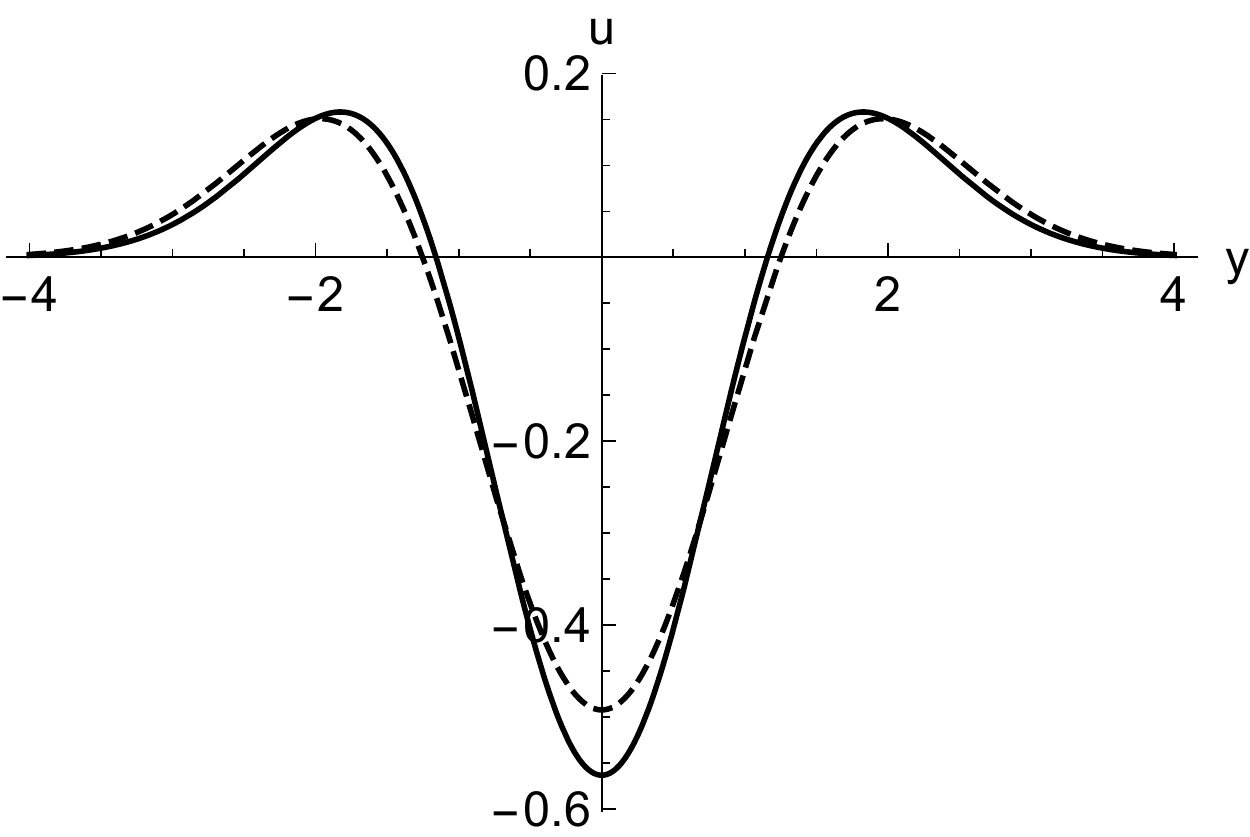}{0.32\textwidth}{Zonal velocity $u$.}
 \fig{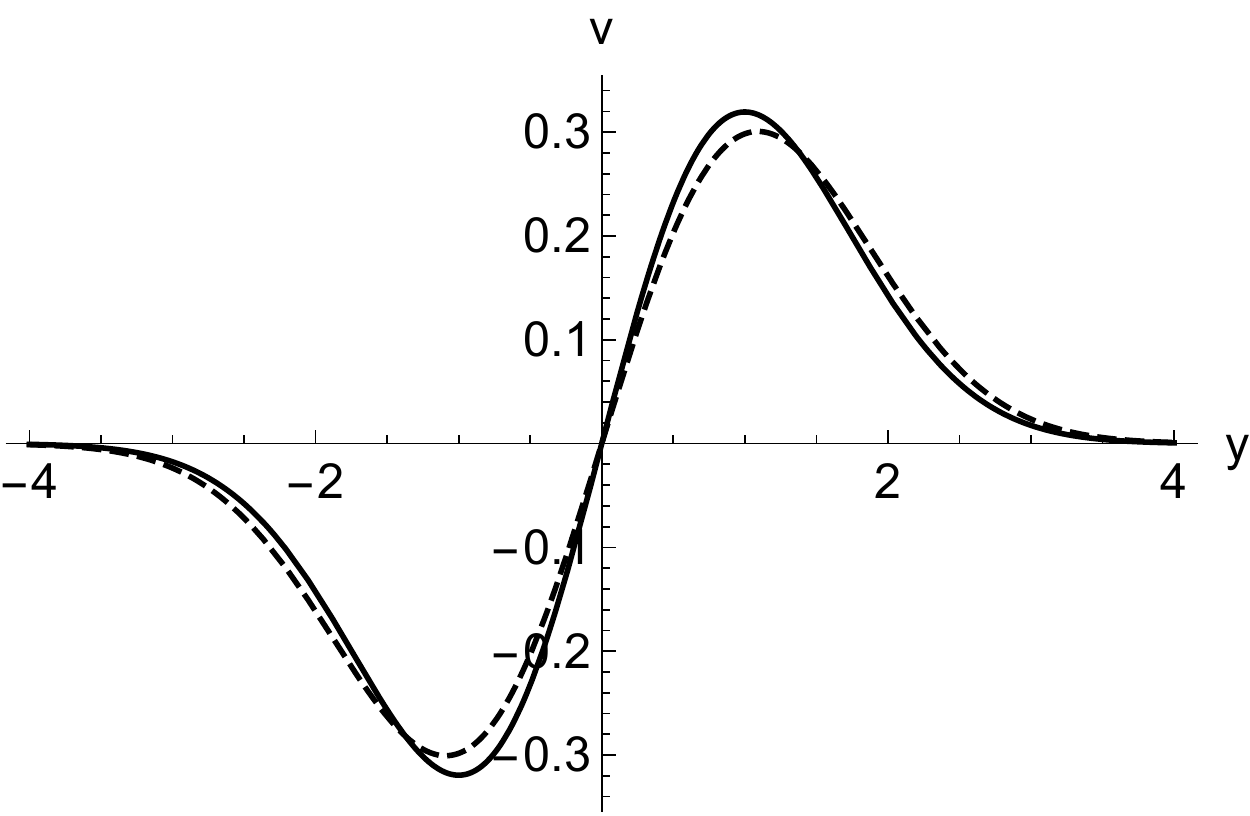}{0.32\textwidth}{Meridional velocity $v$}
 \fig{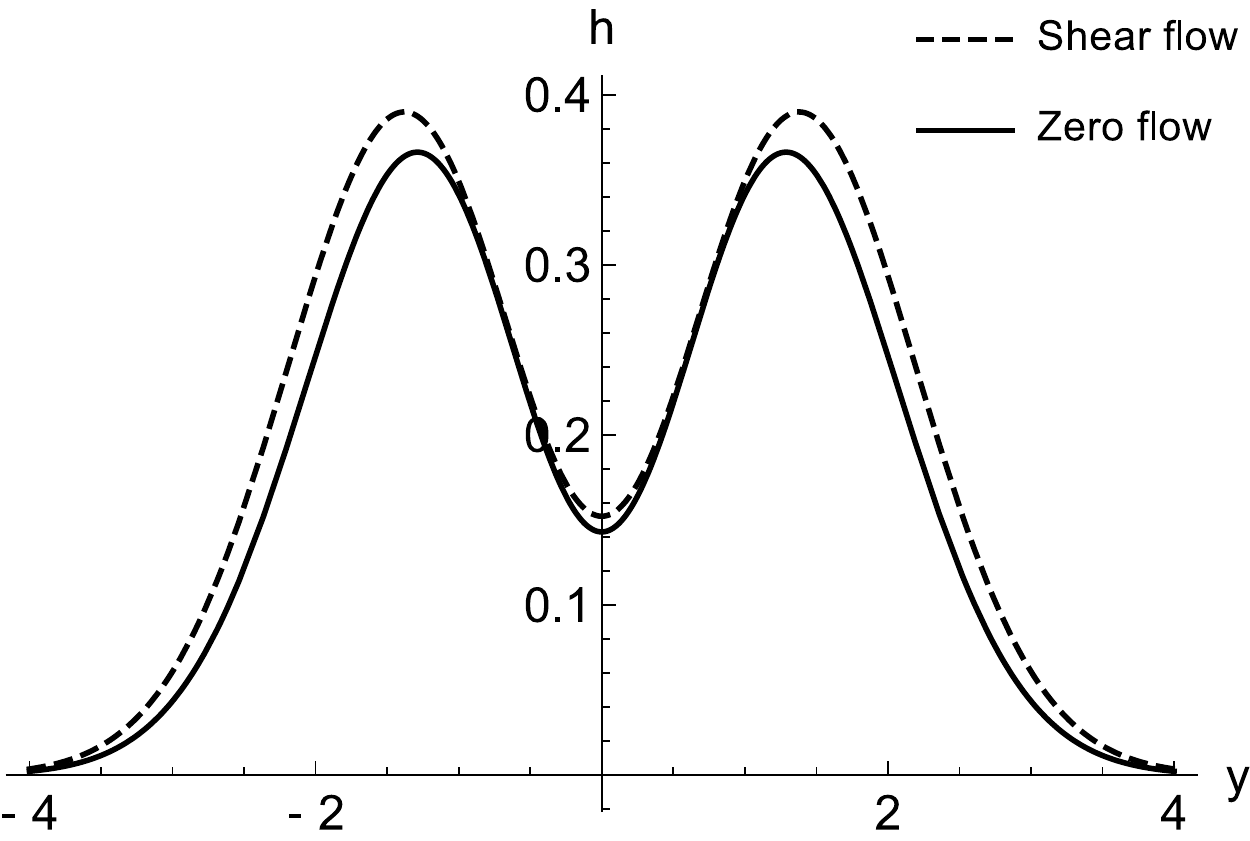}{0.32\textwidth}{Height $h$}
           }
\caption{The meridional structure of the free $n=1$ Rossby mode of Equation \ref{eqn:free-sw-shear}, with and without a background shear flow $\bar{U}(y)=0.1 e^{-y^{2}/2}$.}\label{fig:free-shear-meridional}
\end{figure*}

In this section, we will discuss the effect of the shear flow on the free modes of Equation \ref{eqn:free-sw-shear}, in order to understand changes to the forced response in the next section. In Section \ref{sec:sw-equations} we showed how the forced solution to the shallow-water equation in a uniform background flow ( Equation \ref{eqn:sw-eqns}) can be expressed as a sum of the free modes using Equation \ref{eqn:sw-eqns-projection}. It is not possible to express the forced solution to Equation \ref{eqn:forced-sw} in a shear background flow using such a simple sum of the free solutions to Equation \ref{eqn:free-sw-shear}. However, we can still use the free solutions of Equation \ref{eqn:free-sw-shear} to qualitatively understand the forced solutions that we will find numerically in Section \ref{sec:forced-solutions-subsection}.

We write the free solutions to the shallow water equations as a complex function of latitude $A(y)$. Later, we will write and plot the forced solutions as functions of latitude and longitude, in the form $A(y) e^{i \delta(y) x}$. The phase shift $\delta(y)$ in a shear flow is equivalent to the phase shift $(\omega_{m} - k_{x} \bar{U})$ in a uniform flow (Equation \ref{eqn:project-coeff-flow}). The sign of the eigenvalue $\omega_{m}$ of any mode in shear determines where the free mode appears in the forced solution, as in the previous section for uniform background flow.

The free modes of the beta-plane shallow-water system (Equation \ref{eqn:free-sw-shear}) depend on the background flow and height fields. For zero background flow, the free modes are the same as those discussed in Section \ref{sec:sw-equations}. For an analytic solution with a uniform background flow $\bar{U}_{0}$, the free modes are linearly Doppler-shifted as discussed in Section \ref{sec:form-effect-zonal-flow} \citep{tsai2014three}. For a shear background flow $\bar{U}(y)$, we found the free modes of Equation \ref{eqn:free-sw-shear} using the method described in Appendix \ref{sec:app-ps-method}, with the parameters listed at the start of Section \ref{sec:forced-system} (and equal radiative and damping rates, $\alpha_{rad}=\alpha_{dyn}$).

Figure \ref{fig:shear-flow-eval-shift} shows the real parts of the eigenvalues of the free Kelvin mode and the symmetrical free Rossby modes of Equation \ref{eqn:free-sw-shear}, for a background flow $\bar{U}(y)=U_{0}e^{-y^{2}/2}$ with a variable magnitude $U_{0}$. These are the lowest-order modes excited by the symmetric forcing. The Kelvin mode already has a positive eigenvalue for $U_{0}=0$, which increases and shifts further east as the flow increases, moving the Kelvin mode towards a maximum shift of $+90\degr$ and contributing to the hot-spot shift.

\citet{tsai2014three} suggests that in a uniform background flow the $n=1$ Rossby mode is shifted eastward of the substellar point, adding to the hot-spot shift. Figure \ref{fig:shear-flow-eval-shift} shows that in this shear flow $\bar{U}(y)$, the $n=1$ Rossby mode eigenvalue increases but does not become positive for $U_{0}=1.0$ (corresponding to the jet speed in the forced response plotted later in Figure \ref{fig:shear-2D}). This means that it is shifted eastwards in the forced response, but does not reach the east of the substellar point. However, the higher order Rossby mode eigenvalues do become positive, shifting to the east of the substellar point and contributing to the hot-spot shift. So, the forced response and the hot-spot shift are affected by the higher-order Rossby modes, and not just dominated by the Kelvin and $n=1$ Rossby modes.

That is not to say that the $n=1$ mode is never responsible for the hot-spot shift -- later, we will show that in a spherical geometry the $n=1$ mode shifts close to $+90\degr$ eastwards. It is also possible in the beta-plane system for different input parameters (flow speed, damping rates) to shift the $n=1$ Rossby mode past the substellar point. But, our free mode expansion has shown that the $n=1$ Rossby mode is not the only important mode, and that the higher-order modes are also important to the forced response.

For zero damping, half of these eigenvalues will have positive imaginary parts, and the modes corresponding to them will grow exponentially. Non-zero damping decreases the imaginary part of all the modes, so will make some or all of these modes stable. In general, the free linear system in Equation \ref{eqn:free-sw-shear} will have some unstable modes unless the damping is very large. These unstable modes are similar to those discussed by \citet{wang2014instability}, who show how similar modes can produce superrotation even on a planet without a permanent day-night heating difference.

These unstable modes technically make the linear forced wave problem ill-posed, since the result of any linear initial value problem will be eventually dominated by the most rapidly growing modes rather than the stationary response. Later comparison with nonlinear GCM simulations in Section \ref{sec:gcm-results} will show that the forced response still has considerable explanatory power. This may be because in reality the unstable modes equilibrate due to damping or nonlinear effects, at a sufficiently low amplitude that they take the form of mobile waves propagating across the forced stationary pattern without significantly disrupting its basic structure. Future work should investigate the exact nature of these instabilities, and the effect of damping and shear flow on their growth rates.

The shear flow also affects the latitudinal structure $A(y)$ of the modes. The lowest-order free solutions of Equation \ref{eqn:free-sw-shear} (the Kelvin and Rossby modes), plotted in Figure \ref{fig:free-shear-meridional} and \ref{fig:free-shear-meridional-kelvin}, resemble the free solutions with zero shear flow \citep{matsuno1966quasi}, with their latitudinal structure slightly changed by a weak shear flow $\bar{U}=0.1 e^{-y^{2}/2}$. The shear flow perturbs the solutions by adding higher order meridional structure.  For example, the meridional wind of the Rossby wave in Figure \ref{fig:free-shear-meridional} resembles the $n=1$ parabolic cylinder function added to the $n=3$ function (see Figure \ref{fig:hermite-functions} in Appendix \ref{sec:app-ps-method}). \citet{boyd1978shearI} discusses how a shear flow affects the meridional structure of these modes in more detail.

\subsection{Forced Solutions in Shear Flow}\label{sec:forced-solutions-subsection}

In this section, we discuss solutions of the forced linear shallow-water equations with $\omega=0$ (Equation \ref{eqn:forced-sw}) as an inhomogeneous forced linear problem.

\begin{figure*}
 \gridline{\fig{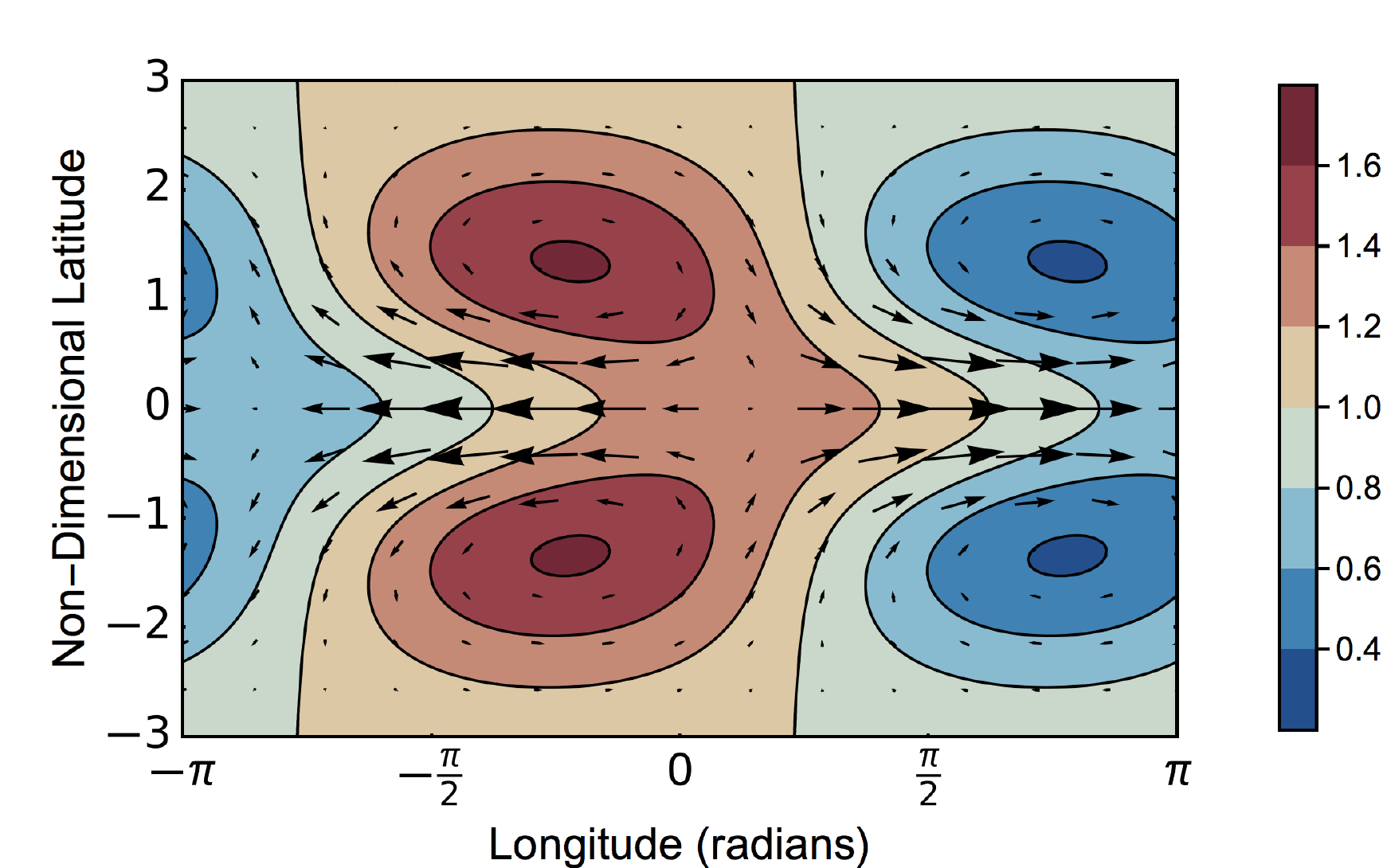}{0.45\textwidth}{Forced solution without a background jet, calculated with the pseudo-spectral method but exactly the same as the analytic solution in the second plot in Figure \ref{fig:motivating-plot}.}
 \fig{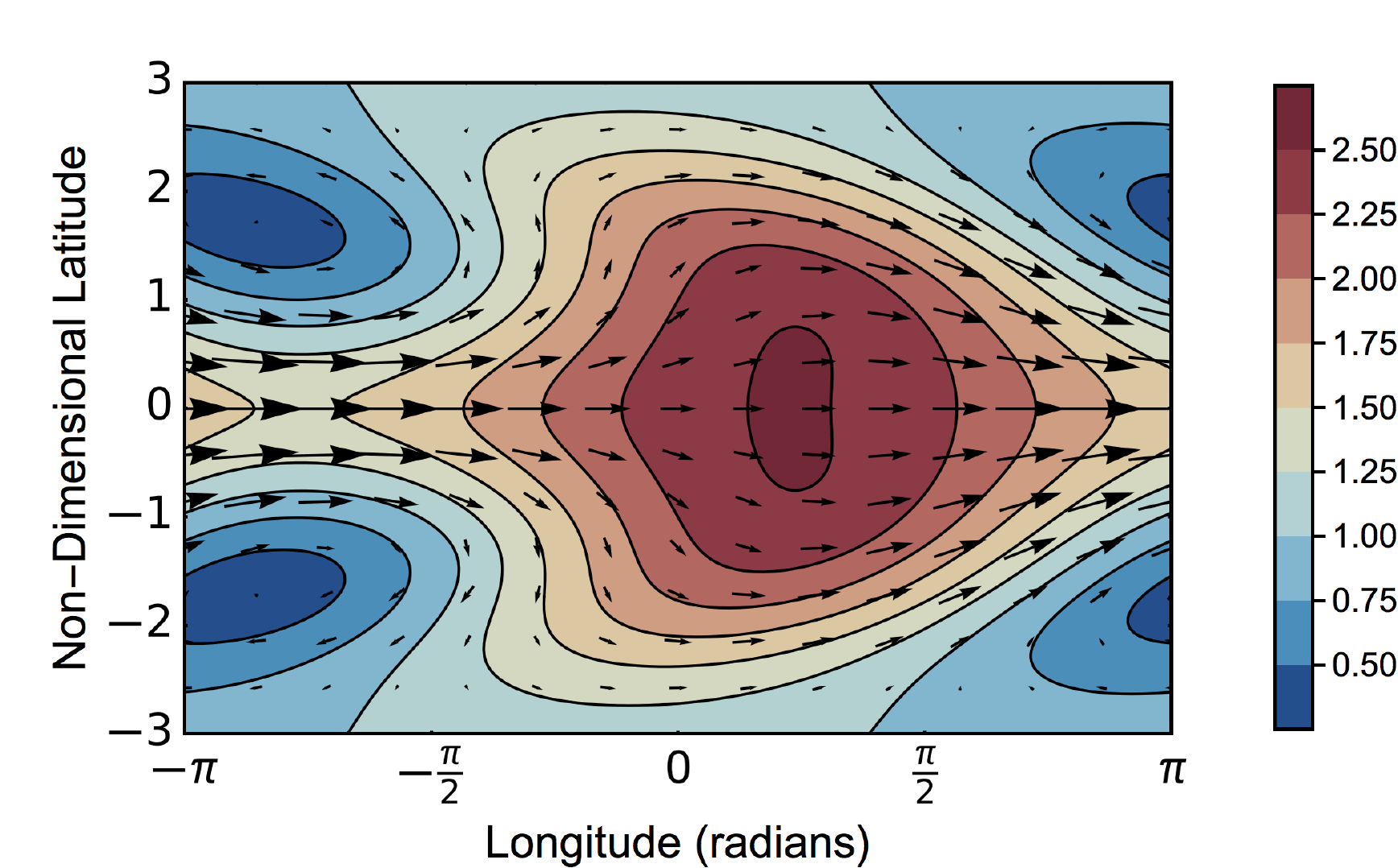}{0.45\textwidth}{Forced solution with a background jet --  now, the high height (hot) region matches our GCM results, as do the the low height (cold) Rossby lobes.}
           }
\caption{Effect of shear on the height field of the solutions to the forced shallow-water equations linearized about an eastward equatorial jet $\bar{U}(y)=e^{-y^{2}/2}$ on the beta-plane, with the parameters listed at the start of Section \ref{sec:forced-system}.}\label{fig:shear-2D}
\end{figure*}

Figure \ref{fig:shear-2D} shows the main result of this paper: how linearizing the equations about a background jet changes the response to stationary forcing. The first plot in the figure is the case without a jet -- the exact forced solution of \citet{matsuno1966quasi}. The case with a jet is our new solution to Equation \ref{eqn:forced-sw}, linearized about a background flow $\bar{U}(y)$ and $\bar{H}(y)$. Both plots were calculated with the method in Appendix \ref{sec:app-ps-method}, but the first plot has identical form to the analytic solution in zero flow (the second plot in Figure \ref{fig:motivating-plot}), as discussed in Appendix \ref{sec:app-beta}.

We suggest that the background shear flow changes the solution in two key ways. First, the forced response is Doppler-shifted eastwards as discussed previously. Second, the jet velocity and height adds to the forced solution, adding to the height along the equator, magnifying the cold Rossby lobes, and combining with the shifted stationary waves to form the hot-spot.

In Section \ref{sec:free-solutions-subsection} we showed how the Doppler-shift from the jet affects the Kelvin mode and the Rossby modes to different extents. Comparing Figure \ref{fig:shear-2D} to Figure \ref{fig:shear-flow-eval-shift} shows how the position of each mode in the forced response depends on its eigenvalue. The Kelvin mode has the most positive eigenvalue in Figure \ref{fig:shear-flow-eval-shift}, so produces a large hot-spot shift on the equator.

Unlike the Kelvin mode, the eigenvalue of the $n=1$ Rossby mode is initially negative. It increases as the flow increases, but does not become positive for $U_{0}=1$, so does not shift past the substellar point. This means that in Figure \ref{fig:shear-2D} the hot-spot shift is smaller at high latitudes. The higher-order Rossby modes are shifted further east, contributing more to the hot-spot shift. However, the modes get progressively weaker at higher orders so affect the total response less. This shift of all the wave modes is dominated by the Kelvin mode and lowest-order Rossby mode. It makes the position of the hot-spot and the cold Rossby lobes match our GCM output in Section \ref{sec:gcm-results}.

The solutions in this section are limited by the linear approximation of the Coriolis parameter and non-dimensionalized y-coordinate on the beta-plane. This makes them inaccurate at high latitudes, and difficult to compare directly to real planets. Solutions on the beta-plane in other studies are similar to solutions in spherical coordinates \citep{showman2011superrotation} \citep{heng2014analytical}, which suggests that its limitations are not too problematic.

The beta-plane system also requires a forcing with the same scale $L_{R}$ as the scale of the y dimension for a simple solution \citep{matsuno1966quasi}. We found that the solution was not well represented by the parabolic cylinder functions if these scales were not similar. This limits our solutions to tidally locked planets where the scale of the forcing (the planetary radius) is comparable to the the equatorial Rossby radius $L_{R}$. This is not appropriate for exoplanets such as tidally locked planets with similar size and rotation rate as Earth. In Section \ref{sec:sphere-solutions}, we will solve the linear shallow-water equations in a shear flow on a sphere, which relaxes some of the constraints of the beta-plane.

\subsection{Jet Acceleration and Equilibrium}\label{sec:disc-accn}

In this section, we show how the zonal acceleration of the linear shallow-water system, given by Equation \ref{eqn:sw-jet-accn}, decreases as the zonal flow increases. This explains why the system should reach an equilibrium rather than accelerating the zonal flow indefinitely.

Figure \ref{fig:forced-response-accn} shows the different physical sources making up the zonal acceleration. The equatorial acceleration is primarily controlled by a balance of acceleration due to the convergence of zonal momentum (the second term in Equation \ref{eqn:sw-jet-accn}), deceleration due to vertical momentum transport (the third term), and deceleration due to drag on the zonal flow (the fourth term). When the mean zonal flow is zero, Figure \ref{fig:forced-response-accn} shows that there is a positive eastward acceleration on the equator which accelerates the flow initially \citep{showman2011superrotation}.

As the mean eastward zonal flow $\bar{U}(y)$ increases (we model this as an eastward jet with Gaussian shape despite the westward acceleration at high latitudes, as discussed previously), the equatorial eastward acceleration in our linear model decreases. \citet{showman2011superrotation} suggested that the shallow-water system should reach equilibrium when the deceleration due to drag on the zonal flow increases sufficiently to balance the other terms. \citet{tsai2014three} suggested that in addition to the increased drag, the acceleration from zonal momentum convergence decreases as the zonal flow shifts the forced response eastwards.

Our solutions also show this decrease in acceleration. Figure \ref{fig:shear-2D} shows that when the background flow is zero, the Rossby and Kelvin components of the forced response are out of phase (i.e. at different longitudes) \citep{matsuno1966quasi}, so the part $h^{\prime}u^{\prime}$ of the zonal momentum convergence term is non-zero. The magnitude of this term depends on the phase difference between the modes \citep{tsai2014three}, and corresponds to the tilt of the forced $u$, $v$, and $h$ fields \citep{showman2011superrotation}. As the background jet flow increases in the forced linear solutions, the Rossby and Kelvin components (and other higher-order terms) shift eastwards, tending towards $+ 90 \degr$. So as the zonal flow increases, the waves become more in phase with each other and the acceleration due to zonal momentum convergence decreases.

\begin{figure}
    \plotone{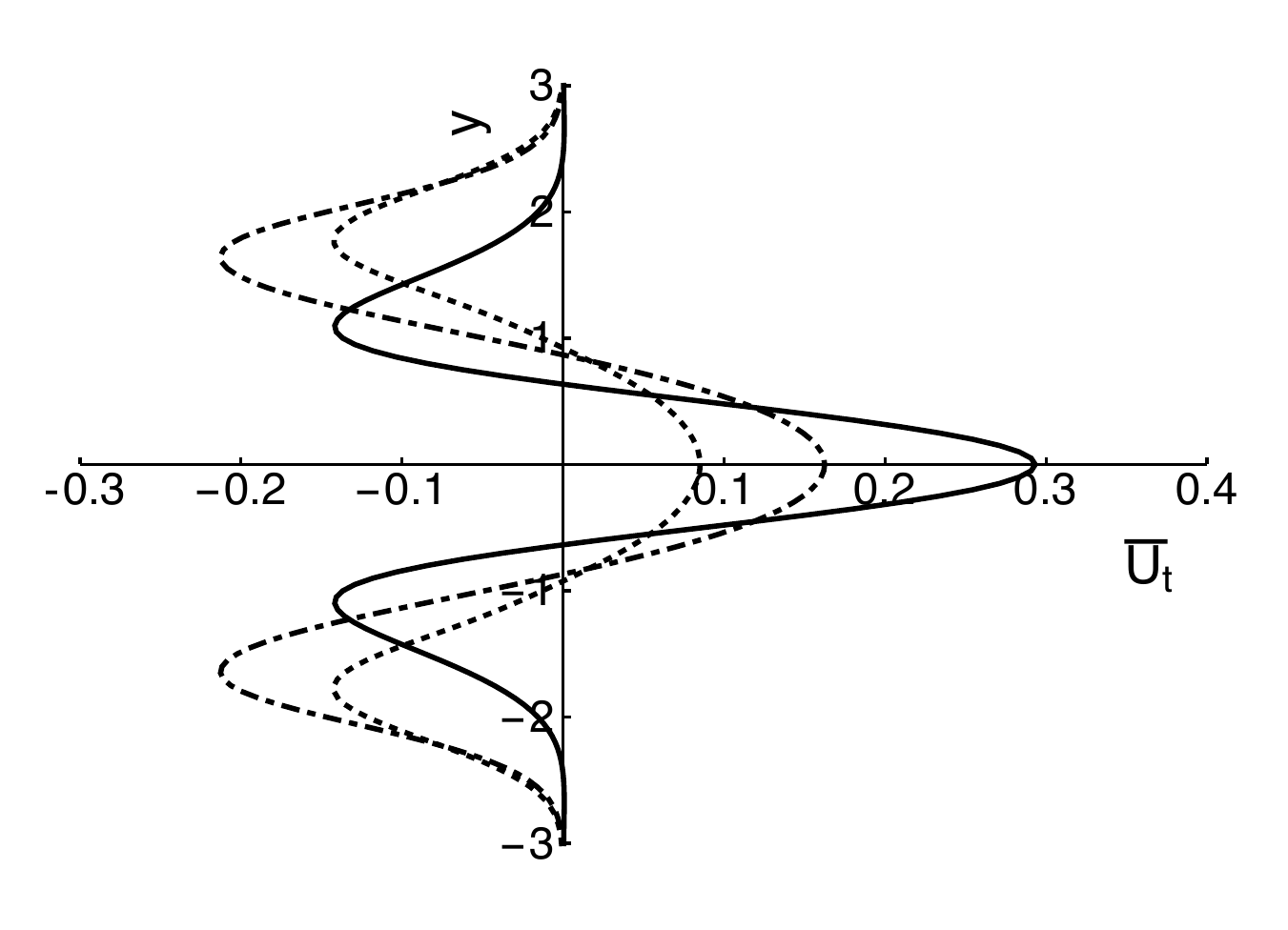}
    \caption{Acceleration due to zonal momentum convergence (the second term in Equation \ref{eqn:sw-jet-accn}) at different equatorial background flow speeds $U_{0}$, for a background flow $\bar{U}=U_{0}e^{-y^{2}/2}$. The solid line has $U_{0}=0$, the dot-dashed line $U_{0}=0.75$, and the dotted line $U_{0}=1.0$}
    \label{fig:accn-vs-u}
\end{figure}

Figure \ref{fig:accn-vs-u} shows the acceleration due to zonal momentum convergence at three different background zonal flow speeds. As the background flow increases, the equatorial acceleration from this term decreases. This combines with the increased drag $-\bar{u}^{*}/\tau_{\mathrm{drag}}$ on the mean zonal flow to give a net equatorial acceleration of zero at high enough flow speeds, in agreement with \citet{tsai2014three}.

With the length and time scales on the beta-plane discussed previously, a non-dimensional velocity of $U_{0}=1$ corresponds to a velocity of order $100 \mathrm{ms}^{-1}$ on an Earth-like planet, and to a velocity of order $1000 \mathrm{ms}^{-1}$ on a Hot Jupiter. These are comparable to the velocities seen in GCM simulations, so the suggestion of \citet{tsai2014three} that equilibrium is reached when the flow and Doppler shift becomes great enough is consistent with our linear shallow-water model.

\subsection{Effect of Damping}\label{sec:effect-of-damping}

In the previous section we assumed that the radiative damping $\alpha_{rad}$ and dynamical damping term $\alpha_{dyn}$ were equal. This allows an analytic solution but is not physically justified. Radiative damping in the shallow-water equations has a clear physical basis, but linear dynamical damping does not, apart from effects like eddy viscosity, MHD damping, or extra nonlinear terms \citep{heng2014analytical}. In this section, we suggest that although the damping rates could be very different in reality, this does not alter the overall form of the solution.

\begin{figure*}
 \gridline{\fig{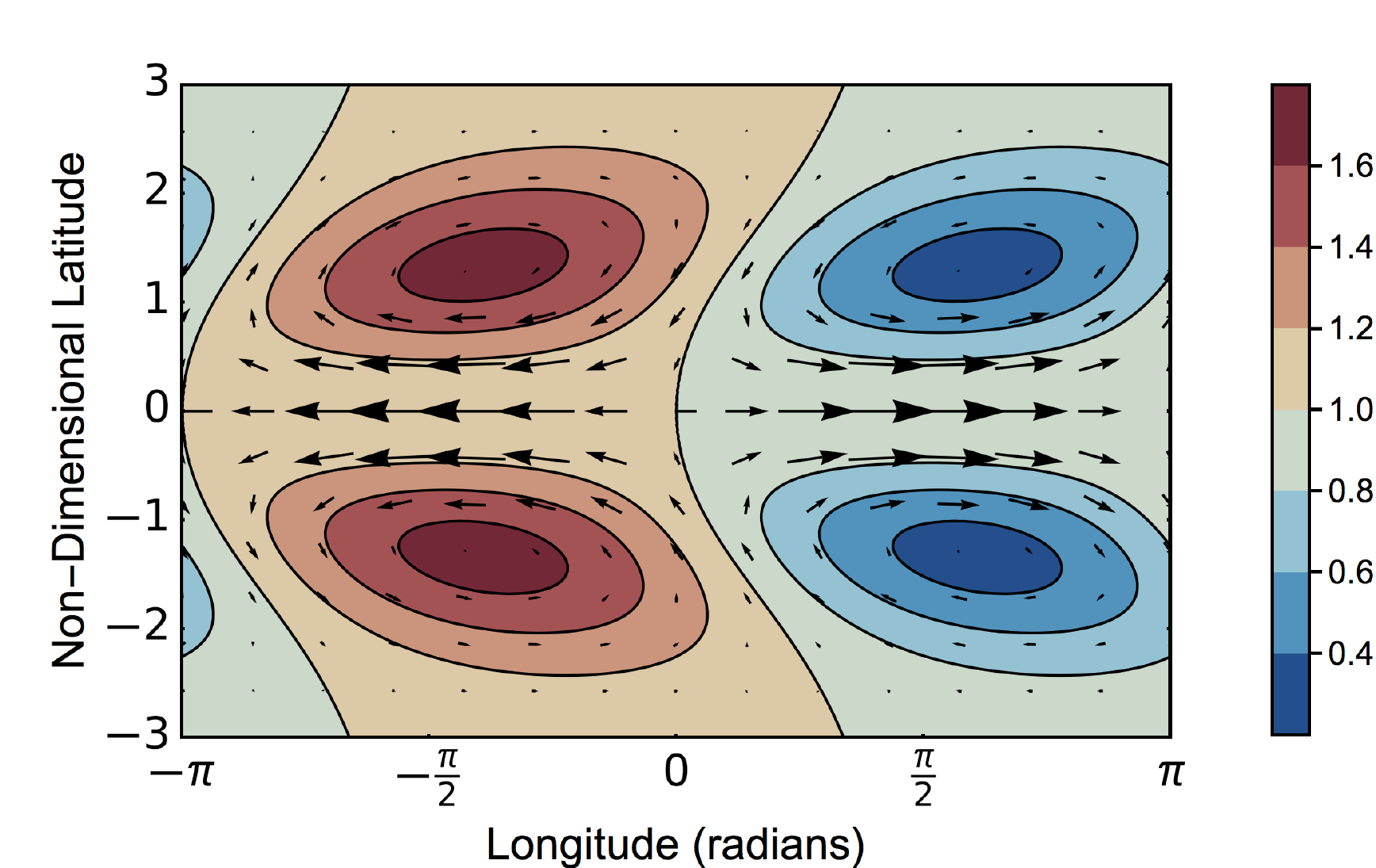}{0.45\textwidth}{Forced response in zero flow with $\alpha_{dyn}=0$.}
 \fig{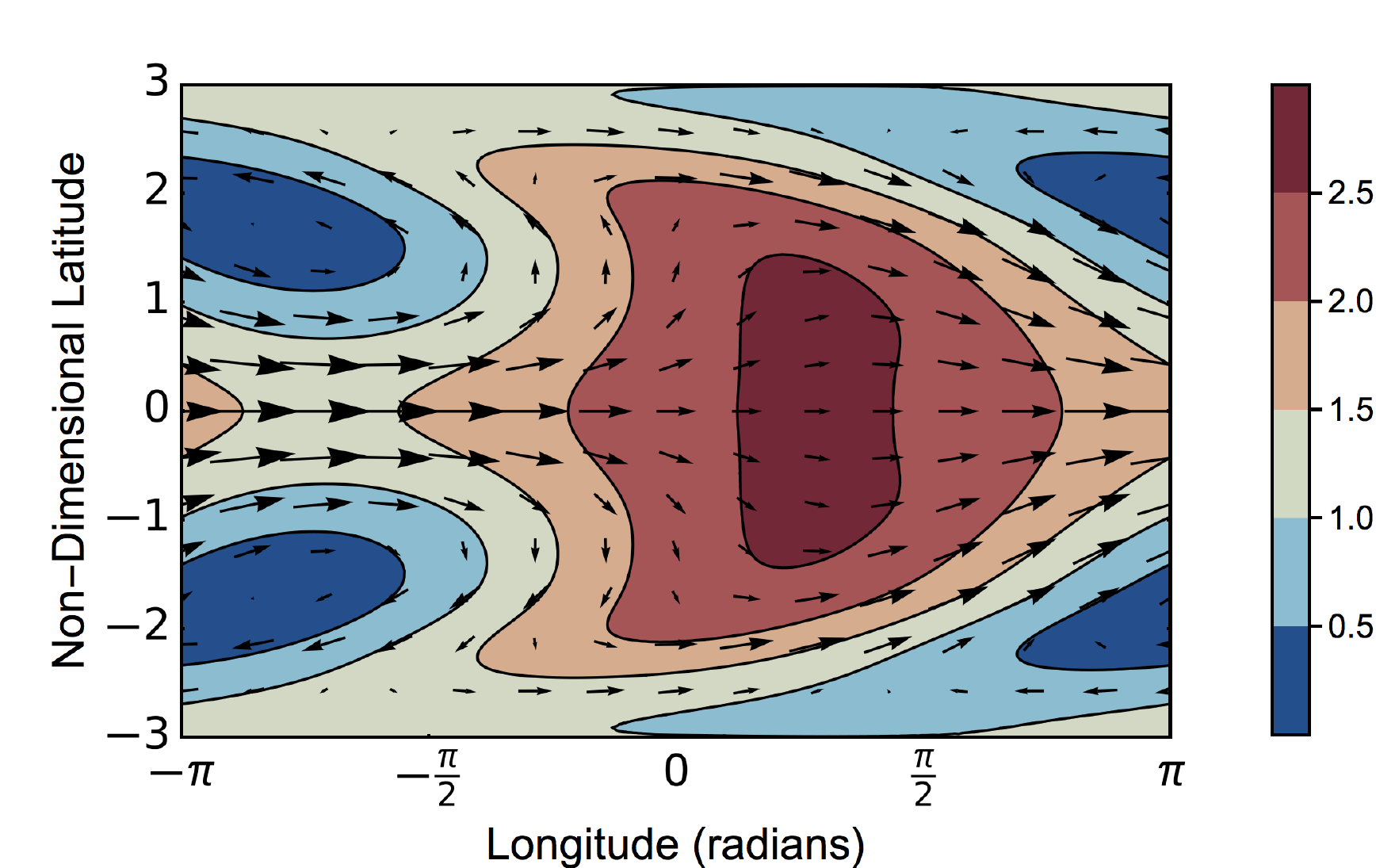}{0.45\textwidth}{Forced response in flow $\bar{U}(y)=e^{-y^{2}/2}$ with  $\alpha_{dyn}=0$.}
}
\caption{The same plots as Figure \ref{fig:shear-2D}, but with $\alpha_{dyn}=0$. The first plot shows the forced wave response in zero background flow, where the Rossby mode now dominates. The second plot shows the forced response on a background jet with $\bar{U}(y)=e^{-y^{2}/2}$ -- despite the lack of dynamical damping and weak Kelvin response, the general form of the solution is the same as in Figure \ref{fig:shear-2D}.}\label{fig:beta-dyn-rad-damping}
\end{figure*}

Figure \ref{fig:beta-dyn-rad-damping} shows the solution with zero dynamical damping. In this first plot, which has no background flow, this appears to weaken the Kelvin wave response relative to the Rossby wave response (compared to Figure \ref{fig:shear-2D}). The Rossby lobes also tilt in the opposite direction to Figure \ref{fig:shear-2D}, matching Figure 6(c) in \citet{showman2011superrotation}. In the second plot, which has a background jet, this effect is less significant as the jet velocity and height dominate on the equator where the Kelvin response would be. So, the dynamical damping is not crucial to the form of the solution, and it is not a large problem that it is not a well defined quantity. In summary, once a strong jet forms, the form of our linear shallow-water solutions is the same regardless of the exact values of the damping $\alpha_{rad}$ and $\alpha_{dyn}$.

\subsection{Effect of Shear Flow on Hot-Spot Location and Global Circulation}

Understanding the mechanism behind the global circulation and eastward hot-spot shift observed on many tidally locked planets \citep{parmentier2017handbook} is vital to interpreting observations of such planets, particularly why the magnitude of the hot-spot shift varies. \citet{showman2011superrotation} suggest that the hot-spot shift is caused by ``eastward group propagation of Kelvin waves'', as the free Kelvin waves in the shallow-water model propagate eastwards (opposite to the Rossby waves) \citep{matsuno1966quasi}. This is similar to the non-periodic Gill model, where the forced response propagates eastwards along the equator \citep{gill1980solutions}. Other explanations have represented the wave dynamics indirectly. For example, \citet{komacek2016daynightI} and \citet{zhang2017dynamics} use a linear model based on balancing an advective timescale against a radiative timescale to explain the day-night contrast and hot-spot shift on tidally locked planets. This model predicted a hot-spot shift on the equator, but did not represent wave dynamics directly so did not include their effect on the hot-spot shift, or predict the height field off the equator.

\citet{tsai2014three} focused on the effect of waves using a linear shallow-water model on a beta-plane with uniform flow $\bar{U}_{0}$, and suggested that the hot-spot shift is instead caused by ``the eastward shift of the Rossby wave with almost
no zonal phase shift of the Kelvin wave''. Our linear model builds on \citet{tsai2014three} by using a non-uniform flow $\bar{U}(y)$ and and a height perturbation $\bar{H}(y)$ in balance with this flow. The height perturbation and sheared flow are crucial to the overall form of the circulation seen in GCM results such as Figure \ref{fig:example-gcm-results}.

The forced linear solution in shear in Figure \ref{fig:shear-2D} can be approximated as the analytic solution in Section \ref{sec:sw-equations}, modified by the eastward equatorial jet in two ways:

\begin{enumerate}
  \item The flow Doppler-shifts the wave response eastwards (see Section \ref{sec:sw-equations})
  \item The flow is balanced by a zonally uniform height perturbation $\bar{H}(y)$
\end{enumerate}

\begin{figure*}
 \gridline{\fig{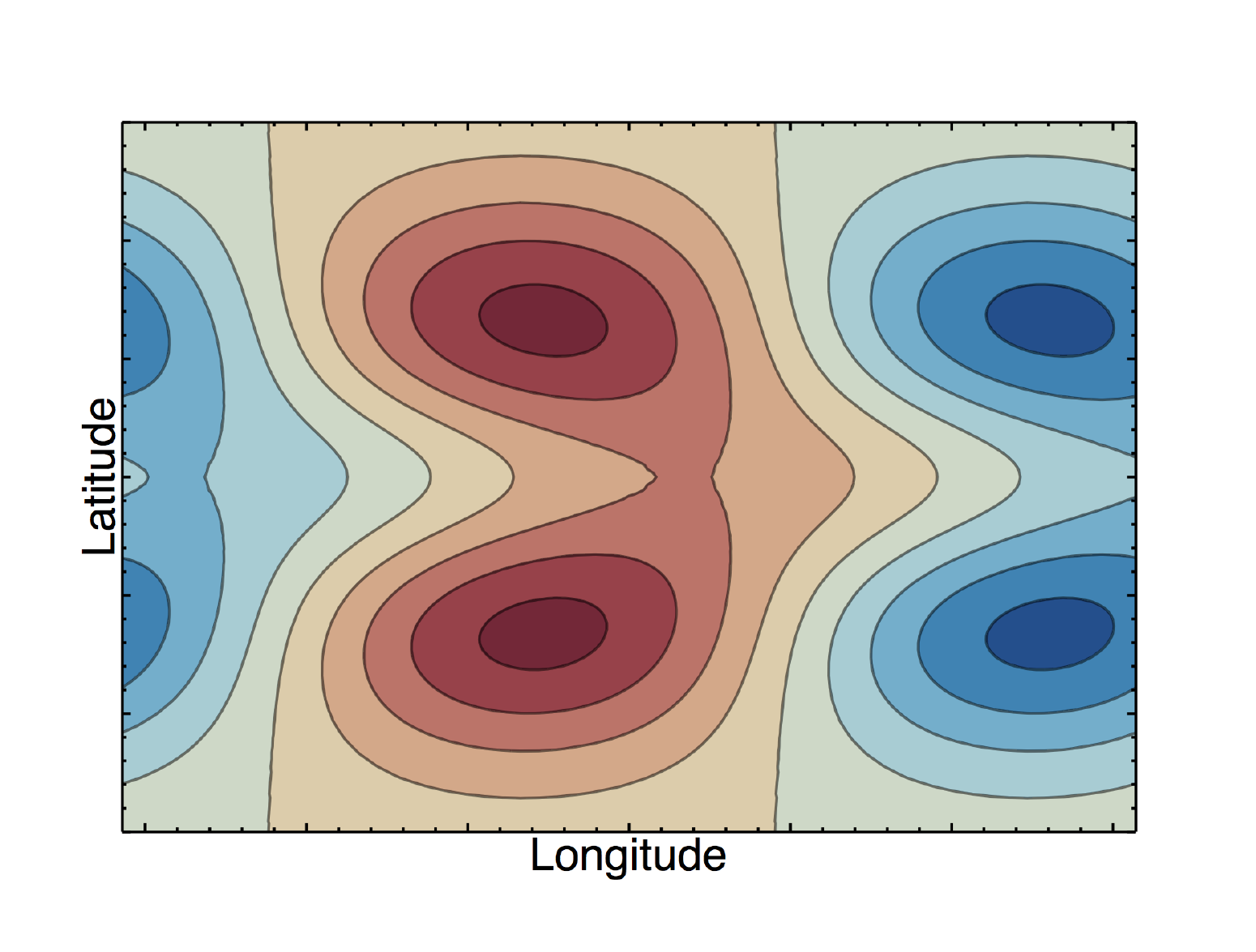}{0.24\textwidth}{(a) Height field of the exact forced solution \citep{matsuno1966quasi}.}
 \fig{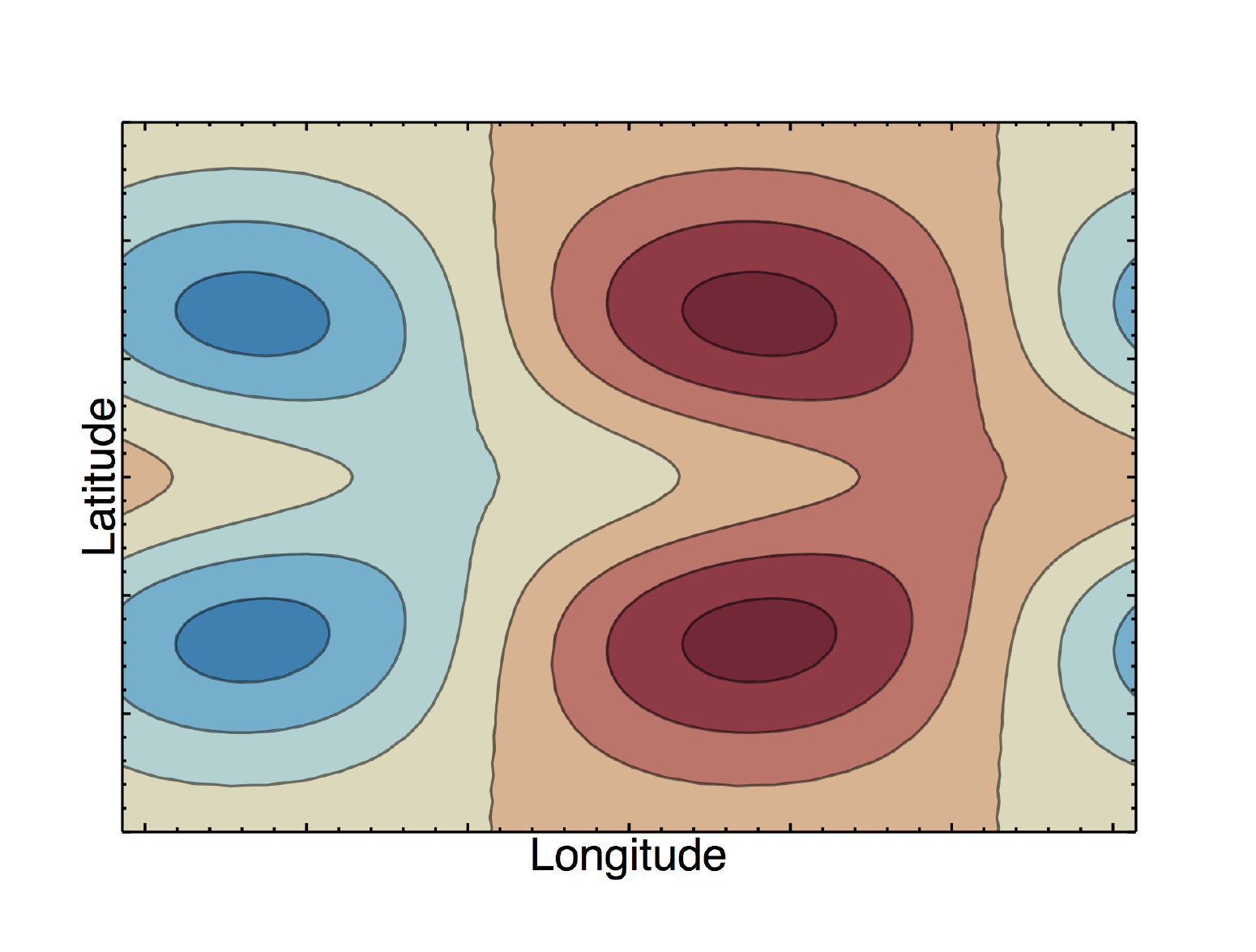}{0.24\textwidth}{(b) Solution Doppler-shifted eastwards by jet.}
\fig{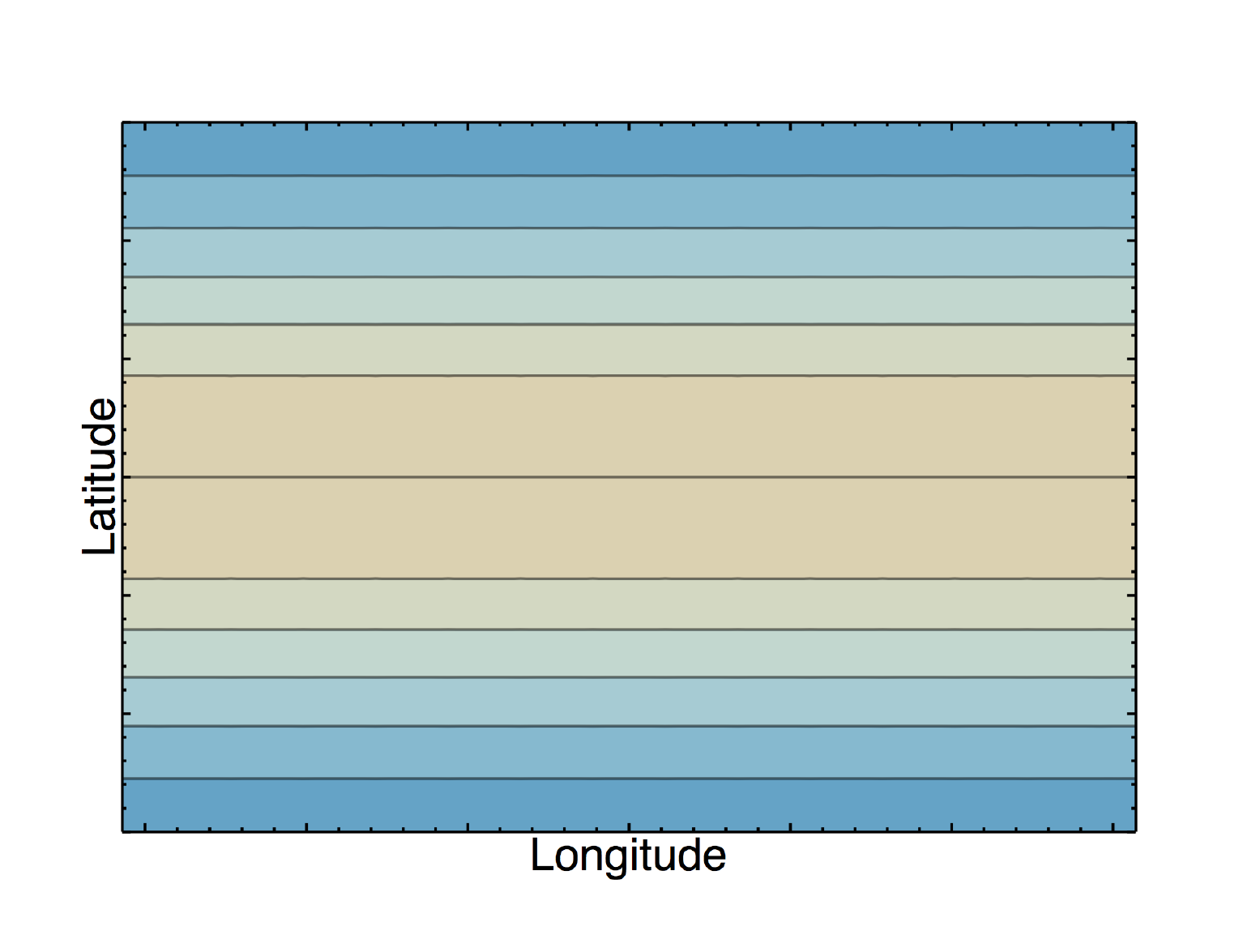}{0.24\textwidth}{(c) Height perturbation from jet.}
\fig{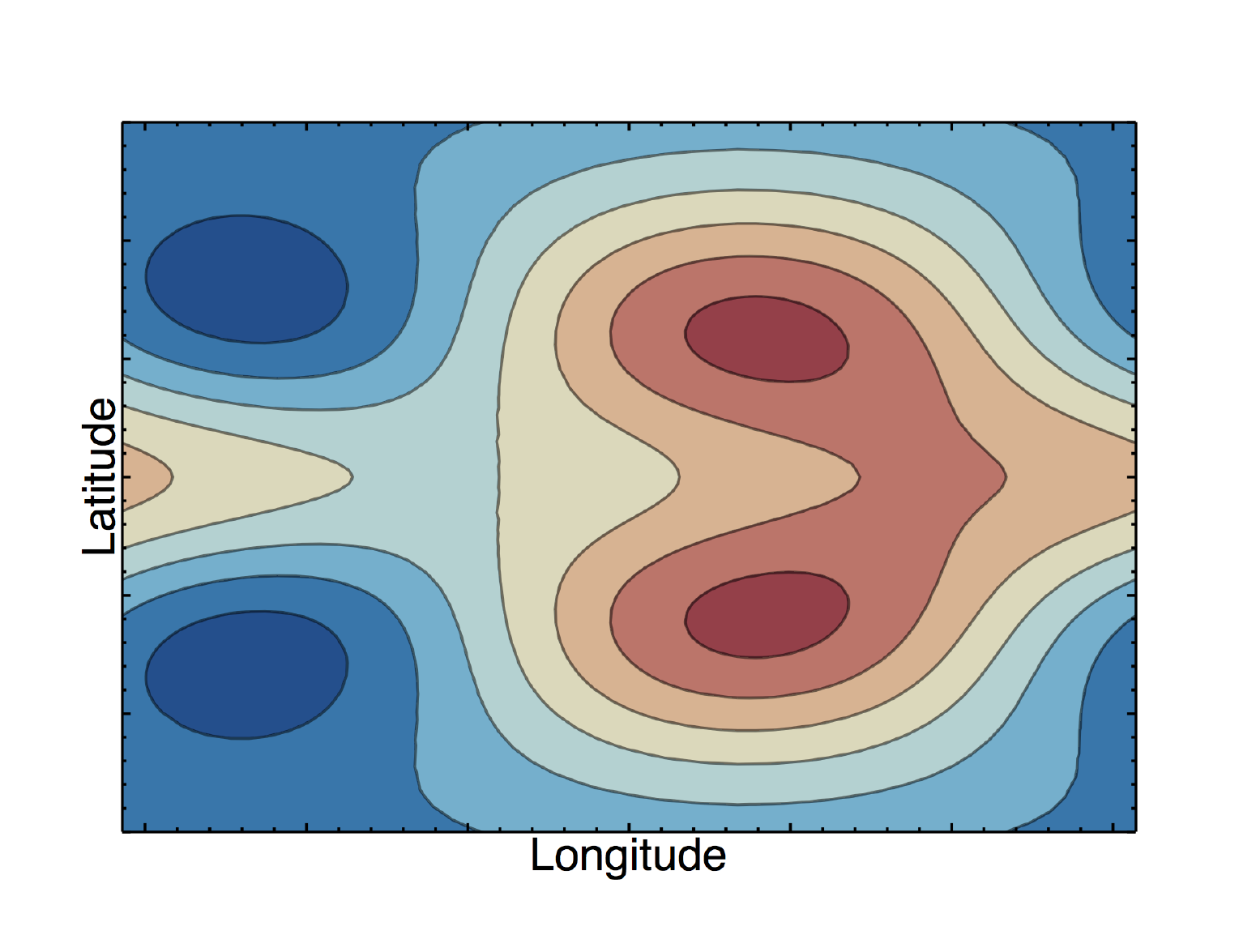}{0.24\textwidth}{(d) Sum of (b) and (c), giving the same form as the full pseudo-spectral solution in Section \ref{sec:forced-system}.}
}
\caption{Explanation of the forced solutions in a shear flow in Section \ref{sec:forced-system} and Figure \ref{fig:shear-2D}, simplified to a sum of analytic solution and mean zonal flow.}\label{fig:first-order-solutions}
\end{figure*}

Figure \ref{fig:first-order-solutions} shows a first-order model of these effects, using the analytic solutions of the shallow-water equations in a uniform flow. It shows how the jet height perturbation completes the global circulation pattern, particularly the hot spot pattern. The hot-spot in the second plot in Figure \ref{fig:first-order-solutions} (without the jet height perturbation) resembles a Rossby wave. Summed with the height perturbation in the third plot, it becomes centered on the equator and matches the form of the full solution in Section \ref{sec:forced-system} and the GCM simulations in Section \ref{sec:gcm-results}. We therefore agree with \citet{tsai2014three} that the hot-spot shift is caused by an eastward Doppler-shift of the stationary Kelvin and Rossby waves towards $+90\degr$, rather than eastward propagating Kelvin waves or pure heat transport from air in the eastward flowing jet.

The analytic solutions in Figure \ref{fig:first-order-solutions} are approximately the first-order terms in the pseudo-spectral expansion in Appendix \ref{sec:app-ps-method}. The pattern in Figure \ref{fig:shear-2D} is more complex than the sum of patterns in Figure \ref{fig:first-order-solutions} as the flow $\bar{U}(y)$ is not constant in $y$, but the eastward shift is still the most important effect. The shear flow introduces a tilt to the stationary Rossby waves, which we suggest is due to the faster flow at the equator producing a stronger Doppler shift there.

The linear model in \citet{showman2011superrotation} does not include a background zonal flow, which may be why it matches their GCM in spin-up but not in equilibrium. \citet{showman2011superrotation} also compare the linear model to numerical solutions of the nonlinear shallow-water equations, which should include the effect of any background flow. The nonlinear solutions in Figure 8 of \citet{showman2011superrotation} do have a mean zonal flow but it appears that either the damping is too strong, or the jet is too weak or narrow, to produce a hot-spot shift. In their case (a), the maximum zonal speed is approximately $100\ \textrm{ms}^{-1}$, which we suggest in Section \ref{sec:sw-equations} is not large enough for a significant phase shift on a Hot Jupiter. Their case (b) has a maximum zonal wind of approximately $700\ \textrm{ms}^{-1}$, which is large enough for a phase shift, but the jet is so narrow that this shift only happens very close to the equator, shown by the narrow eastward ``bump'' in the hot-spot in the top right panel of their Figure 8. If the cases are sufficiently strongly damped, then a zonal flow may still not produce a hot-spot shift. Equation \ref{eqn:project-coeff-flow} shows that the projection coefficient will be mostly real if $\alpha$ is much larger than $k_{x}\bar{U}$, so the maximum of the wave response will be in phase with the forcing (at the substellar point, with no hot-spot shift).

\section{Linearized Shallow-Water Equations in Shear Flow on a Sphere}\label{sec:sphere-solutions}

Linearizing the shallow-water equations about a jet on an equatorial beta-plane preserves the link to the intuitive, exact solutions of \citet{matsuno1966quasi}.  However, the beta-plane approximation is very inaccurate at high latitudes, and is not well suited to represent the effect of parameters like rotation rate on the latitudinal extent of the waves and circulation. The $y$ coordinate in the previous sections is non-dimensionalized to the equatorial Rossby radius of deformation.

Converting the $y$ coordinate on the beta-plane to a fraction of planetary radius shows how the solution varies with latitude, which works well for planets where the equatorial Rossby radius is smaller than the planetary radius. This is why the previous beta-plane solutions resemble both the solutions in spherical coordinates in this section, and the GCM simulations below. However, this conversion leads to serious inaccuracies for planets with a larger equatorial Rossby radius than planetary radius, as the \citet{matsuno1966quasi} beta-plane solution assumes that the scale of the heating (the size of the tidally locked planet in our case) is the same as the equatorial Rossby radius.

In this section, we solve the shallow-water equations linearized about an equatorial jet in spherical coordinates. This represents the latitudinal dependence of our solutions better, and lets us compare them directly to GCM simulations. We use the same pseudo-spectral method as before, with the Legendre polynomials as our basis set. We emphasize that the beta-plane solutions are still the most useful for understanding the global circulation, as they expand the forced solutions in term of the free modes of the shallow-water system in zero background flow \citep{matsuno1966quasi}. The spherical solutions lose the link to these intuitive solutions, but do represent the Coriolis force properly, and let us relate our results to real planets more closely.

The non-dimensional linearized spherical shallow-water equations are:

\begin{widetext}
\begin{equation}\label{eqn:sphere-sw-eqns}
  \begin{gathered}
  \frac{\partial u}{\partial t} + \alpha_{dyn} u + \frac{1}{\cos \phi} \frac{\partial (\bar{U}u)}{\partial \theta} + \frac{v}{\cos \phi} \frac{\partial}{\partial \phi}(\bar{U} \cos \phi) - v\sin\phi + \frac{G}{\cos \phi} \frac{\partial h}{\partial \theta} = 0\\
  2\bar{U}u\tan\phi + u \sin \phi + \frac{\partial v}{\partial t} + \alpha_{dyn} v + \frac{1}{\cos\phi}\frac{\partial \bar{U} v}{\partial \theta} + G \frac{\partial h}{\partial \phi} = 0 \\
  \frac{1+\bar{H}}{\cos \phi} \frac{\partial u}{\partial \theta} + \frac{1}{\cos \phi} \frac{\partial}{\partial \phi} ((1+\bar{H})v \cos\phi) + \frac{\partial h}{\partial t} + \alpha_{rad} h + \frac{\bar{U}}{\cos \phi} \frac{\partial h}{\partial \theta} = F
  \end{gathered}
\end{equation}
\end{widetext}

where $\theta$ is the longitude, $\phi$ is the latitude, $m$ is the zonal wavenumber, F is the forcing, $\Omega$ is the angular speed of the planet, and $G=gH_{0}/a^{2}\Omega^{2}$ where $H_{0}$ is the scale height. The equations have been non-dimensionalized with a length scale equal to the planet radius $a$ and a time scale of $\Omega^{-1}$ \citep{dikpati2001spherical}.

As before, the zonal flow $\bar{U}(\phi)$ is balanced by a height field $\bar{H}(\phi)$. Substituting $\bar{U}(\phi)$ into the second line of Equation \ref{eqn:sphere-sw-eqns} gives a height field in gradient wind balance with the zonal flow:

\begin{equation}\label{eqn:spherical-H-definition}
  \frac{\partial  \bar{H(\phi)}}{\partial \phi} = -\frac{1}{G} ( (\bar{U}(\phi))^{2}\tan{\phi} + 2 \bar{U}(\phi) \sin{\phi} )
\end{equation}

The background flow is set to be:

\begin{equation}
    \bar{U}(\phi)=U_{0}\cos(\phi)e^{-(\phi/\phi_{0})^{2}}
\end{equation}

The $\cos(\phi)$ factor is required as the flow must be zero at the poles for Equation \ref{eqn:spherical-H-definition} to be valid. We assume the perturbations have the form $f(\phi) \exp [ i(m \phi - \omega t)]$. We set $\partial / \partial t = 0$ and insert the forcing $Q$, to find the response to stationary forcing:

\begin{widetext}
\begin{equation}\label{eqn:forced-sphere-sw}
  \begin{pmatrix}
  \alpha_{dyn} + i m\bar{U} / \cos\phi & \frac{\partial\bar{U}\cos\phi}{\partial \phi}-\sin\phi & \frac{i m G}{\cos\phi} \\
  2\bar{U}\tan\phi + \sin\phi & \alpha_{dyn} + \frac{i m\bar{U}}{\cos\phi} & G \frac{\partial}{\partial \phi} \\
  \frac{i m (1+\bar{H})}{\cos\phi} & \frac{1}{\cos \phi} \frac{\partial(1+\bar{H})\cos\phi)}{\partial \phi} & \alpha_{rad} + \frac{i m \bar{U}}{\cos\phi}
  \end{pmatrix}
  \begin{pmatrix}
  u \\
  v \\
  h
  \end{pmatrix}
  =
  \begin{pmatrix}
  0 \\
  0 \\
  Q(\phi)
  \end{pmatrix}
\end{equation}
\end{widetext}

We use the same method as in Appendix \ref{sec:app-ps-method}, with the Legendre polynomials rather than the parabolic cylinder functions \citep{wang2016hough}. We solve the equations with y-coordinate $\mu = \sin \phi$, in the domain $-1 < \mu < 1$. This matches the domain of the Legendre polynomials and reduces most of the trigonometric terms and derivatives in the calculation to powers of $\mu$. We rescale the height variable $h$ in the calculation to avoid singularities at the poles, which we explain in Appendix \ref{sec:app-spherical}.

Figure \ref{fig:spherical-tests} shows the non-dimensional spherical solutions for the forced problem with no shear, the Doppler-shifting of this solution, and the total forced solution in a shear flow. They have the same form as Figure \ref{fig:shear-2D}, showing that the beta-plane approximation produces much the same results as the spherical geometry. It appears that the $n=1$ Rossby mode is shifted past the substellar point in this case, unlike the beta-plane system discussed in the previous section. The beta-plane system could be made to match the spherical solution more closely by choosing different values for parameters like zonal flow speed or damping rates.

In the figures in this section, we used the same damping rates as before, $\alpha_{rad}=\alpha_{dyn}=0.2$, which are also used in \citet{matsuno1966quasi} and \citet{showman2011superrotation}. The time and length scales specified above require that the radius is $a=1$ and the angular frequency is $\Omega=1$. We set the jet width $\phi_{0}=\pi/3$, but the exact value of this is not vital to the form of the solution (unless it is unphysically narrow, or so broad as to be uniform). An eastward shift of approximately $90 \degr$ requires $U_{0} \approx 0.5$, so we set this as our maximum flow speed.

We set the forcing $Q_{0}=0.5$ so that the background jet height and forced wave response had comparable magnitude, which is more illuminating than either one dominating the other. This value of $Q_{0}$ is comparable to those used in \citet{showman2011superrotation}, and we will show the effect of varying it later.

We will consider Earth-like tidally locked planets with rotation rates in the range 1 to 10 days, which have values of $G=gH/a^{2}\Omega^{2}$ varying from 0.3 to 30. For comparison, Hot Jupiters with rotation rates of 1 to 2 days have a value of $G\sim 0.1$. In Figure \ref{fig:spherical-tests} we set $G=1$. Later we will show how varying all of these parameters changes the forced solution and often makes either the jet or the wave response dominate.

\begin{figure*}
 \gridline{\fig{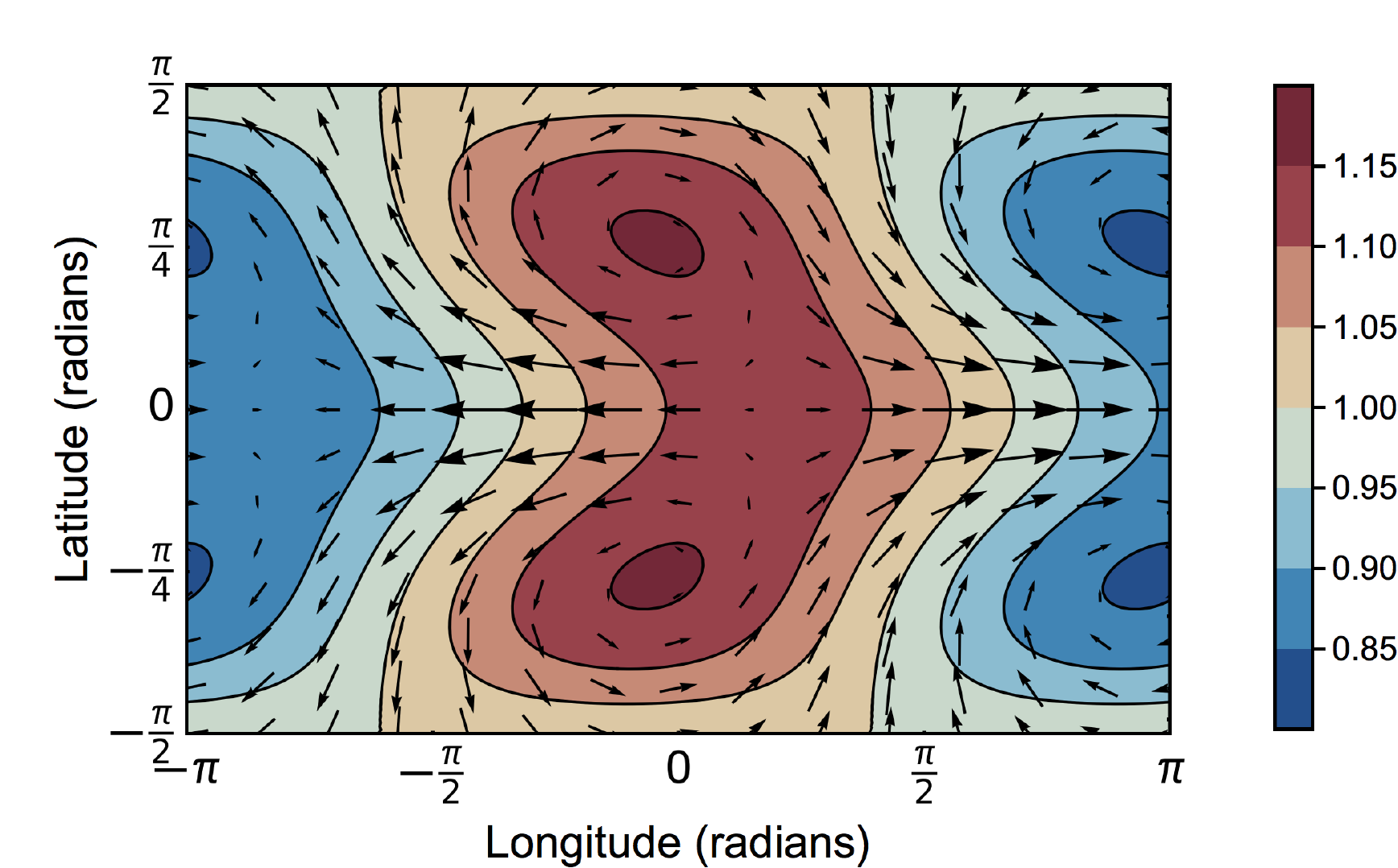}{0.33\textwidth}{Height field of the forced solution with $\bar{U}(\phi)=0$.}
 \fig{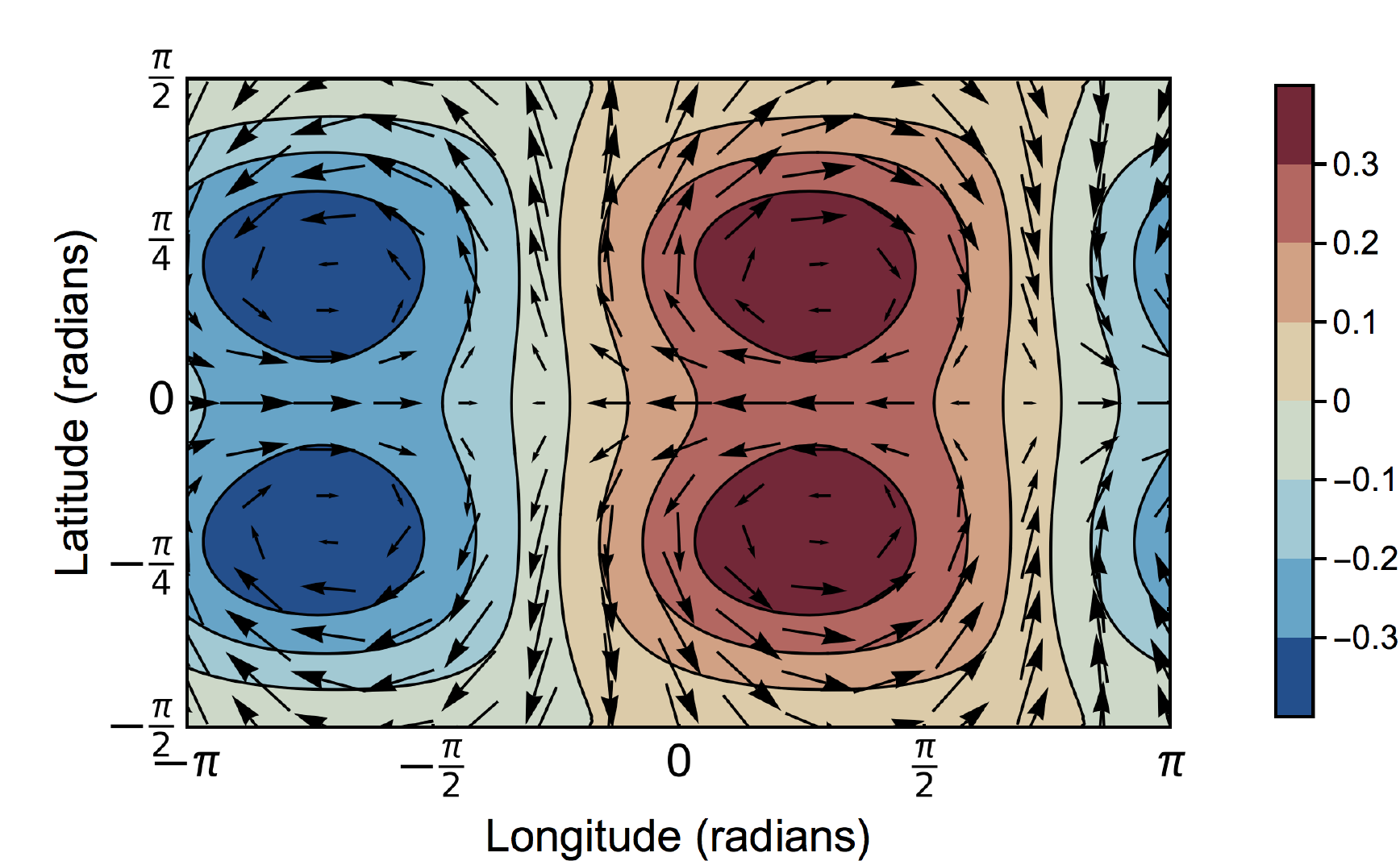}{0.33\textwidth}{Forced wave solution with a background flow, minus the flow itself.}
 \fig{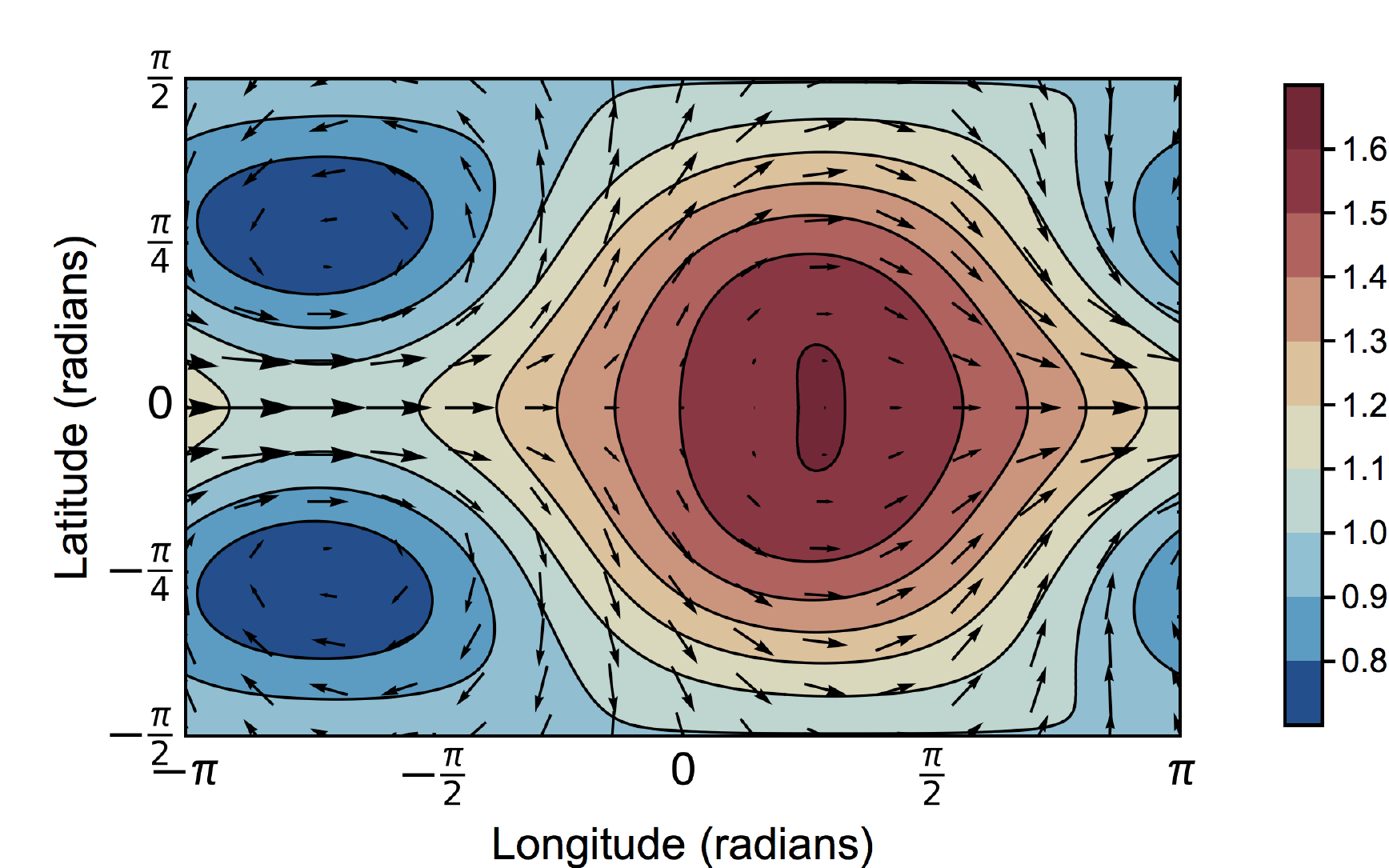}{0.33\textwidth}{Forced wave solution with a background flow.}
           }
\caption{Forced linear shallow-water solutions in spherical geometry with and without background shear flows. The parameters of the model are $\bar{U}=0.5 \cos\phi \exp(-(\phi/\phi_{0})^{2})$, $\phi_{0}=\pi / 3$, $\alpha_{dyn}=\alpha_{rad}=0.2$, $G=1$, $Q=0.5\cos(\phi)$, as discussed in Section \ref{sec:sphere-solutions}.}\label{fig:spherical-tests}
\end{figure*}

\section{Scaling Relations}\label{sec:scaling-relations}

In this section, we will simplify our linear shallow-water model to a 1D model of the most dominant terms on the equator. In Section \ref{sec:1d-scaling}, we use this 1D model to predict simple scaling relations between the planetary parameters (damping rate, jet speed), and the phase and amplitude of the equatorial height variation (i.e. the hot-spot shift and day-night contrast). In Section \ref{sec:2d-scaling} we will show that these simple relations qualitatively describe the behaviour of our pseudo-spectral solutions to the full 2D linear equations.

\subsection{Scaling Relations in 1D Model}\label{sec:1d-scaling}

The pseudo-spectral solutions to Equation \ref{eqn:forced-sw} such as Figure \ref{fig:shear-2D} show the forced response of a shallow-water system with particular parameters such as damping rate and jet speed, but do not provide a direct relation between these parameters and observable quantities such as the position of the atmospheric hot-spot. In this section, we calculate the on-equator height response on the beta-plane, where $\beta=0$ so we can ignore the effects of waves in the linear shallow-water equations. This gives a simpler 1D model which predicts a direct scaling relation between the equatorial height maximum and the planetary parameters.

For $y=0$ (on the equator), retaining only damping and advection terms, the third line of Equation \ref{eqn:shear-sw-equations} simplifies from

\begin{equation}
  \frac { \partial H ^ { \prime } u } { \partial x } + H^ { \prime } \frac { \partial v } { \partial y } - y \bar{ U } ( y ) v + \frac { \partial \bar { h } } { \partial t } + \alpha h + \frac { \partial \bar { U } ( y ) h } { \partial x } = Q_{0}e^{-y^{2}/2}
\end{equation}

to

\begin{equation}
   \frac { \partial \bar { h } } { \partial t } + \alpha h + \frac { \partial  U_{0} h } { \partial x } = Q_{0}
\end{equation}

Then as before, we set $\partial/ \partial t = 0 $ and $\partial/ \partial x =i k_{x}$:

\begin{equation}\label{eqn:on-equator-height}
-\alpha h (y=0) + i k_{x} { U }_{0} h (y=0) = Q_{0}
\end{equation}

So then the real part of the full solution $h (x,y=0)=h (y=0)e^{i k_{x}x}$ is:

\begin{equation}\label{eqn:phi-curve}
  h (x,y=0)=\frac{Q_{0}}{\alpha^{2}+k_{x}^{2}U_{0}^{2}}(-\alpha \cos(k_{x}x)+k_{x}U_{0}\sin{k_{x}x})
\end{equation}

This corresponds to a sinusoidal height perturbation on the equator. The magnitude of this height perturbation is approximately:

\begin{equation}
  h_{0}=\frac{\alpha+k_{x}U_{0}}{\alpha^{2}+k_{x}^{2}U^{2}}Q_{0}
\end{equation}

The position of the hot-spot $x_{0}$ is the maximum of this height perturbation, where $\partial h / \partial x = 0$:

\begin{equation}\label{eqn:x0}
  x_{0}=\frac{1}{k_{x}}\tan^{-1}(k_{x}\frac{U_{0}}{\alpha})
\end{equation}

Note that this is the same as the approximation to the hot-spot shift from \citet{zhang2017dynamics}, $\lambda_{s}=\tan^{-1}(\frac{\tau_{rad}}{\tau_{adv}})$, if we set $\tau_{rad}=1/\alpha$ and $\tau_{adv}=k_{x}/U_{0}$. This is because their prediction for the hot-spot shift is calculated from Equation 41 in \citet{zhang2017dynamics}:

\begin{equation}
    \frac { \partial T } { \partial t } + \frac { 1 } { \tau _ { \mathrm { adv } } } \frac { \partial T } { \partial \lambda } = \frac { T _ { \mathrm { eq } } ( \lambda ) - T } { \tau _ { \mathrm { rad } } }
\end{equation}

This is equivalent to our Equation \ref{eqn:on-equator-height}, if we set $T\equiv h $, $\tau_{adv}=k_{x}/U_{0}$, $\lambda \equiv x$, and $( T _ { \mathrm { eq } } ( \lambda ) - T ) /  \tau _ { \mathrm { rad }  }=Q_{0}$. \citet{zhang2017dynamics} use a more complex day-night forcing difference than our $Q(x,y)$ to solve this equation, which is why our prediction for the hot-spot shift only matches the approximate solution to their equation.

This 1D model predicts a hot-spot shift varying from $0\degr$ to $90\degr$ east of the substellar point. This is similar to the linear shallow-water model of \citet{tsai2014three}, where the uniform background flow could shift the maximum of the wave response to a maximum of $90\degr$ east. With zero zonal flow, the 1D model predicts a hot-spot shift at $0\degr$, the substellar point. This is only the case in our 2D linear solutions if the damping is strong, i.e. if the damping rate $\alpha$ is much greater than the Rossby and Kelvin wave frequencies. At higher zonal flow $U_{0}$ the 1D model predicts a hot-spot shift approaching $90\degr$. This can agree with the 2D model -- see Figure \ref{fig:sp-alpha-effect}, where the second plot has very low damping -- but in general, wave terms and damping terms prevent such a large hot-spot shift (as in Figure \ref{fig:shear-2D}).

So, the 1D model balancing advection and damping is useful for intuition and basic scaling relations, and can predict the hot-spot shift of a tidally locked planet to first order, but is not accurate when wave effects become important. The 2D model is more useful when considering the full disk-integrated phase curve, where the off-equator response is very different to the 1D one-equator model.

\subsection{Scaling Relations in 2D Model}\label{sec:2d-scaling}

In the previous section, we used a 1D model to calculate simplified predictions of how observables such as hot-spot shift and day-night contrast should scale with planetary parameters such as damping rate. In this section, we will test the predictions from the 1D model and our discussion in previous sections.

The solutions such as those in Figure \ref{fig:shear-2D} used a representative set of non-dimensional parameters, which roughly matched a typical Earth-sized tidally locked planet or Hot Jupiter. In this section, we identify the free parameters of the beta-plane solution and the solution in spherical coordinates, and vary these parameters to test the predictions of the previous section.

\begin{figure*}
 \gridline{\fig{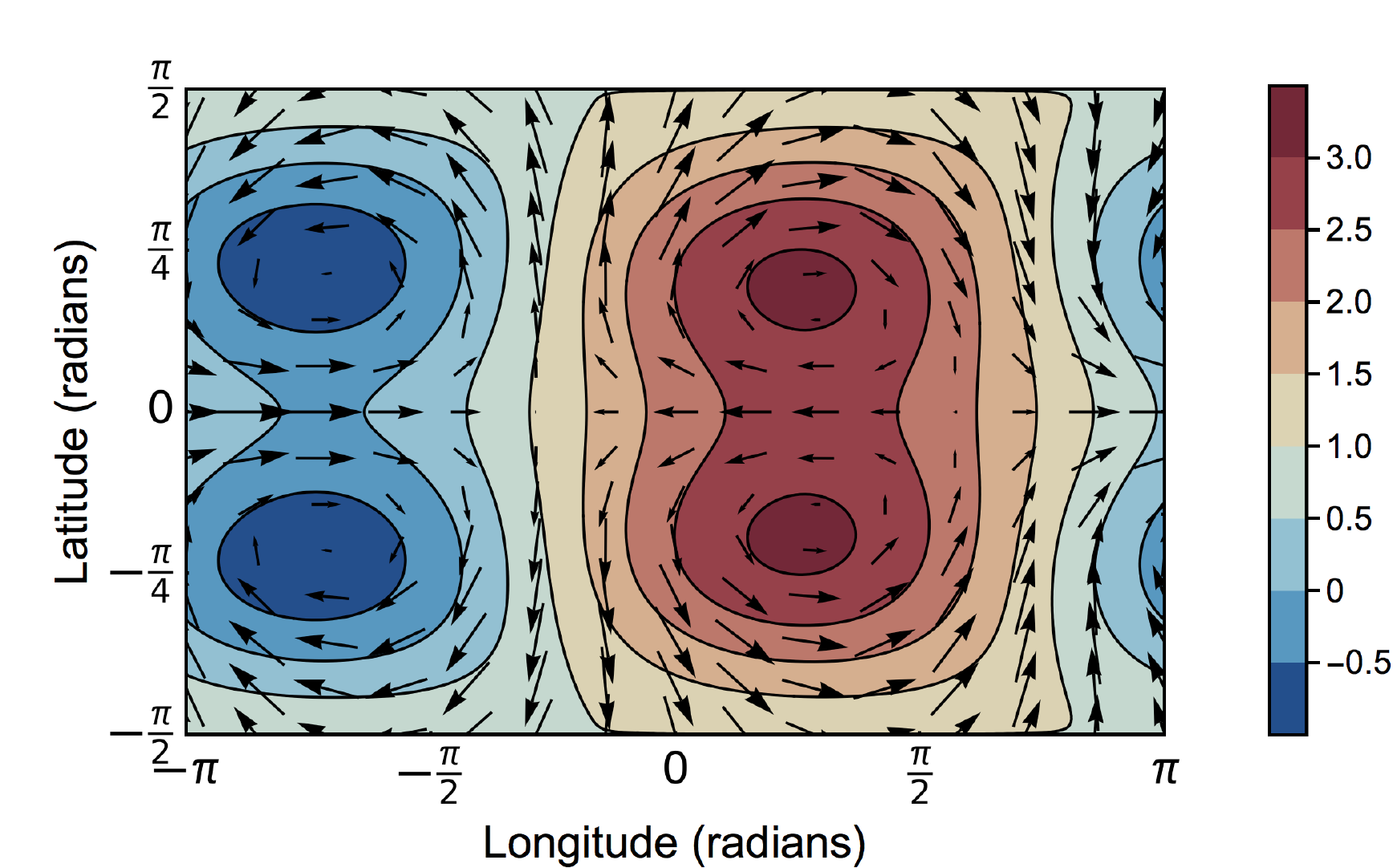}{0.45\textwidth}{$Q_{0}=2.5$, a high forcing strength gives a strong response to forcing and a wave-dominated response.}
 \fig{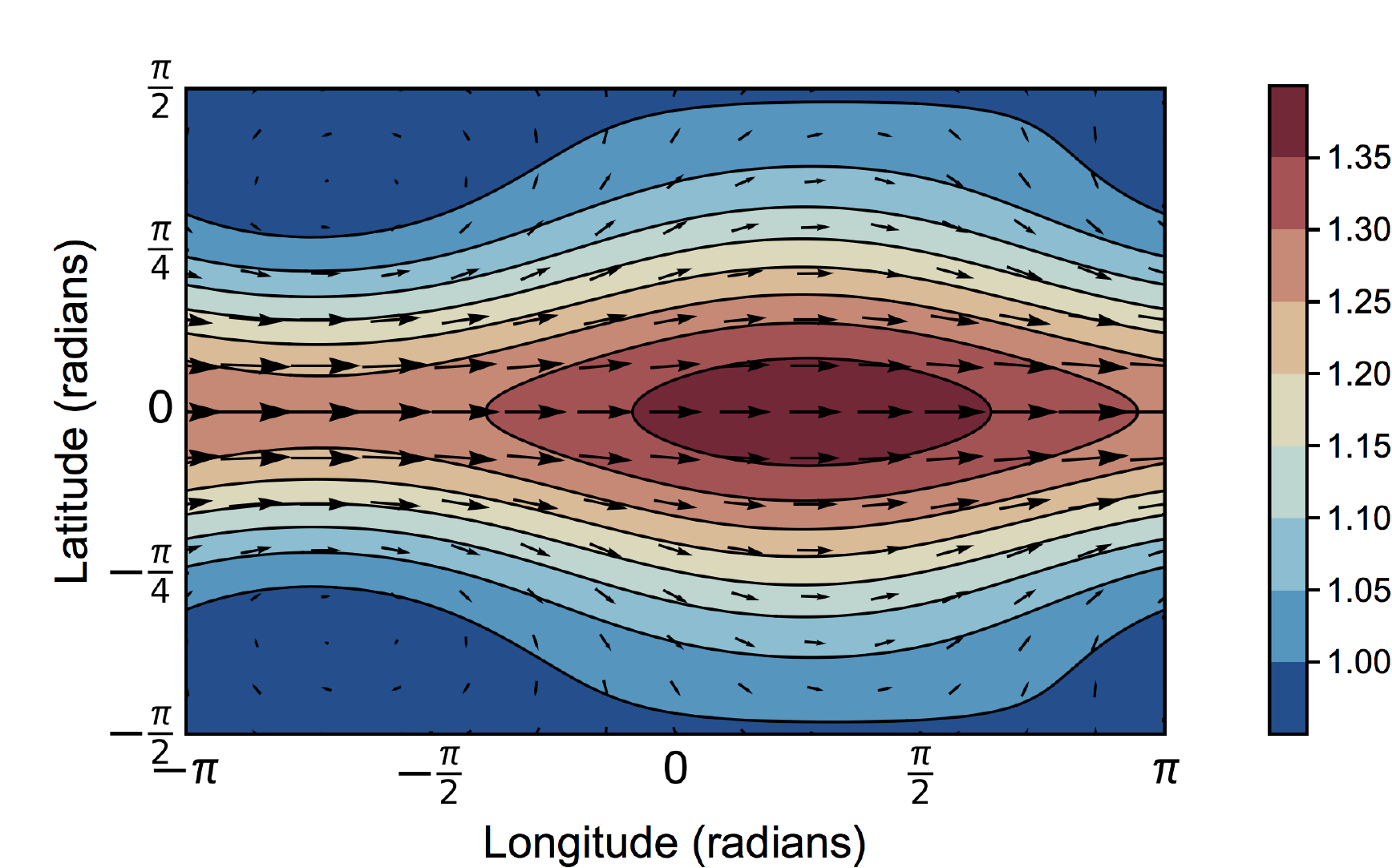}{0.45\textwidth}{$Q_{0}=0.1$, a low forcing strength gives a weak response to forcing and a jet-dominated response.}
}
\caption{Height fields of the response to forcing in spherical coordinates for different values of forcing strength $Q_{0}$, which controls the relative magnitude of the response to forcing and the background jet height and velocity.}\label{fig:sp-q-effect}
\end{figure*}

The free parameters on the beta-plane are the background jet strength and  width $U_{0}$ and $y_{0}$, the forcing strength $Q_{0}$, and the radiative and dynamical damping rates $\alpha_{rad}$ and $\alpha_{dyn}$. The solution in spherical geometry also depends on these parameters. The jet and forcing strengths $U_{0}$ and $Q_{0}$ affect the relative magnitude of the jet and wave responses without changing the actual wave solutions itself, as we show below for the case in spherical geometry. Figure \ref{fig:sp-q-effect} shows how changing the relative magnitude of the forcing $Q_{0}$ and the jet $U_{0}$ changes the solution in spherical coordinates. For strong forcing, the wave response dominates and the height field is very asymmetric, resulting in a large day-night contrast. For weak forcing or a strong jet, the jet velocity and height field dominates, leading to a more zonally homogeneous circulation.

\begin{figure*}
 \gridline{\fig{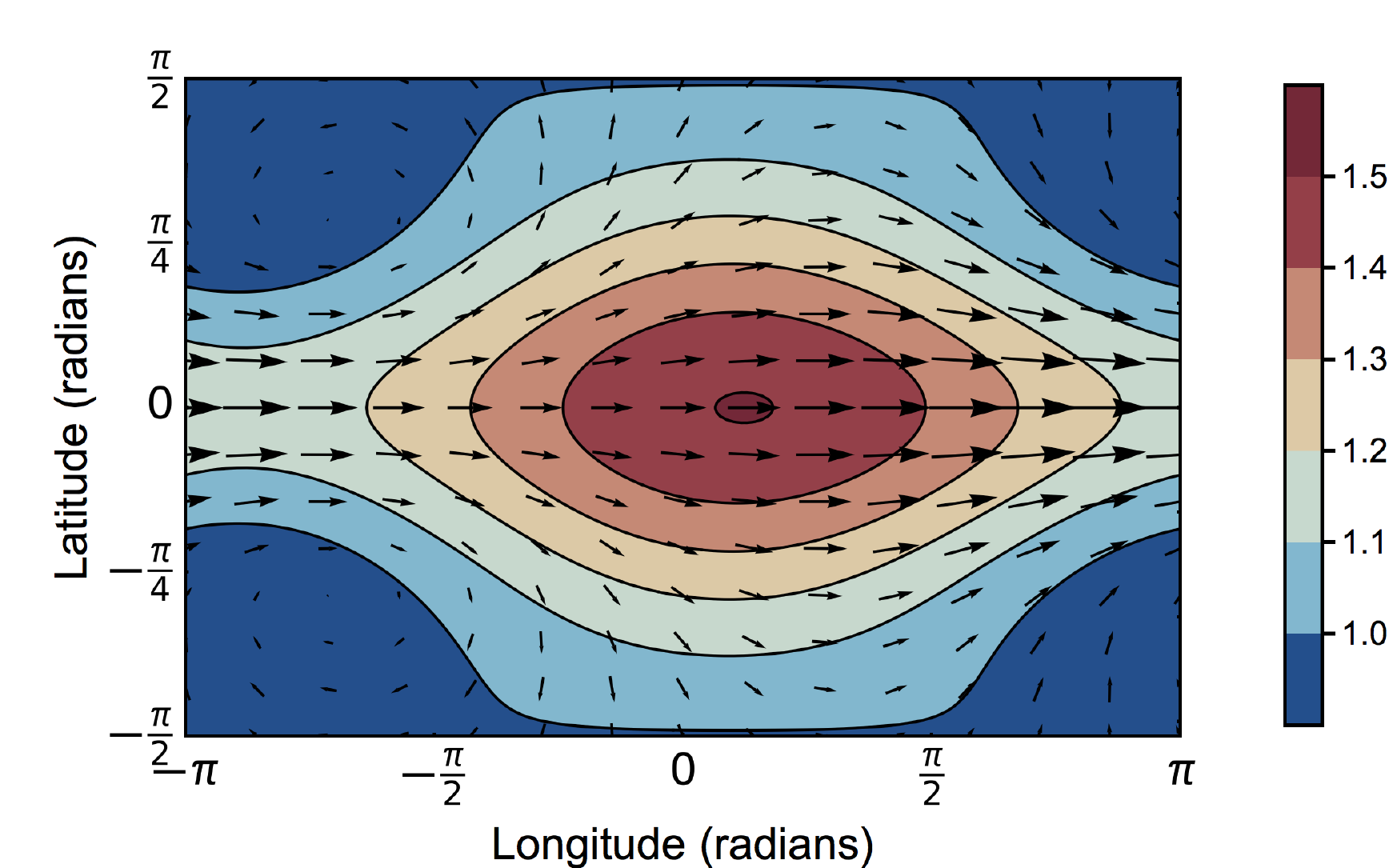}{0.45\textwidth}{$\alpha_{rad}=\alpha_{dyn}=1$, a high damping gives a weaker response to forcing which is more in phase with the forcing $Q$.}
 \fig{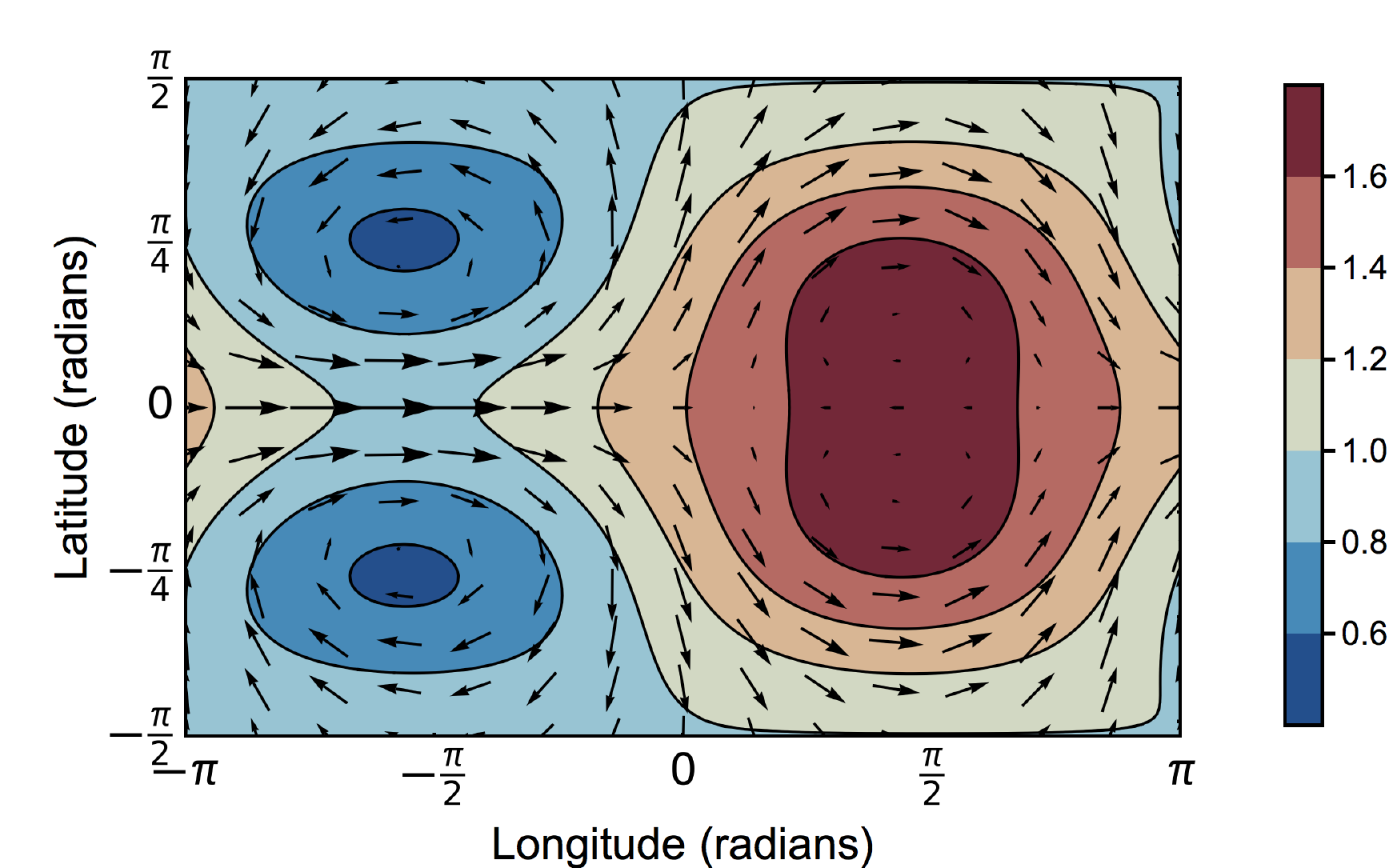}{0.45\textwidth}{$\alpha_{rad}=\alpha_{dyn}=0.04$, a low damping gives a stronger response to forcing which is shifted further eastwards.}
}
\caption{Height fields for high and low damping rates $\alpha$. The damping rate affects the strength of the forced wave response and the shift of the response due to the background jet flow.}\label{fig:sp-alpha-effect}
\end{figure*}

Section \ref{sec:1d-scaling} predicts that increasing the damping rate $\alpha$ should decrease the hot-spot shift and decrease the day-night contrast. Figure \ref{fig:sp-alpha-effect} shows that varying the damping $\alpha$ has this effect in the full solutions (both $\alpha_{dyn}$ and $\alpha_{rad}$ are set to be equal here). We can interpret this effect using Equation \ref{eqn:project-coeff-flow} -- increasing the damping increases the real part of the projection coefficient, so the maximum of the forced solution is more in phase with the maximum of the forcing. The damping also affects the relative strength of the Rossby and Kelvin parts of the forced response. For strong damping, the Kelvin part dominates, leading to the single maximum centered on the equator. For weak damping, the Rossby part dominates, leading to the two maxima off the equator.

The linear shallow-water equations in a spherical geometry described in Section \ref{sec:sphere-solutions} also have the additional parameter $G=(c/a\Omega)^{2}=gH/a^{2}\Omega^{2}$. This extra parameter is due to the extra degree of freedom added by constraining the latitudinal direction, unlike on the beta-plane.

The additional free parameter $G$ in the spherical geometry describes the effect of rotation rate, planetary radius, and gravity wave speed. It is the square of the ratio of the Rossby radius of deformation to the planetary radius, and as such represents the relative latitudinal scale of the waves in the system to the size of the planet. It is also equivalent to the square of the WTG parameter, which \citet{pierrehumbert2016nondilute} used to characterise the global circulation of tidally locked planets.

\begin{figure*}
 \gridline{\fig{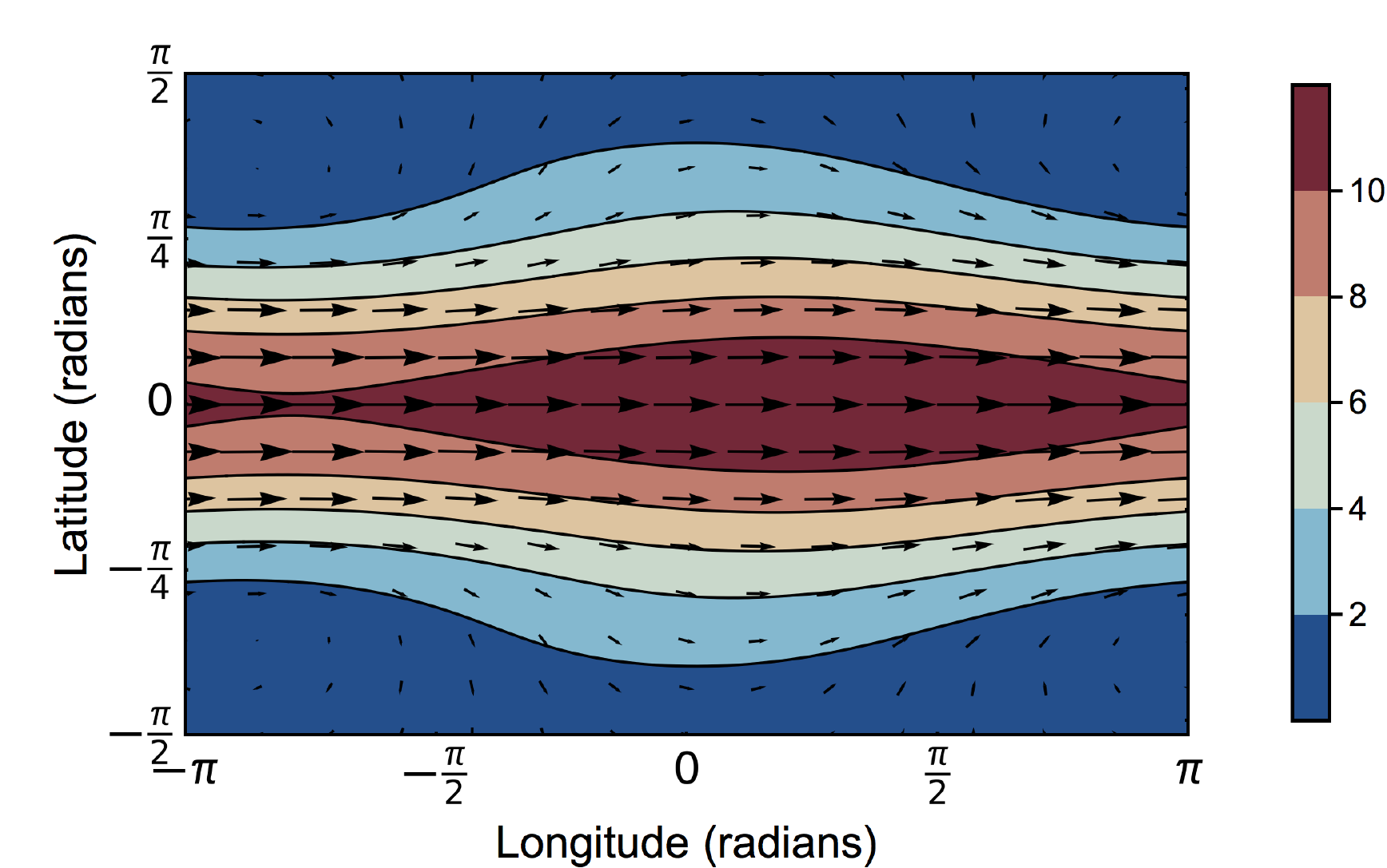}{0.45\textwidth}{$G=1/30$, for an orbital period of 8 hours (or a low gravity or scale height, or a high radius).}
 \fig{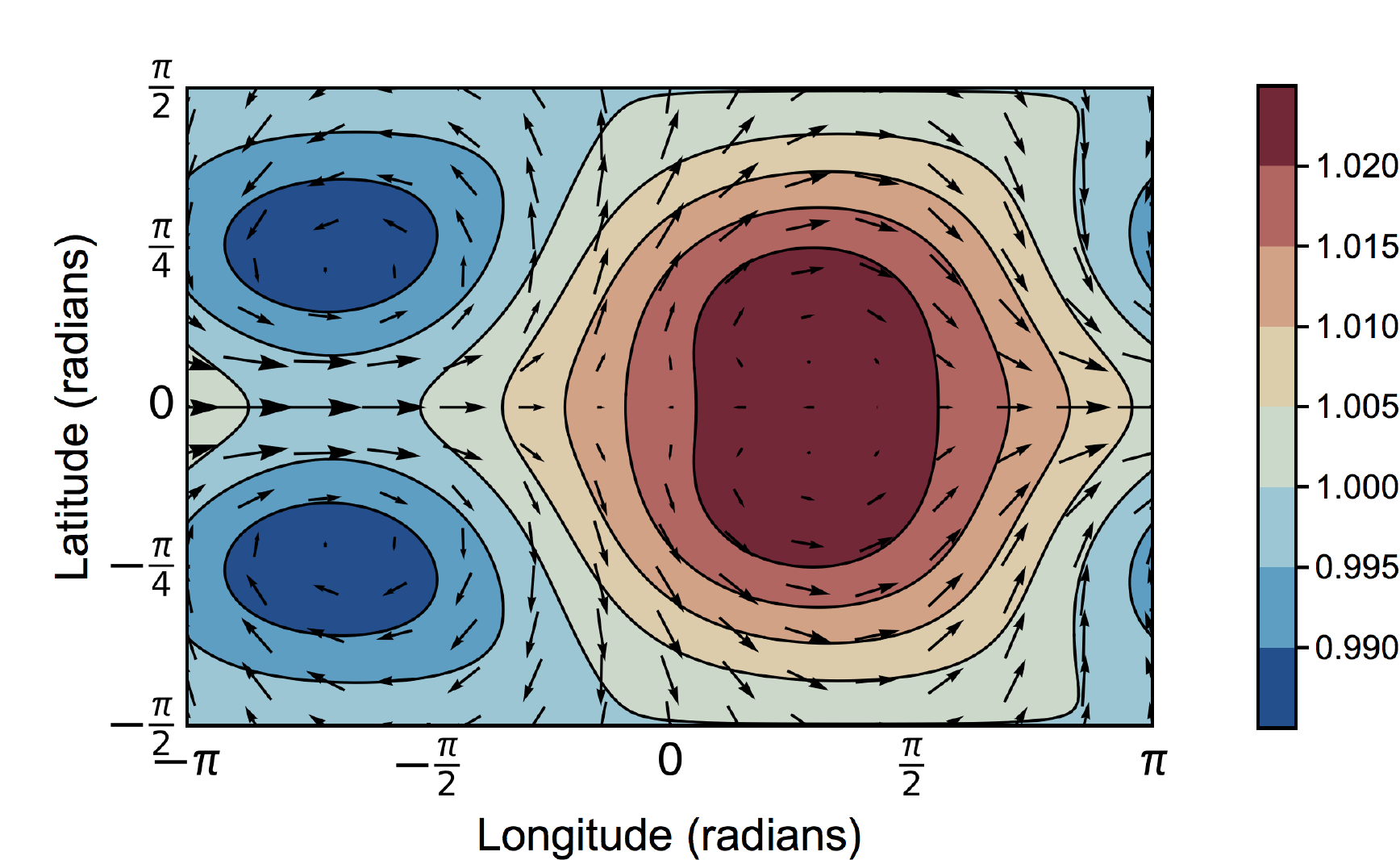}{0.45\textwidth}{$G=30$, for an orbital period of 10 days (or a high gravity or scale height, or a low radius).}
}
\caption{Height fields for low and high $G$. A low value of $G$ corresponds to a stronger jet height perturbation and a more zonally homogeneous global circulation with a smaller latitudinal extent, as discussed in Section \ref{sec:sphere-solutions}.}\label{fig:g-effect}
\end{figure*}

Equation \ref{eqn:spherical-H-definition} shows that $G$ is inversely proportional to the background jet height perturbation. Figure \ref{fig:g-effect} shows that for $G>>1$ (a low rotation rate $\Omega$) waves dominate the forced response, giving a large day-night contrast. For $G<<1$ the jet velocity and height dominate, giving a more zonally symmetric system. Our description of the effect of $G$ matches the qualitative predictions of previous studies, which interpreted changes in atmospheric dynamics on these planets by comparing the equatorial Rossby radius and the planetary radius \citep{koll2015phasecurves} \citep{showman2015circulation} \citep{carone2015regimes}.

This 2D linear model does not give such a straightforward prediction of the hot-spot shift and day-night contrast as the 1D model, but does capture the important off-equator behaviour and the effect of the planetary waves. It would be ideal for comparison with observational methods which retrieve 2D maps of the planets \citep{rauscher2018phase}.

\section{Comparison of Shallow-Water Model and GCM Results}\label{sec:gcm-results}

In this section, we discuss the results of tests comparing 3D simulations in our GCM Exo-FMS \citep{hammond2017climate} to the 2D pseudo-spectral solutions described in this paper. Exo-FMS is based on the GFDL Flexible Modelling System. Our tests had a 96 by 144 by 40 grid, a timestep of 100 seconds, and grey-gas radiation. The tests took at most 50 Earth days to reach their equilibrium circulation pattern and at most 300 days to reach a steady energy balance. Our ``equilibrium'' results in this section were an average from 500 to 1000 days. Our ``spin-up'' results were taken over the first 100 days.

We simulated Earth-sized planets with 1 bar atmospheres with the thermodynamical properties of nitrogen, longwave optical depth 1, shortwave optical depth 0, and the same incoming stellar radiation as on Earth and a 10 day orbital period unless otherwise specified. We chose a 10 day orbital period with these parameters for the default test as this appears to give a pattern where the jet height and velocity (the wavenumber-zero component of the mean GCM output) is comparable to the wavenumber-one wave response, similar to Figures \ref{fig:shear-2D} and \ref{fig:spherical-tests}. Very high or very low rotation rates lead to a less illuminating global circulation dominated either by a zonally homogeneous jet or a zonally inhomogeneous wave reponse. We will show the effect of varying the 10 day orbital period later.

\begin{figure}
\plotone{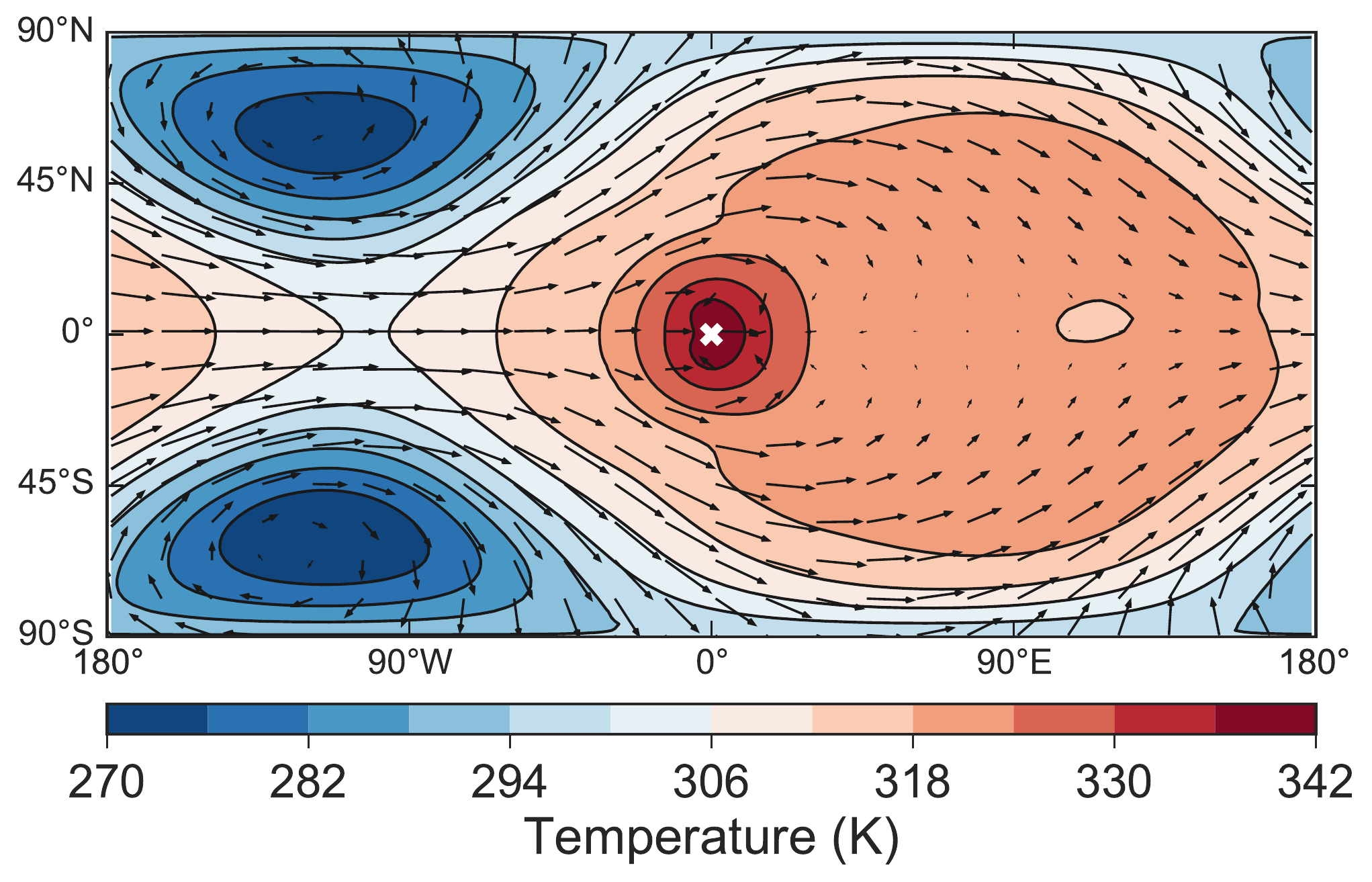}
\caption{Mean temperature and velocity fields from the GCM simulation discussed in Section \ref{sec:gcm-results}, corresponding to the half-surface pressure level. The substellar point is the white cross at $0 \degr$ latitude and $0 \degr$ longitude. This simulation has similar parameters to the linear solution in Figure \ref{fig:g-effect}.}\label{fig:example-gcm-results}
\end{figure}

Figure \ref{fig:example-gcm-results} shows the mean temperature and wind pattern in our simulations of a planet with the parameters given above, and a 10 day orbital period. The temperature distribution and wind direction qualitatively match our linear shallow-water solutions. The maximum zonal wind speed is about $100\textrm{ms}^{-1}$ The GCM results differ from the shallow-water model in the higher temperature at the substellar point, which may be due to rising hot air from the substellar point, which our shallow-water model does not account for.

Some GCM tests with high rotation rates have mid-latitude instabilities resembling travelling planetary waves, or baroclinic or barotropic instabilities \citep{instability2012polichtchouk} \citep{showman2015circulation} \citep{pierrehumbert2018review}. The time-averaged circulation of most of these planets still resemble the stationary wave pattern in Figure \ref{fig:example-gcm-results}. We suggest that the instabilities lead to equilibrated travelling waves which move through the stationary wave pattern without greatly changing it (as discussed in Section \ref{sec:free-solutions-subsection}), which is why our linear shallow-water model matches the mean global circulation pattern.

The shallow-water solutions and our plots of the mid-atmosphere in the GCM do not represent the temperature of every level of the atmosphere. GCM simulations of terrestrial planets have a surface temperature which is closely coupled to the incoming stellar radiation. Hot Jupiters have a varying temperature distribution with height, with an upper atmosphere dominated by radiative balance, sometimes with a temperature inversion. The mid-atmosphere circulation, where the wave patterns show up most clearly, is still very important to understanding observations and climate. Phase curve observations of Hot Jupiters normally find a hot-spot shift \citep{crossfield2015observations}, showing that the observations are of a level of the atmosphere where the jet and waves are important. On terrestrial planets, the strong mid-atmosphere circulation does affect the surface temperature, cooling the day-side and warming the night-side. The global circulation may have an even stronger effect on the climate if clouds are present \citep{kopparapu2017moist}.

\subsection{Equilibrium Global Circulation}

\begin{figure*}
 \gridline{\fig{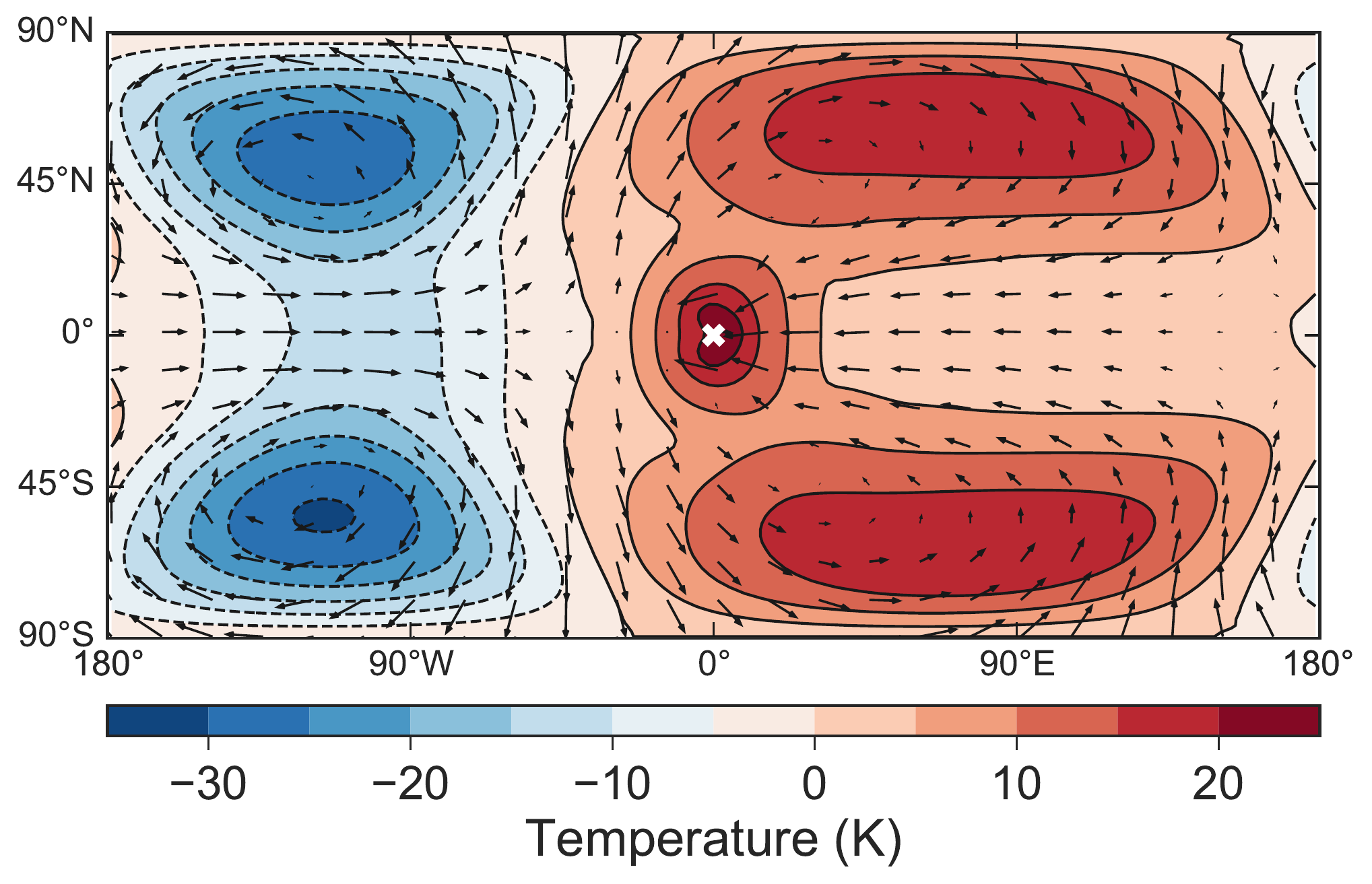}{0.45\textwidth}{Temperature and velocities minus zonal mean.}
\fig{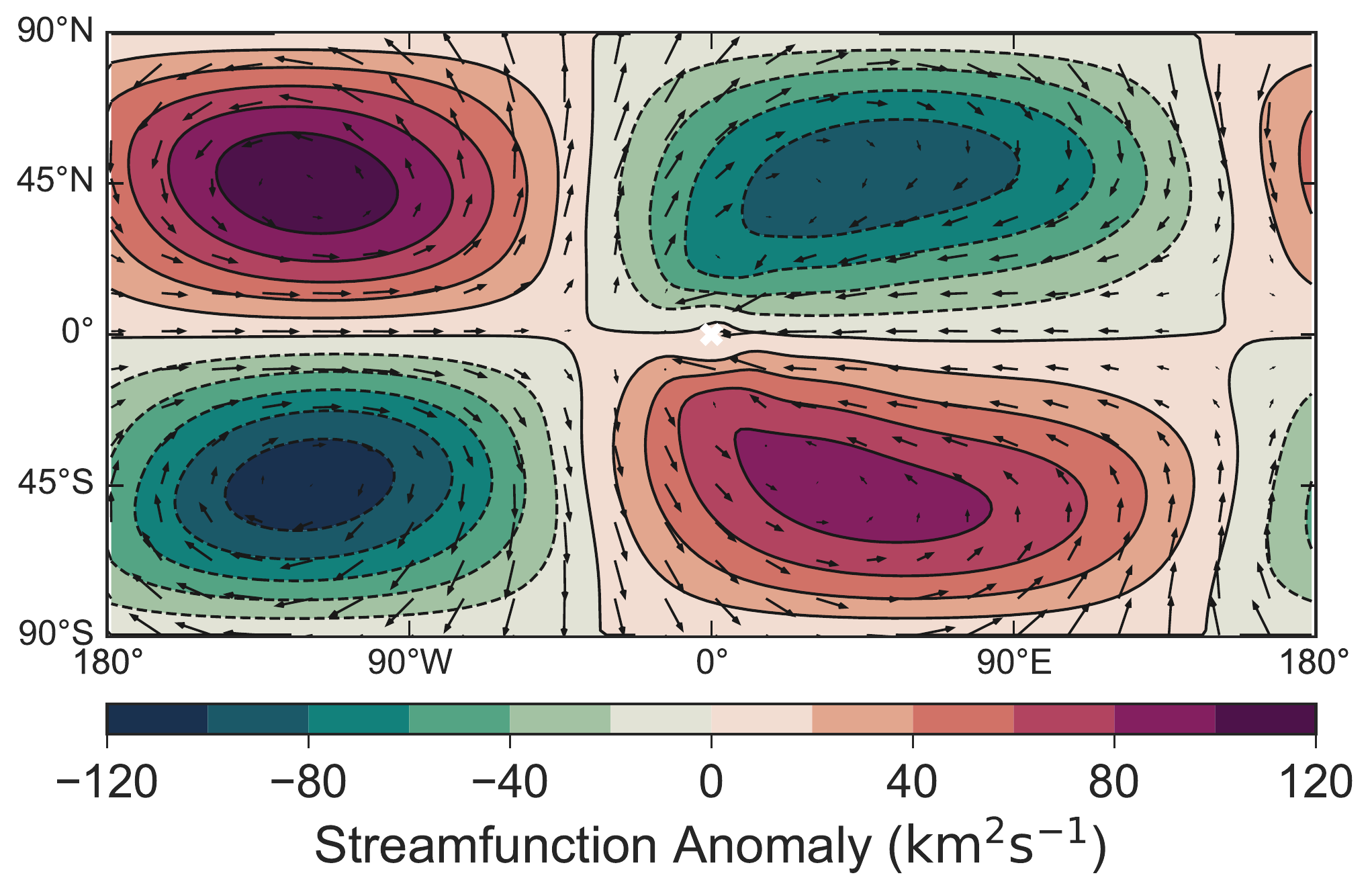}{0.45\textwidth}{Streamfunction and velocities minus zonal mean.}
}
\caption{Time-mean eddy fields (with zonal means subtracted) from the GCM simulation discussed in Section \ref{sec:gcm-results}, corresponding to the half-surface pressure level.}\label{fig:eddy-gcm-results}
\end{figure*}

The plots in Figure \ref{fig:eddy-gcm-results} show the time-averaged eddy streamfunction and eddy velocities of our simulations of the planet discussed above, with a 10 day orbital period. In Sections \ref{sec:sw-equations} and \ref{sec:forced-system}, we showed how the forced planetary waves are Doppler-shifted eastwards in the linear shallow-water model. This is clearest in the pattern of the eddy velocities in the GCM results in Figure \ref{fig:eddy-gcm-results}. The pattern of velocities is $180 \degr$ out of phase with the second plot in Figure \ref{fig:motivating-plot}, but is in phase with the third plot, which has been Doppler-shifted eastwards by a background flow.

This explains why in Figure \ref{fig:example-gcm-results} the equatorial $u$ velocity is lower at about $+90 \degr$, as the local Rossby wave velocity opposes it, and higher at about $-90 \degr$, as the local Rossby wave velocity adds to it. Studies such as \citet{carone2015regimes} and \citet{pierrehumbert2018review} have shown this pattern of Rossby wave velocities, which differs from the Rossby wave velocities in the non-shifted linear shallow water solutions \citep{matsuno1966quasi} \citep{showman2011superrotation}.

\subsection{Spin-up of Global Circulation}

The linear shallow-water model predicts that the Rossby and Kelvin components of the forced response should shift eastwards as the eastward zonal flow increases, up to a maximum of $+ 90 \degr$. In this section, we test if this is the case in the spin-up of our GCM. We initialized a tidally locked planetary atmosphere from rest with the parameters listed previously (idealized Earth-like conditions, with a 10 day orbital period). To compare the GCM results to the linear shallow-water results, we decomposed its daily output into Kelvin and Rossby components, following the method of \citet{boulanger1995propagation}.

\begin{figure*}
 \gridline{\fig{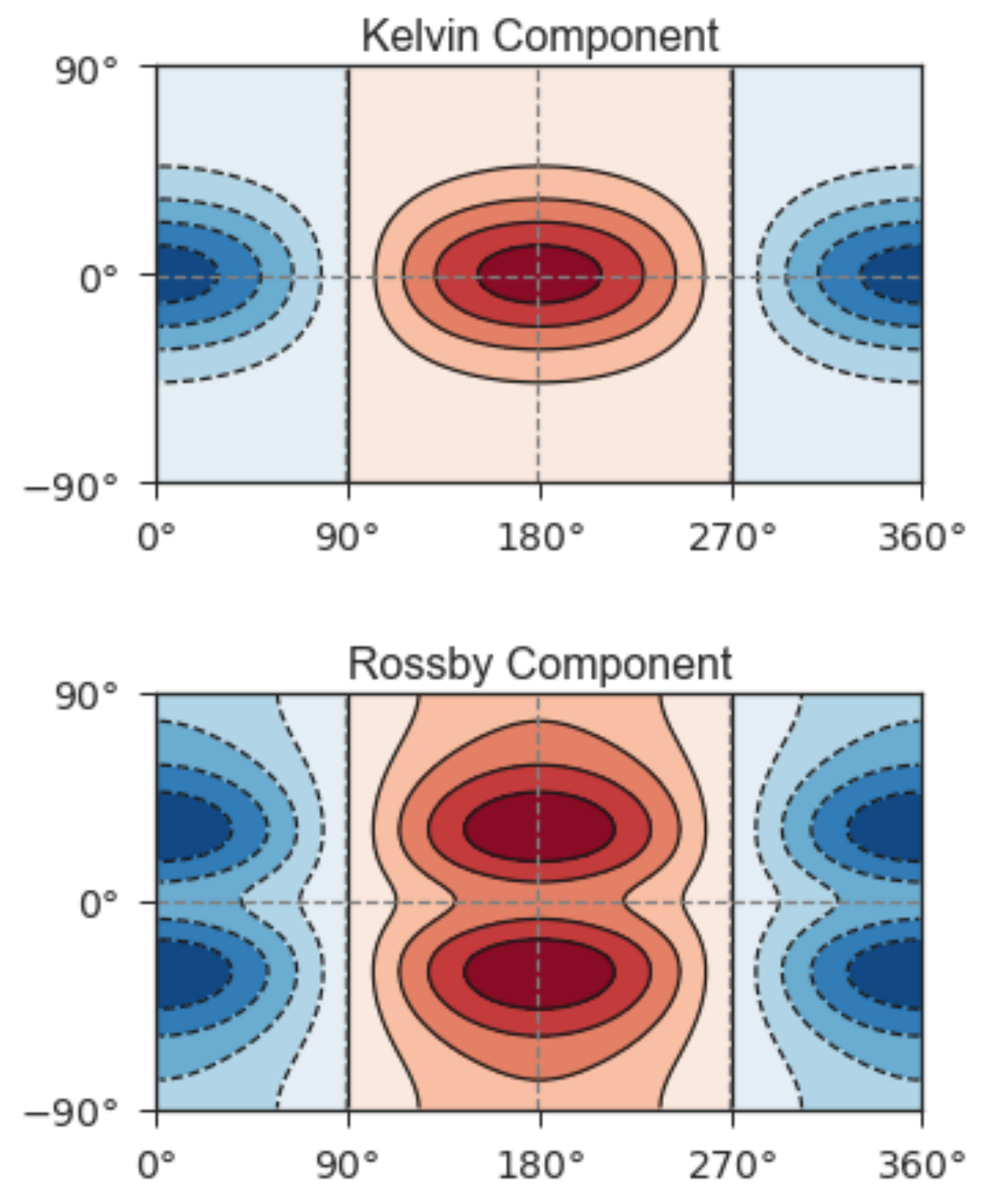}{0.33\textwidth}{0 days.}
\fig{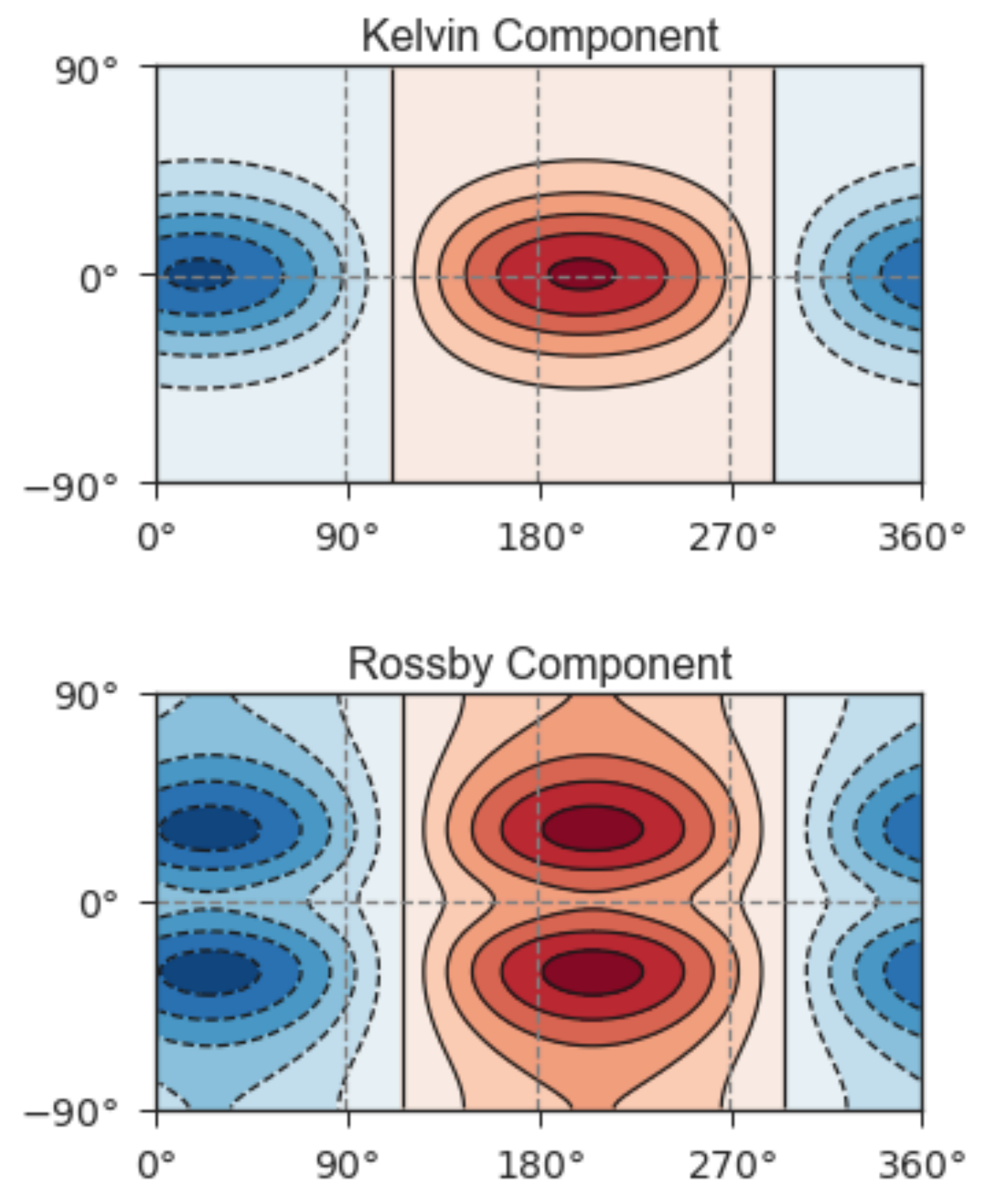}{0.33\textwidth}{30 days.}
\fig{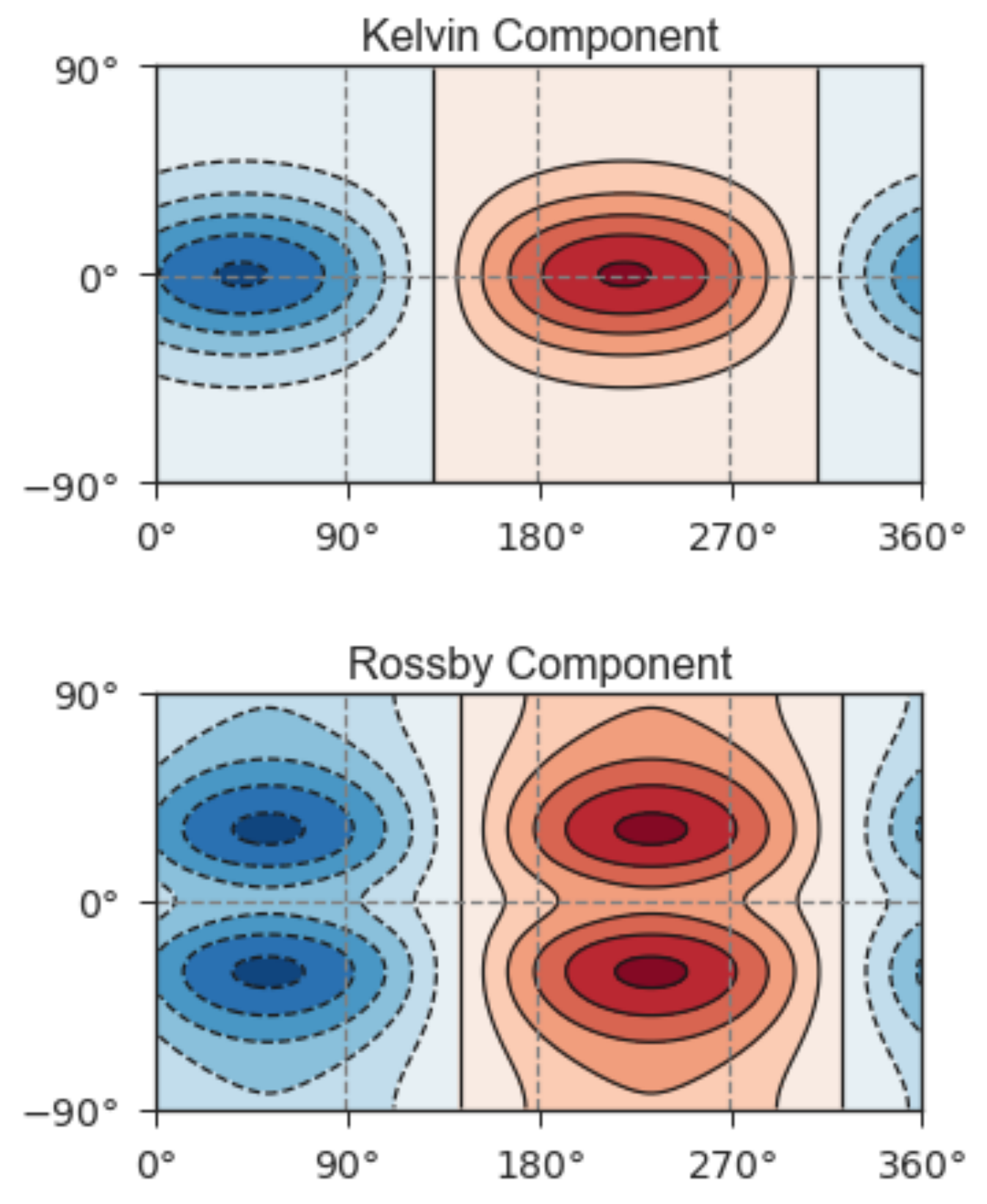}{0.33\textwidth}{60 days.}
}

 \caption{The Rossby and Kelvin components of the height field at half-surface pressure as the GCM is spun-up from rest and uniform temperature.}\label{fig:spin-up-snapshots}
\end{figure*}

Figure \ref{fig:spin-up-snapshots} shows the Kelvin and Rossby components of the GCM height field at 0, 30, and 60 days after spin-up, during which time the mean zonal flow had accelerated from rest to a maximum of approximately $90 \mathrm{ms}^{-1}$ eastwards on the equator. The initial Kelvin and Rossby waves are centered at the substellar point, rather in the western hemisphere as might be expected for a Matsuno-type pattern in zero background flow. We suggest this is because the radiative and dynamical damping is strong enough in the GCM that the Kelvin and Rossby waves are mostly in phase with the stellar forcing in the absence of background flow. This is shown by Figure 3 of \citet{showman2011superrotation}, where the strongly damped Matsuno-type solutions have their maxima close to the substellar point. Figure 10 of \citet{showman2011superrotation} is a snapshot of their GCM during its spin-up phase before the jet forms, which also has a Rossby wave maximum at the substellar point, rather than to its west.

Although the initial position of the waves is not exactly as expected, Figure \ref{fig:spin-up-snapshots} shows that as the jet forms, the Rossby and Kelvin waves shift eastwards and reach an equilibrium at about towards $+ 60 \degr$ east. This matches the prediction of our linear shallow-water model.

\subsection{Global Circulation Regimes}

\begin{figure*}
 \gridline{\fig{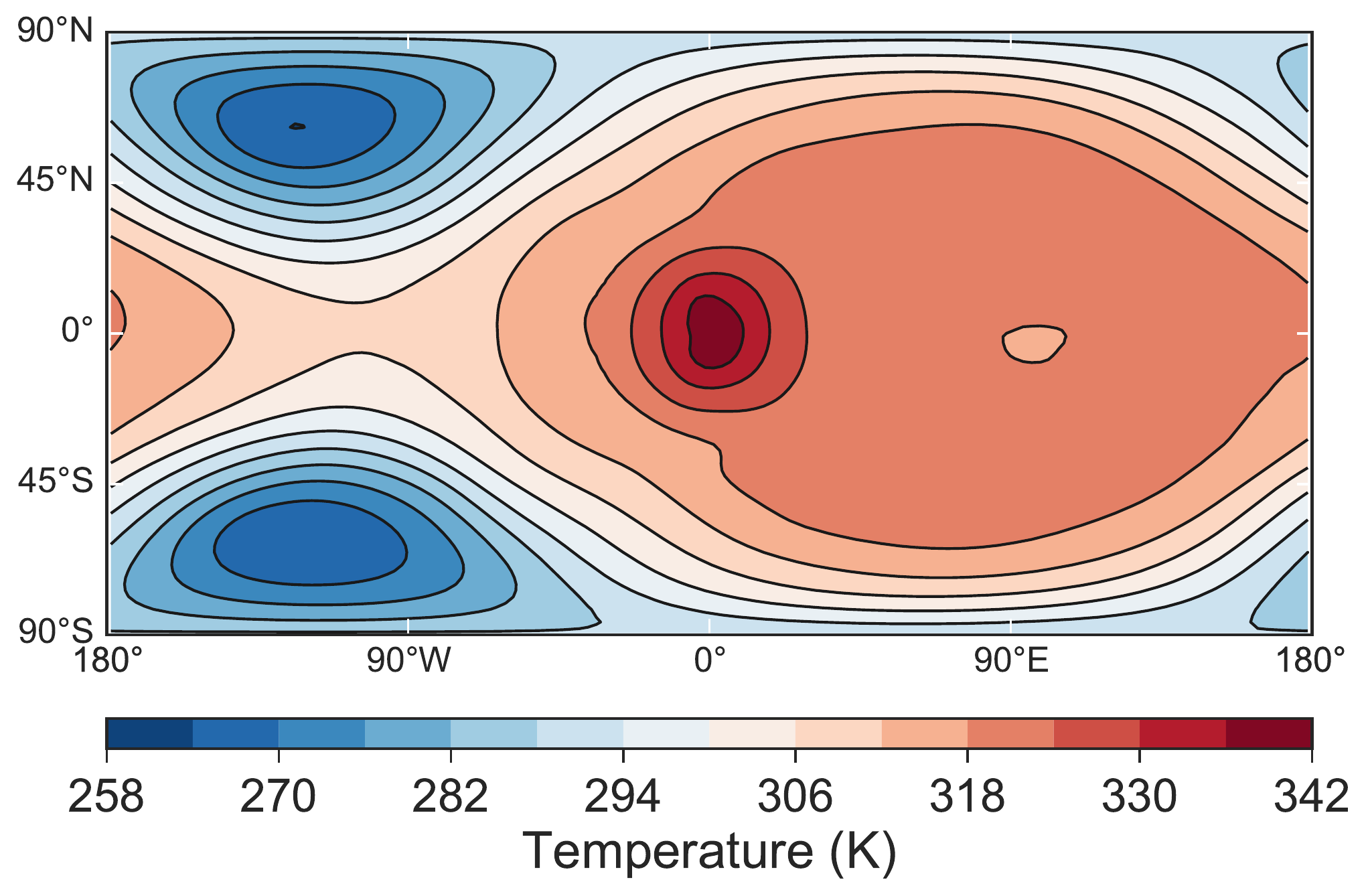}{0.33\textwidth}{10 day period.}
 \fig{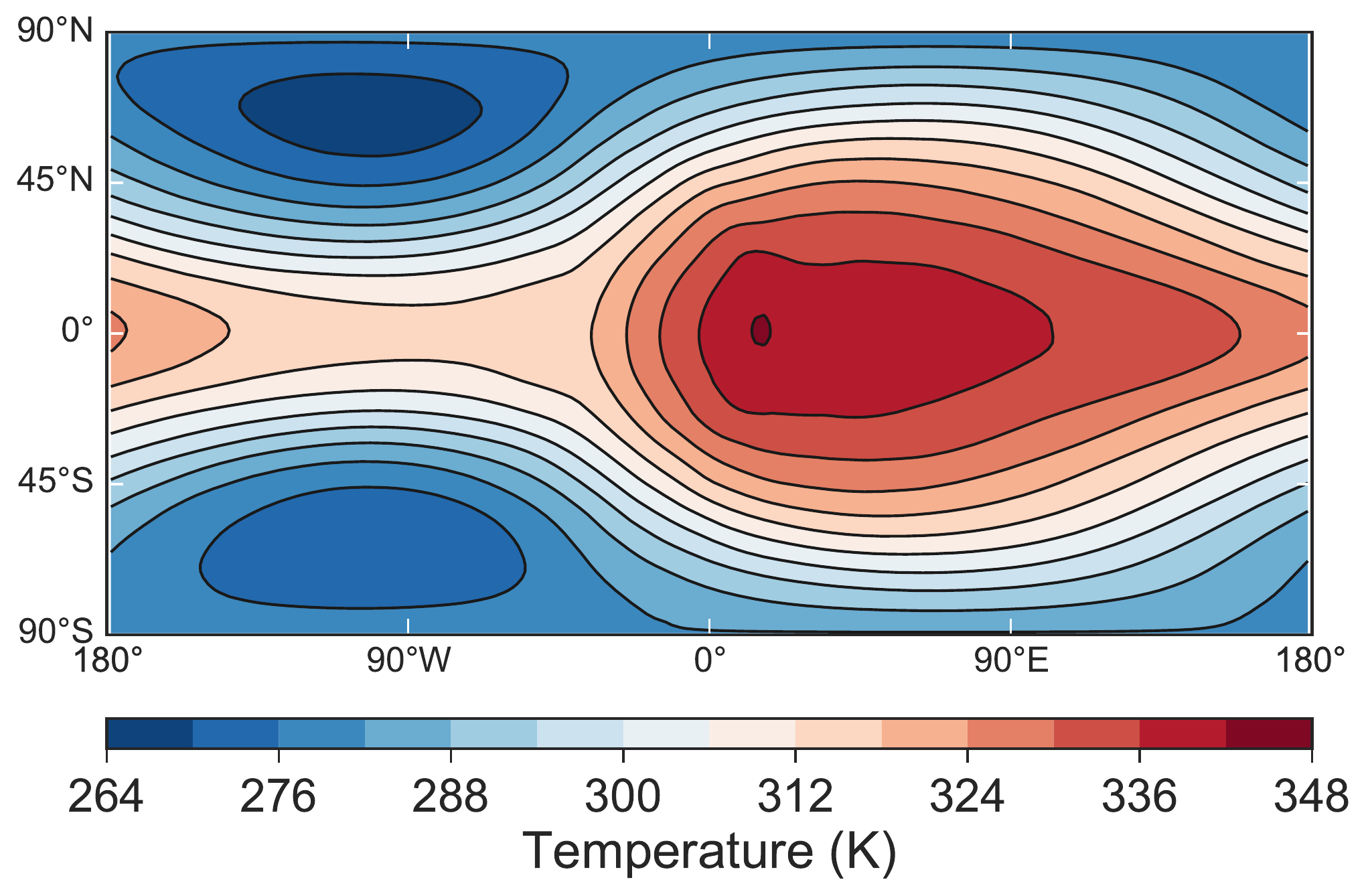}{0.33\textwidth}{5 day period.}
 \fig{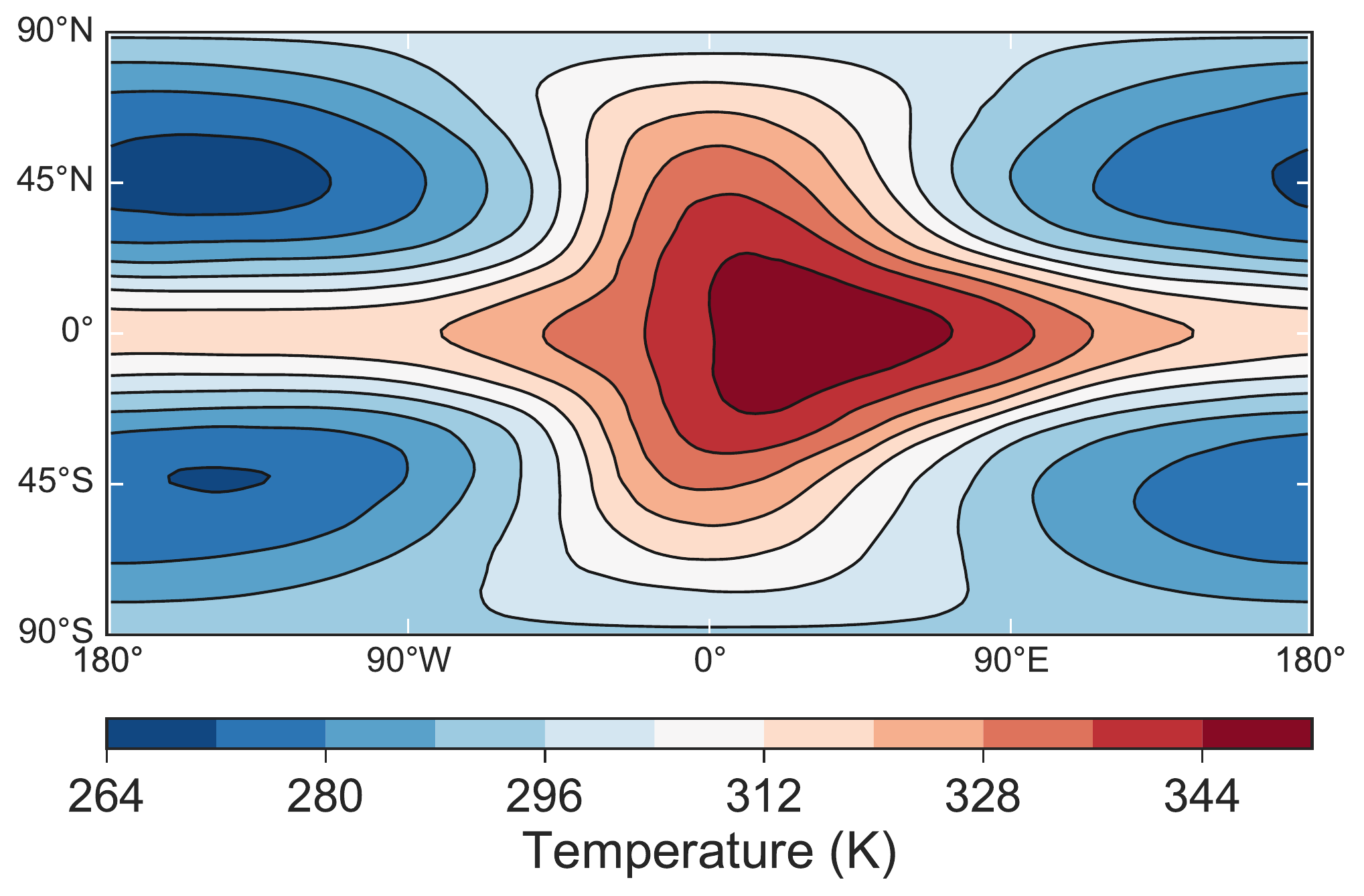}{0.33\textwidth}{2 day period.}
           }
 \caption{Global mid-atmosphere temperature distributions for GCM tests of an Earth-like tidally locked planet, showing how rotation rate affects the global circulation.\label{fig:exofms-regimes-rotation}}
\end{figure*}

\begin{figure*}
 \gridline{\fig{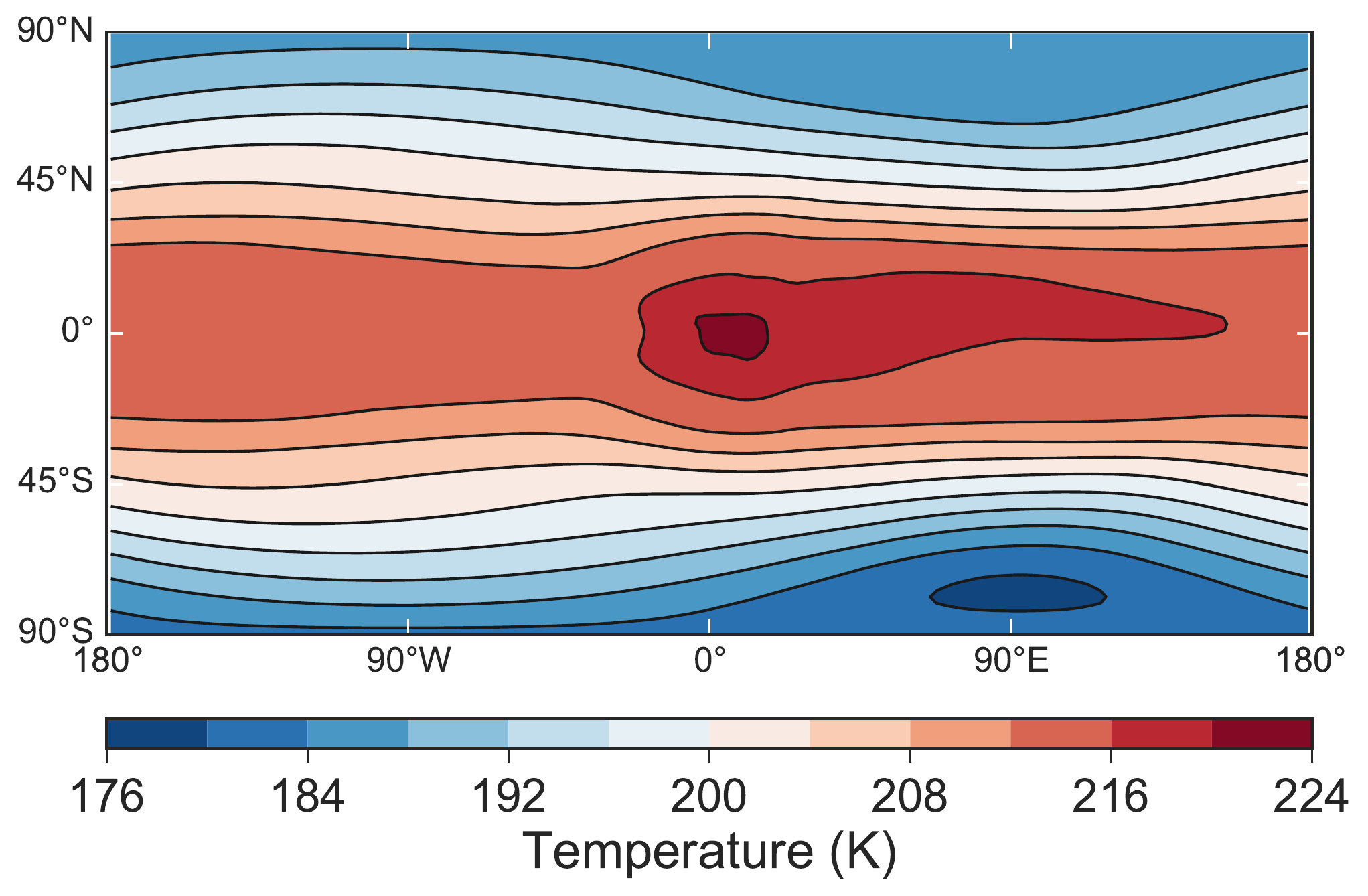}{0.33\textwidth}{Low temperature.}
 \fig{gcm-med-5.pdf}{0.33\textwidth}{Medium temperature.}
 \fig{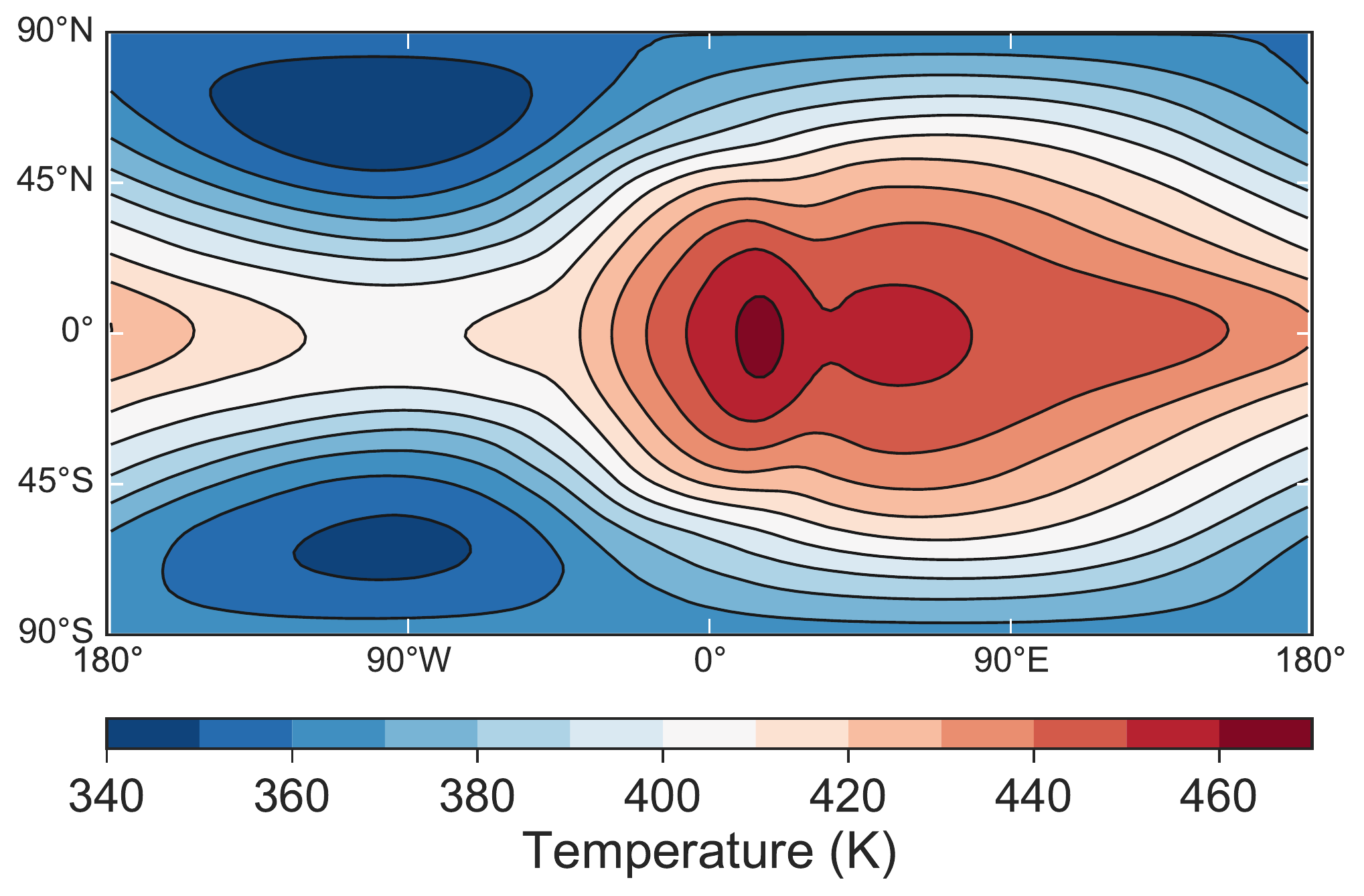}{0.33\textwidth}{High temperature.}
           }
 \caption{Global mid-atmosphere temperature distributions for GCM tests of an Earth-like tidally locked planet, showing how instellation affects the global circulation.\label{fig:exofms-regimes-forcing}}
\end{figure*}

Figures \ref{fig:exofms-regimes-rotation} and \ref{fig:exofms-regimes-forcing} show the mean mid-atmosphere temperature of a suite of GCM tests with variable stellar forcing and planetary rotation rate. Studies such as \citet{showman2015circulation}, \citet{carone2015regimes}, \citet{haqqmisra2018regimes}, and \citet{pierrehumbert2018review} varied the parameters of simulated tidally locked planets and identified changes in the global circulation.

We suggest that we can qualitatively explain these circulation patterns with our linear shallow-water solutions, using the ideas discussed in Section \ref{sec:scaling-relations}. Figure \ref{fig:exofms-regimes-rotation} shows the effect of rotation rate. At high rotation rates, the jet height (and temperature perturbation) dominates as it is proportional to $\Omega$ via $G$, as shown in Figure \ref{fig:g-effect}. This gives a more zonally homogeneous circulation. At high jet speed the wave response will be more Doppler-shifted and the hot-spot shift will be larger. This gives a smaller day-night contrast and larger hot-spot shift.

Second, the effect of stellar forcing in Figure \ref{fig:exofms-regimes-forcing}. At high forcing (instellation) the wave response dominates as shown in Figure \ref{fig:sp-q-effect}, so the wave response and the day-night contrast are large. The temperature is also higher, so the radiative damping is stronger and the wave response moves in phase with the forcing (see Section \ref{sec:sw-equations}). This leads to a smaller hot-spot shift as well as the large day-night contrast -- i.e. a zonally inhomogeneous circulation in phase with the stellar forcing, with strong temperature gradients.

These results match the scaling relations of the 1D and 2D models discussed in Section \ref{sec:scaling-relations}. There, we predicted that planets with higher damping $\alpha$ would have a smaller hot-spot shift and larger day-night contrast. This is the case in Figure \ref{fig:exofms-regimes-forcing}, where the hotter planet has a higher radiative damping rate. We also predicted that planets with a higher rotation rate should have a more zonally homogeneous circulation pattern, which we see in Figure \ref{fig:exofms-regimes-rotation}.

These results are similar to the work of \citet{komacek2016daynightI}, \citet{komacek2017daynightII}, and \citet{zhang2017dynamics}, who predicted scaling relations for observables such as hot-spot shift and day-night contrast, based on balancing jet transport against radiative and dynamical damping. Our wave-based approach to the circulation predicts some of the same scaling relations for different reasons. \citet{zhang2017dynamics} predicted that an atmosphere with a low heat capacity and short radiative timescale has a large day-night contrast and small hot-spot shift, which makes sense as the hot air transported in their jet model will cool quickly on the night-side.

We make the same prediction, but for a different reason. A short radiative timescale $t_{rad}$ corresponds to a strong damping term $\alpha_{rad} = 1 / t_{rad}$. Figure \ref{fig:sp-alpha-effect} shows how this weakens the forced wave response and brings the response in phase with the forcing (as discussed in Section \ref{sec:sw-equations}), reducing the hot-spot shift. For a long radiative timescale and weak radiative damping $\alpha_{rad}$, the Doppler-shift due to the jet will dominate the effect of the forcing in Equation \ref{eqn:project-coeff-flow} and the hot-spot shift will be large.

Some tidally locked planets may not match the linear shallow-water model. Very slowly rotating planets are dominated by an overturning day-night circulation without an eastward jet. They may be limited by nonlinear constraints on geostrophic adjustment \citep{pierrehumbert2018review}, and the linear theory in this paper will not apply.

\section{Conclusions}\label{sec:conclusions}

We modelled the global circulation of a tidally locked planet using a shallow-water model linearized about an eastward equatorial jet and its associated height perturbation. This built on previous studies which linearized about a state of rest, or a uniform flow with uniform height. We found that the non-uniform flow and height were crucial to forming the global circulation pattern seen in GCM simulations.

We solved these shallow-water equations using a pseudo-spectral method, first on the beta-plane for simplicity, and then on a sphere to properly represent the Coriolis parameter and the latitude coordinate. Both models showed that the shear flow shifted the wave response eastwards, explaining the hot-spot shift seen in simulations. They also showed that the non-uniform height perturbation of the jet was crucial to the overall form of the global circulation. We varied the parameters of the model such as stellar forcing and planetary rotation rate, to show how they affect the scaling of observables such as hot-spot shift and day-night contrast. We showed how these scaling relations match both a simpler one-dimensional model, and the results of simulations in a GCM.

In this study, we only considered planets that orbit their host star closely and are exactly tidally locked. Thermal tides could prevent such a planet from reaching an exactly tidally locked state. \citet{penn2017phasecurve} used a shallow-water model to investigate planets which are rotating slightly faster or slower than if they were tidally locked, and showed that this affects the position of the hot-spot. It will be important to consider other possible orbital and climate states for any observed exoplanets, rather than assuming tidal locking.

Further work could build on this linear model by adding the effect of vertical structure \citep{tsai2014three} or vertical shear \citep{boyd1978shearI}. Instabilities seen in the GCM simulations could be investigated further by recreating them in a shallow-water model and finding how they scale with planetary properties.

In summary, this paper described a linear shallow-water model of the atmospheric circulation of a tidally locked planet. It included the shear and height perturbation of the eastward equatorial jet which forms on such planets. It matches the results from 3D general circulation models better than previous shallow-water models, especially the position of their shifted hot-spot and night-side cold spots.

\acknowledgments
We thank the anonymous reviewer for their time and valuable comments, which greatly improved the paper. M.H. was supported by an S.T.F.C. studentship. This work also received support from European Research Council project EXOCONDENSE.

\appendix

\section{Pseudo-spectral Method}\label{sec:app-ps-method}

In this appendix, we discuss how we solved the linearized shallow-water equations using a pseudo-spectral collocation method \citep{boyd2000spectral}. Defining a linear ordinary differential equation or system of equations:

\begin{equation}
  L u = q
\end{equation}

$L$ is a differential operator acting on the variable $u$, and $q$ is the forcing or eigenvalue term. The solution is written as a sum of a series of basis functions:

\begin{equation}\label{eqn:pseudospectral_sum}
  u(x) = \Sigma a_{n} \psi_{n}(x)
\end{equation}

For a system of equations rather than a single equation, $L$ is a matrix and $u$ and $q$ are vectors. We impose the condition that the differential equation is satisfied at $N$ ``collocation points'', the positions of which depend on the set of basis functions.

This is equivalent to specifying that the ``residual'' -- the difference between the exact solution and the pseudo-spectral series solution -- is zero at these points. This provides $N$ equations to solve for the $N$ unknowns $a_{n}$, which gives the matrix equation:

\begin{equation}\label{eqn:ps_matrix}
  \textbf{H} \textbf{a} = \textbf{f}
\end{equation}

\subsection{Solving One Equation}

 \citet{boyd1978shearI} solves the linearized shallow-water equations, by reducing them to a single equation for a single variable, and applying the pseudo-spectral method. In this paper, we solve the entire system of shallow-water equations at once with the method in Appendix \ref{sec:app-systems}, but explain the method for a single equation here as it naturally leads to the second method \citep{boyd2000spectral}.

The matrix elements $H_{ij}$ in equation \ref{eqn:ps_matrix} are evaluated using the operator $L$ at the collocation points $x_{i}$ and for every mode $\phi_{j}$, and the vector elements $f_{i}$ are the terms $q$ evaluated at the collocation points $x_{i}$:

\begin{equation}\label{eqn:ps_H}
  H_{ij} = L \phi_{j}(x_{i})
\end{equation}

\begin{equation}
  f_{i} = q(x_{i})
\end{equation}

This is then solved using a standard linear algebra routine to find $a_{n}$, and the solution $u(x)$ is reconstructed using Equation \ref{eqn:pseudospectral_sum}.

\subsection{Solving Systems of Equations}\label{sec:app-systems}

The pseudo-spectral method can also be applied to systems of linear ordinary differential equations. For a system of forced, time-independent equations:

\begin{equation}
  \textbf{L} \textbf{u} = \textbf{q}
\end{equation}

The condition that the differential equation is satisfied at the collocation points gives the equivalent matrix equation to Equation \ref{eqn:ps_matrix}:

\begin{equation}
  \textbf{H} \textbf{a} = \textbf{f}
\end{equation}

$\textbf{H}$ is an $M \times N$ square matrix with elements:

\begin{equation}
  H^{kl}_{ij} = L^{kl}\phi_{j}(x_{i})
\end{equation}

i.e. the operator $L^{kl}$ which acts on the $l$th variable in the $k$th equation, applied to the $j$th basis function and evaluated at the $i$th collocation point. $\textbf{f}$ is a vector made up of $N$ subvectors $f_{i}$, which are the forcing terms in each equation evaluated at each collocation point.

\begin{equation}
    \textbf{H} =
  \begin{pmatrix}
    \begin{pmatrix}
H_{ij} & \dots \\
\vdots & \ddots
    \end{pmatrix}^{kl} & \dots \\
  \vdots & \ddots
  \end{pmatrix}
  \begin{pmatrix}
    \begin{pmatrix}
    \alpha_{i} \\
    \vdots
    \end{pmatrix} \\
  \vdots
  \end{pmatrix}
  =
  \begin{pmatrix}
    \begin{pmatrix}
    f_{i} \\
    \vdots
    \end{pmatrix} \\
  \vdots
  \end{pmatrix}
\end{equation}

$\textbf{H}$ is the same as the matrix in Equation \ref{eqn:ps_H} with the elements $H_{ij}$ replaced by submatrices $H^{kl}_{ij}$. Solving this system returns the coefficients of the basis functions, and the solutions are:

\begin{equation}\label{eqn:ps-coeff-solutions}
 u(y) = \sum_{n=0}^{N} a_{n} \phi_{n}
; \quad
 v(y) = \sum_{n=0}^{N} b_{n} \phi_{n}
; \quad
 h(y) = \sum_{n=0}^{N} c_{n} \phi_{n}
\end{equation}

This gives a linear matrix equation with one solution corresponding to the coefficient vectors $a_{n}$, $b_{n}$, $c_{n}$ of the forced solution.

Without forcing, the shallow-water equations define an eigensystem where the eigenvalue is the frequency $\omega$.

\begin{equation}
  \textbf{L} \textbf{u} = \omega \textbf{P} \textbf{u}
\end{equation}

The pseudo-spectral equation is then:

\begin{equation}
  \textbf{H} \textbf{a} = \omega \textbf{R} \textbf{a}
\end{equation}

 $\textbf{R}$ is an $M \times N$ square matrix with elements:

\begin{equation}
  R^{kl}_{ij} = P^{kl}\phi_{j}(x_{i})
\end{equation}

i.e. the eigenvalue operator $P^{kl}$ acting on the $l$th variable in the $k$th equation, applied to the $j$th basis function and evaluated at the $i$th collocation point.

\begin{equation}
    \textbf{H} =
  \begin{pmatrix}
    \begin{pmatrix}
H_{ij} & \dots \\
\vdots & \ddots
    \end{pmatrix}^{kl} & \dots \\
  \vdots & \ddots
  \end{pmatrix}
  \begin{pmatrix}
    \begin{pmatrix}
    \alpha_{i} \\
    \vdots
    \end{pmatrix} \\
  \vdots
  \end{pmatrix}
  =
  \omega
  \begin{pmatrix}
    \begin{pmatrix}
R_{ij} & \dots \\
\vdots & \ddots
    \end{pmatrix}^{kl} & \dots \\
  \vdots & \ddots
  \end{pmatrix}
  \begin{pmatrix}
    \begin{pmatrix}
    \alpha_{i} \\
    \vdots
    \end{pmatrix} \\
  \vdots
  \end{pmatrix}
\end{equation}

This gives an eigenvalue matrix equation, with $N$ eigenvalues and eigenvectors, corresponding to the frequencies and coefficient vectors $a_{n}$, $b_{n}$, $c_{n}$ for each free mode. Not all $N$ modes must be physically realistic, so we identify the spurious modes by inspecting the eigenvalues for different values of $N$.

\begin{figure*}
 \gridline{\fig{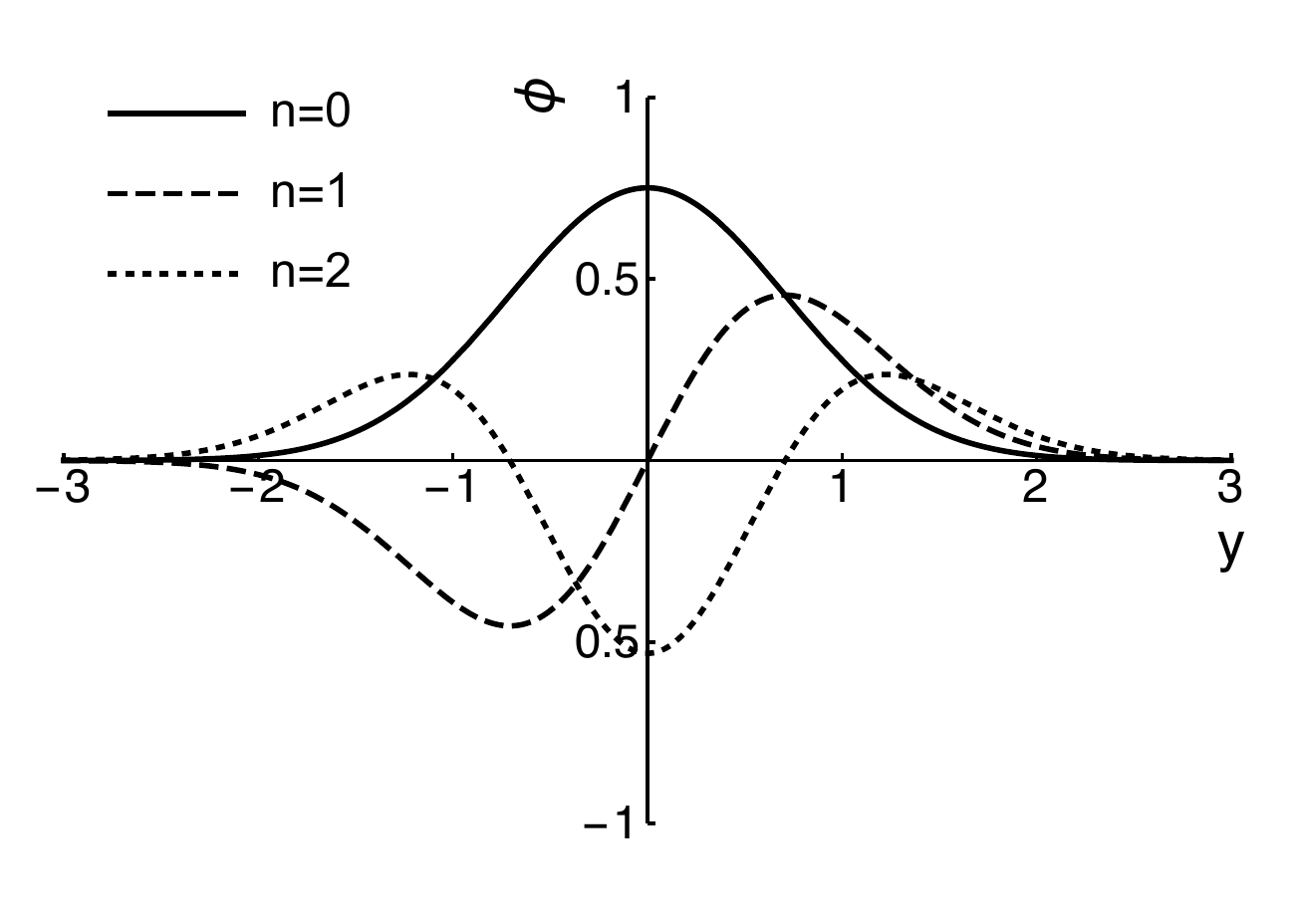}{0.4\textwidth}{Parabolic Cylinder Functions.}\label{fig:hermite-functions}
 \fig{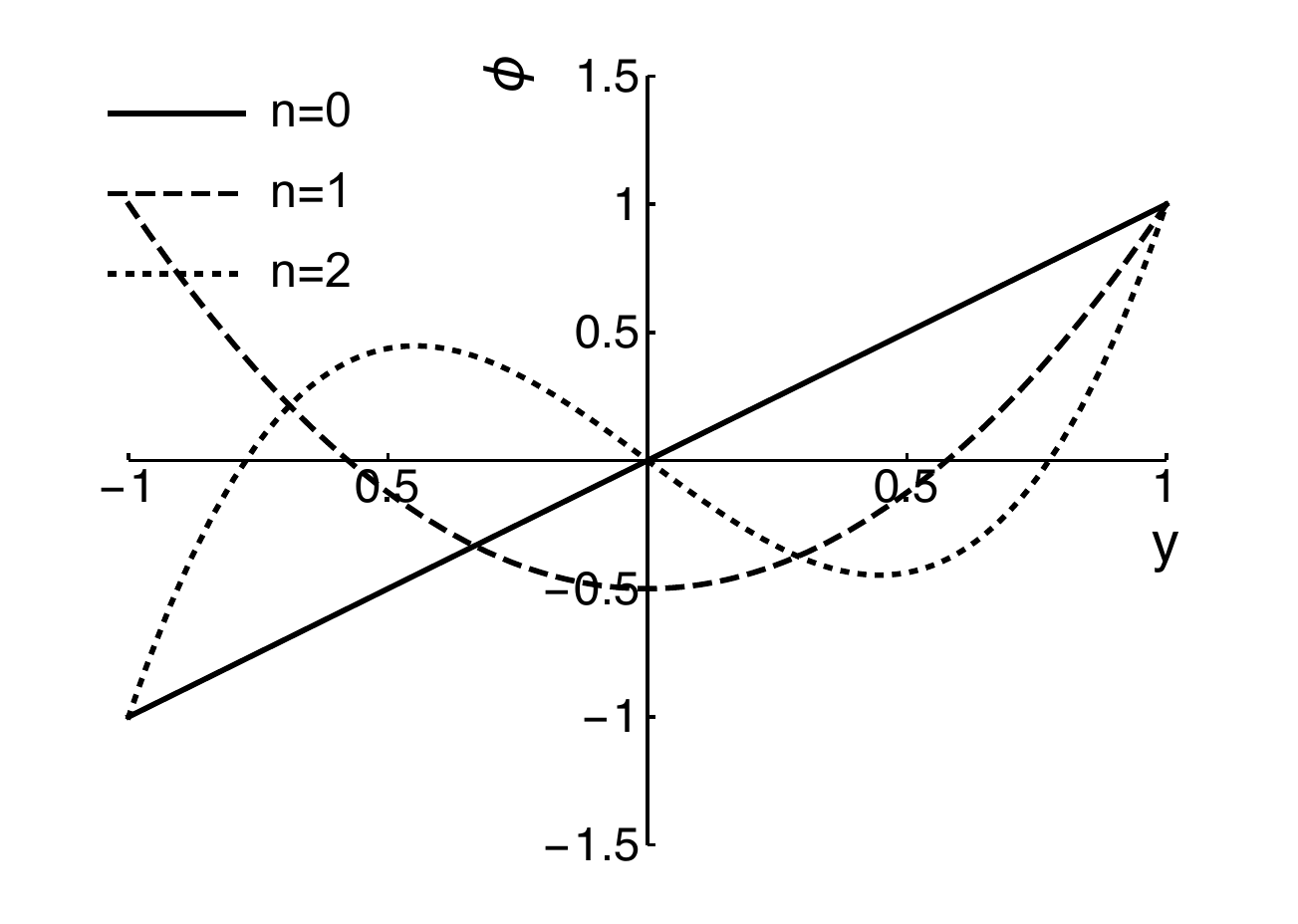}{0.4\textwidth}{Legendre Polynomials.}\label{fig:legendre-polynomials}
           }
\caption{Basis functions used in beta-plane and spherical coordinates.}\label{fig:basis-functions}
\end{figure*}

\subsection{Beta-plane solutions}\label{sec:app-beta}

We use the parabolic cylinder functions $\psi_{n}(y)$ \citep{showman2011superrotation} as defined in Equation \ref{eqn:hermite-functions} as a basis set for the pseudo-spectral method on the beta-plane (Equation \ref{eqn:forced-sw}), as they are the exact free solutions of \citet{matsuno1966quasi} \citep{boyd2000spectral}.

Their collocation points are at their zeros (which are just the zeros of the Hermite polynomials $H_{n}$). Figure \ref{fig:hermite-functions} shows the first few parabolic cylinder functions.

\begin{equation}\label{eqn:hermite-functions}
  \psi_{n}(y) = e^{-y^{2} / 2} H_{n}(y)
\end{equation}

\begin{figure*}
 \gridline{\fig{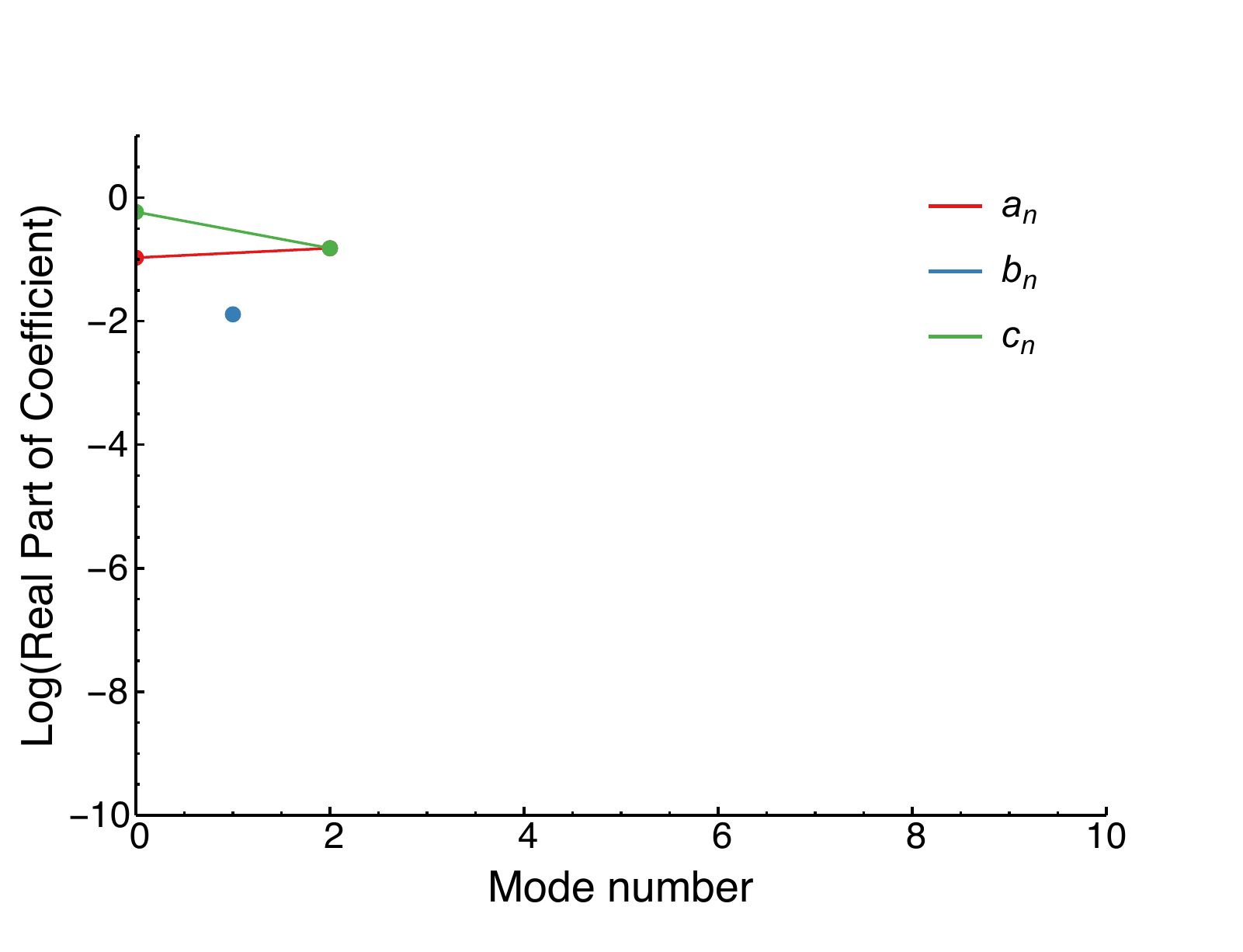}{0.4\textwidth}{Coefficients for $\bar{U}(y)=0$.}
 \fig{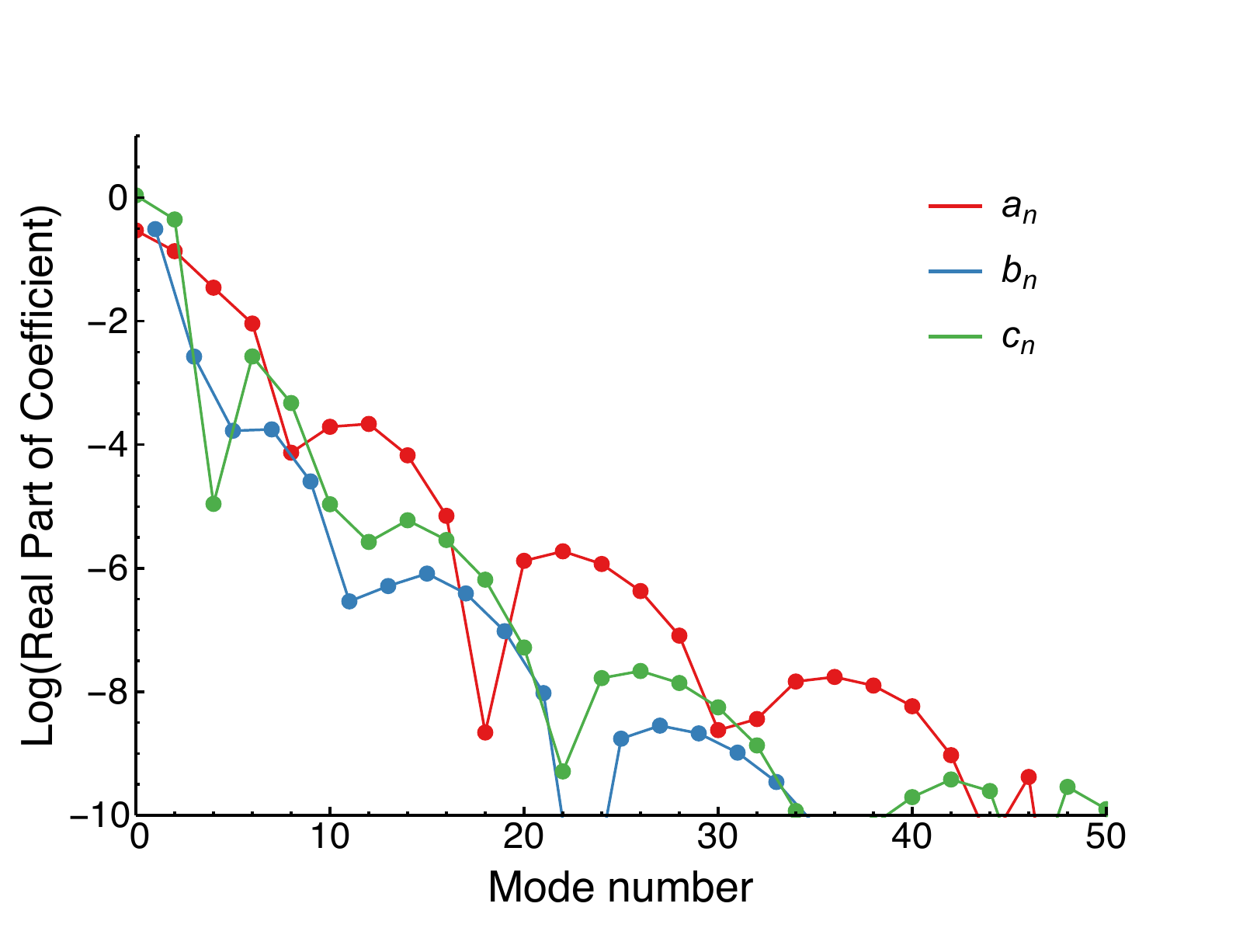}{0.4\textwidth}{Coefficients for $\bar{U}(y)=\exp(-y^{2}/2)$.}
           }
\caption{Coefficients of the pseudo-spectral solution on the beta-plane coordinates with and without a background jet (the plots in Figure \ref{fig:shear-2D}). The method identifies the exact solution in the first case, and converges rapidly to an accurate solution in the second case. }\label{fig:accuracy-hermite}
\end{figure*}

Figure \ref{fig:accuracy-hermite} shows the magnitude of the coefficients (Equation \ref{eqn:ps-coeff-solutions}) of the pseudo-spectral solution of the shallow-water equations linearized about a jet on a beta-plane (plotted in Figure \ref{fig:shear-2D}). The first plot shows that when the background jet flow is zero, only modes up to $n=2$ are non-zero. This is the analytic solution from \citet{matsuno1966quasi}, which the pseudo-spectral method identifies because we have used the free modes (the parabolic cylinder functions) as our basis functions.

For non-zero jet speed (corresponding to Figure \ref{fig:shear-2D}), the pseudo-spectral series solution does not terminate, but the coefficients for the 30th mode are about eight orders of magnitude smaller than the largest mode. The beta-plane solutions in this paper were all calculated with at least 30 modes.

\subsection{Spherical solutions}\label{sec:app-spherical}

We use the Legendre polynomials as a basis set for the pseudo-spectral method in a spherical geometry (Equation \ref{eqn:sphere-sw-eqns}). Figure \ref{fig:legendre-polynomials} shows the first few Legendre polynomials. Our collocation points are the zeros of these functions.

As discussed in Section \ref{sec:sphere-solutions}, Equation \ref{eqn:sphere-sw-eqns} has a singularity at the the poles, which we avoided by using a rescaled height $\gamma$, where $\gamma = h / \cos\phi$ \citep{iga2005spherical}. We replaced $h$ with $\gamma \cos \phi$ in Equation \ref{eqn:sphere-sw-eqns}, solved as normal, then multiplied the solution for $\gamma$ by $\cos\phi$ to recover the solution for $h$.

\begin{figure*}
 \gridline{\fig{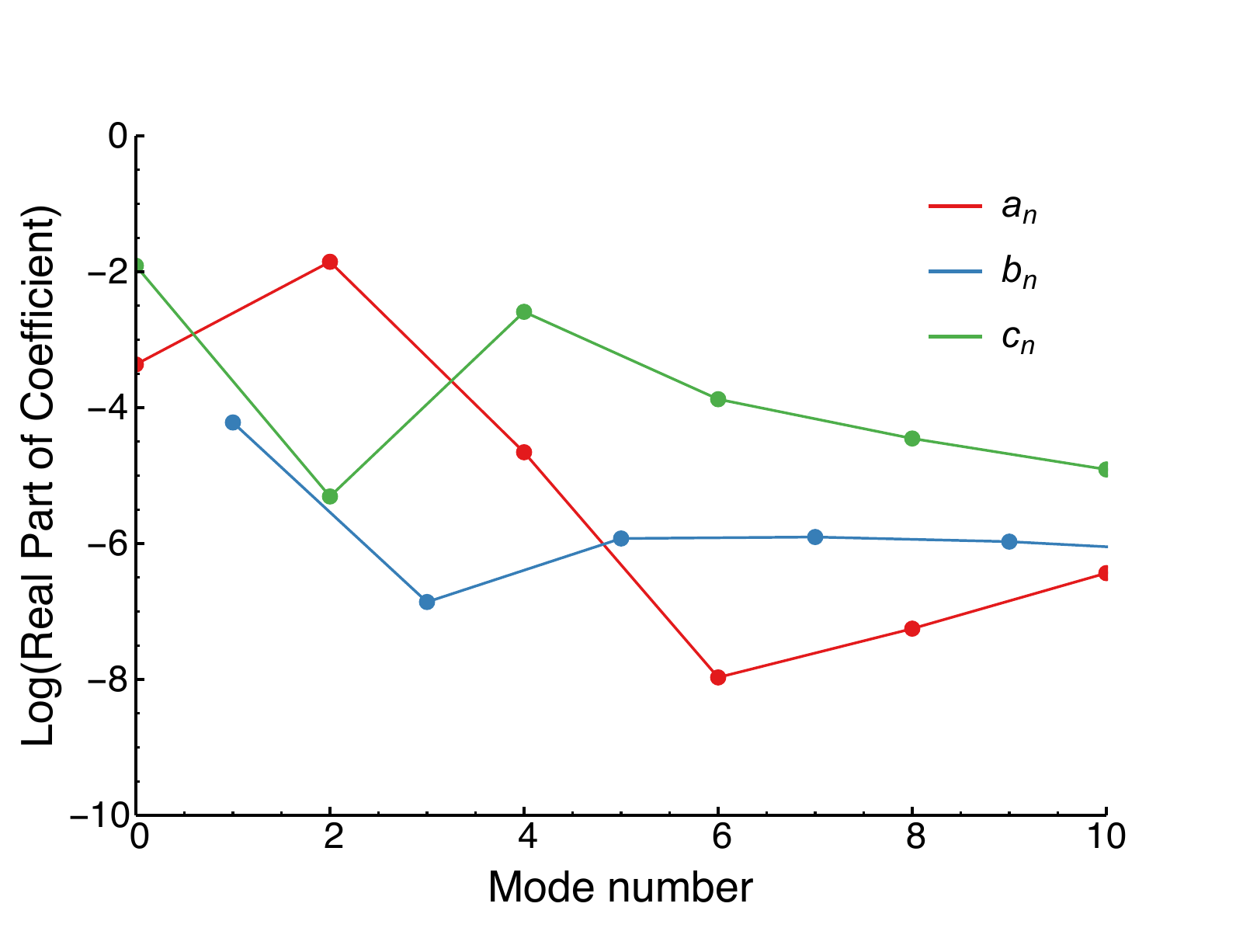}{0.4\textwidth}{Coefficients calculated with height $h$.}
 \fig{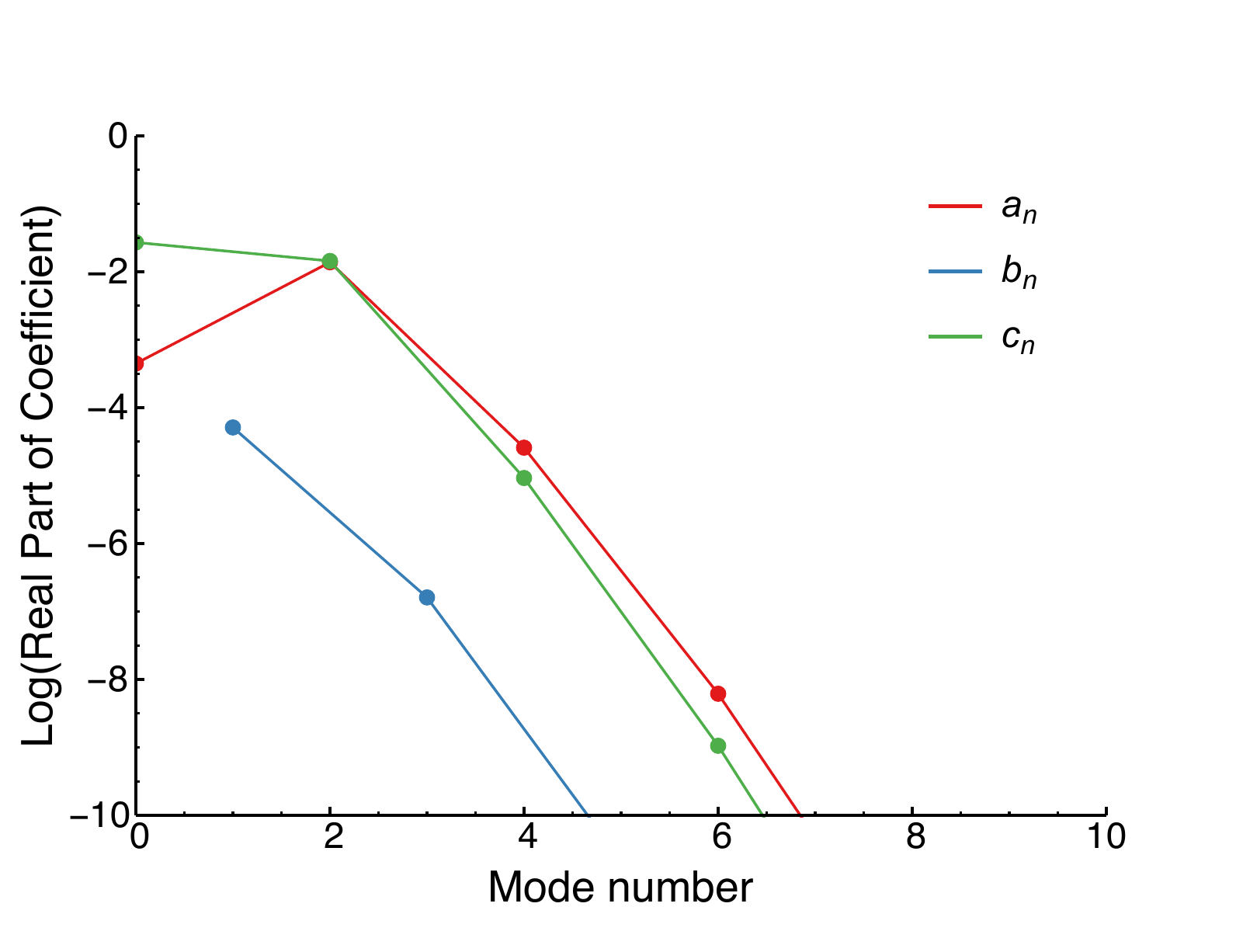}{0.4\textwidth}{Coefficients calculated with rescaled height $\gamma = h / \cos\phi$.}
           }
\caption{Coefficients of the pseudo-spectral solution in spherical coordinates (the first plot in Figure \ref{fig:spherical-tests}), with the height variable $h$ and the rescaled height $\gamma$. Rescaling the height makes the method converge to a smooth solution at the poles.}\label{fig:accuracy-spherical}
\end{figure*}

Figure \ref{fig:accuracy-spherical} shows how rescaling the $h$ variable made the solutions converge much more quickly. In fact, the solutions without a rescaled $h$ variable never reached a smooth solution at the poles.

\newpage
\bibliography{tl-dyn-refs.bib}

\end{document}